\documentclass[final]{siamltex}

\usepackage{amssymb}
\usepackage{amsmath}
\usepackage{subfigure}
\usepackage{lineno}
\usepackage{graphicx}

\title{A Simple Finite Difference Method for Time-Dependent, Variable Coefficient Stokes Flow on Irregular Domains}

\begin{document}

\maketitle

\begin{abstract}
We present a simple and efficient variational finite difference method for simulating time-dependent Stokes flow in the presence of irregular free surfaces and moving solid boundaries. The method uses an embedded boundary approach on staggered Cartesian grids, avoiding the need for expensive remeshing operations, and can be applied to flows in both two and three dimensions. It uses fully implicit backwards Euler integration to provide stability and supports spatially varying density and viscosity, while requiring the solution of just a single sparse, symmetric positive-definite linear system per time step. By expressing the problem in a variational form, challenging irregular domains are supported implicitly through the use of natural boundary conditions. In practice, the discretization requires only centred finite difference stencils and per-cell volume fractions, and is straightforward to implement. The variational form further permits generalizations to coupling other mechanics, all the while reducing to a sparse symmetric positive definite matrix. We demonstrate consistent first order convergence of velocity in $L^1$ and $L^{\infty}$ norms on a range of analytical test cases in two dimensions. Furthermore, we apply our method as part of a simple Navier-Stokes solver to illustrate that it can reproduce the characteristic jet buckling phenomenon of Newtonian liquids at moderate viscosities, in both two and three dimensions.
\end{abstract}



\pagestyle{myheadings}
\thispagestyle{plain}
\markboth{FINITE DIFFERENCE STOKES FLOW ON IRREGULAR DOMAINS}{FINITE DIFFERENCE STOKES FLOW ON IRREGULAR DOMAINS}

\section{Introduction}
\label{introduction}

The equations of Stokes flow describe the motion of fluids where the nonlinear advection terms present in the Navier-Stokes equations have been eliminated or ignored. This approximation is important in many physical scenarios where inertia is essentially negligible ($Re \ll 1$), however it typically leads to the \emph{steady-state} Stokes equations. In this paper we are instead interested in the \emph{time-dependent} or unsteady case, motivated by its use as a building block for solving the Navier-Stokes equations. That is, a number of methods apply operator splitting to the Navier-Stokes equations in order to treat the advective terms separately from the time-dependent Stokes equations.

The goal of this paper is to derive and validate a simple, efficient, and stable method for solving the time-dependent Stokes flow equations in the presence of irregular free surfaces and solid boundaries, while supporting spatially varying viscosity and density coefficients. We solve these equations on the classic staggered Cartesian grid using a finite difference approach. However, in order to support irregular domains which arise frequently in the presence of evolving liquid interfaces and moving objects, we must determine a proper discretization for non-grid-aligned (ie., embedded) boundary conditions. The free surface condition poses a particular challenge because it involves a delicate coupling between the boundary normal, the pressure, and the components of the deviatoric stress tensor. Our approach will be to take inspiration from the finite element method, and express the Stokes flow problem in a variational form relating velocity, pressure, and deviatoric stress. We will show that this yields a hybrid approach that is discretized with simple finite differences, but which implicitly enforces complex embedded boundaries through the natural boundary conditions of the problem. We believe this to be the first method that can provide a fully implicit discretization of this problem on Cartesian grids, while simultaneously yielding a symmetric positive-definite linear system and correctly handling both variable coefficients and the difficult free surface case.

\section{Background}

While there exist finite element and finite volume methods that can be applied to the Stokes equations on irregular domains, these methods require meshes comprised of well-shaped elements that align with the physical boundaries. Frequent remeshing becomes necessary in settings with rapidly evolving boundaries, and as others have noted, such remeshing is a challenging and expensive problem in its own right \cite{Bedrossian2010}. Unstructured meshes often also incur a performance penalty as compared to regular grids due to the overhead costs of manipulating more complex data structures. We therefore pursue the simpler and more efficient embedded finite difference approach, allowing boundaries to cut arbitrarily through an underlying regular Cartesian grid.

Viscous free surface flows have long posed challenges for finite difference methods. The core problem is ensuring that the region external to the liquid applies zero force on the surface itself through the application of appropriate boundary conditions. The seminal marker-and-cell (MAC) paper of Harlow and Welch \cite{Harlow1965} discussed these difficulties, and dispensed with the free surface viscous stress conditions for simplicity. However, they noted that at lower Reynolds numbers the more accurate condition becomes necessary. Hirt and Shannon proposed a simple correction to the normal stress \cite{Hirt1968}, which Nichols and Hirt subsequently improved to address the tangential stress \cite{Nichols1971}. Because an explicit discretization of viscous terms suffers from a stringent time step restriction of $\Delta t < O(\rho \Delta x^2 / \mu)$ that becomes problematic at high viscosities, Pracht incorporated the preceding ideas into a fully implicit integration scheme to improve stability \cite{Pracht1971}. All of these schemes assume that the surface normal is either aligned with coordinate axes, or at a $45^{\circ}$ angle, in order to explicitly design conditions for these special cases. This effectively rasterizes or voxelizes the domain into a ``stairstep" approximation that may not realistically reflect the physical geometry. This can limit the generality and accuracy of the approach, and the case-by-case analysis somewhat complicates the implementation. Nevertheless, Tom\'{e} et al.\ developed a three-dimensional extension that convincingly simulates a range of free surface phenomena \cite{Tome1994,Tome2004}, while using explicit time integration and a pressure projection method that splits the Stokes equations into separate pressure and viscous components (see \cite{Guermond2006} for an exploration of the issues raised by this type of splitting). Oishi et al.\ later implemented a more stable implicit time integration scheme \cite{Oishi2008} within the same framework to enable larger time steps and improve efficiency. Aside from our own work, this is the only other implicit finite difference method that has addressed the phenomenon of viscous jet buckling with proper free surface conditions. However, it requires the same case-by-case analysis as its predecessors, as well as the use of the less robust bi-conjugate gradient method to solve a large, sparse, non-symmetric linear system. A few authors have also advocated a semi-implicit treatment of spatially varying viscosity, in which terms that couple different velocity components are treated explicitly, and non-coupled terms are treated implicitly \cite{Li2000,Sussman2007,Stewart2008}.

In computer animation, staggered grid schemes have also been used extensively for free surface flows, again using a pressure projection approach \cite{Carlson2002,Falt2003,Rasmussen2004}. Recently, Batty and Bridson showed that improper viscous free surface conditions in these methods tend to destroy angular momentum \cite{Batty2008}. To address this, they proposed a fully implicit scheme for variable viscosity that yields symmetric positive-definite linear systems, and presented animations of rotating and buckling viscous liquids. While that method shares much with the present work, it differs in several key respects. First, it uses a pressure projection approach to split up the Stokes equations which entails weaker coupling between pressure and viscous terms. Secondly, it enforces a simplified free surface boundary condition that incorrectly neglects the interaction between pressure and viscous stresses at the surface. Thirdly, the present approach yields a linear system in terms of stresses, rather than velocities. Finally, the current work provides numerical experiments indicating that our scheme is indeed convergent.

There exists a broad family of methods that seek to accurately enforce boundary conditions of various kinds on irregular domains, while solving equations on Cartesian finite difference grids. We will collectively refer to this family, which includes the current work, as embedded boundary methods (although other authors have sometimes used the term immersed boundary methods \cite{Mittal2005}). As noted above, the motivation for embedded approaches is to avoid the substantial computational cost of generating and manipulating high quality, fully unstructured, boundary-conforming meshes as required by finite element or finite volume methods. Common examples of embedded boundary methods include the traditional immersed boundary methods (IBM) \cite{Peskin2002}, ghost fluid methods (GFM) \cite{Fedkiw1999}, immersed interface methods (IIM) \cite{Leveque1997}, matched interface and boundary methods (MIB) \cite{Zhou2005}, and cut-cell methods \cite{Johansen1998}. These have been quite effective for a number of problems and some can achieve higher order accuracy. Nevertheless, we are not aware of any symmetric fully implicit embedded boundary method with a comparably simple implementation that can address the variable coefficient unsteady Stokes equations with free surfaces. As an example, recent work that primarily emphasizes the use of the ghost-fluid approach for simulating turbulent atomization \cite{Desjardins2008}, Desjardins et al.\ revert to a diffuse interface method for viscous terms, stating that this choice was motivated by the complexity of viscous GFM discretizations \cite{Kang2000} and the difficulty of achieving an implicit formulation.

Among existing embedded boundary methods, the most closely related to our approach are staggered grid methods that use standard centred-difference stencils, scaled by simple diagonal weighting matrices to support irregular boundaries. These methods yield symmetric positive-definite linear systems, but have primarily been applied to Poisson and diffusion problems to date. First, a simple ghost fluid method has been used to enforce Dirichlet boundary conditions \cite{Chan1970,Gibou2002,Enright2003,Ng2009} for pressure projection methods and for implicit integration of spatially constant viscosity with solid boundaries. Secondly, a finite volume-like technique has been proposed for handling Neumann (and Robin) boundary conditions in the context of pressure projection methods for solid-fluid interaction \cite{Purvis1979,Roble2005,Ng2009,Papac2010}. Work by Batty et al.\ on solid-fluid coupling and viscous flows \cite{Batty2007,Batty2008} also falls into this category, although they are derived from variational principles rather than ghost fluid or finite volume concepts. The current work directly extends and unifies these last two methods, while providing numerical experiments that validate the approach.

Robinson-Mosher et al.\ have developed a related symmetric positive-definite formulation of time-dependent Stokes flow, focusing on solid-fluid coupling \cite{Robinson-Mosher2010}. However, their viscosity discretization is limited to voxelized (axis-aligned) solid geometry and constant viscosity, and does not address the free surface boundary condition we consider. Moreover, our derivation identifies their algebraic transformation as a change of variables from a velocity-pressure formulation to a pressure-stress formulation, lending physical insight to the method, and recovers a more fundamental variational form of the mechanics that offers improved specialized solvers.

There are of course a host of methods for treating the more common symmetric indefinite form of the Stokes equations for steady and unsteady problems \cite{Benzi2005}. Among these are several methods carefully designed to be optimal for the Stokes problem, in the sense that their convergence rates are independent of the size of the discrete simulation mesh, with ``appropriate" choices of preconditioners or smoothers \cite{Elman1998}. Nonetheless, such methods face additional challenges in addressing boundary conditions and variable coefficients, particularly for the case of irregular embedded boundaries. For example, Wan et al.\ applied substantial modifications to a geometric multigrid method to efficiently handle embedded interfaces for the simpler Poisson equation \cite{Wan2004}. Similarly, many fast solvers for the Stokes problem rely on a simplification (discussed in section \ref{simplerstokes}) that decouples the viscous contributions into three independent Poisson-type problems, enabling the use of existing fast Poisson solvers as sub-components. This simplification does not readily apply in the free surface or variable viscosity settings. We therefore prefer to pursue a symmetric positive-definite formulation for which effective black-box iterative solvers and preconditioners are more readily available. The development of special-purpose solvers and preconditioners for the positive-definite formulation is a potentially exciting research direction, which we defer to future work.

\section{Time-Dependent Stokes Flow}
The governing equations for time-dependent Stokes flow are:
\begin{eqnarray}
\rho \vec{u}_t & = & - \nabla p + \nabla\cdot\mathbf{\tau} \label{fullstokes1}\\
\nabla \cdot \vec{u} & = & 0 \label{fullstokes2}\\
\tau & = & 2 \mu\left(\frac{\nabla \vec{u} + (\nabla \vec{u})^T}{2}\right) \label{fullstokes3}
\end{eqnarray}
where $\rho$ is the fluid density, $\vec{u}$ is the fluid velocity, $p$ is the pressure, $\tau$ is the symmetric deviatoric stress tensor, $\mu$ is the dynamic viscosity coefficient, and the subscript $t$ indicates a time derivative.  We allow both $\rho$ and $\mu$ to vary spatially.  At solid boundaries, the no-slip condition applies, which dictates that the fluid velocities match those of the solid boundary, $\vec{u}_{BC}$:
\begin{equation}
\label{solidbc}
\vec{u} = \vec{u}_{BC}
\end{equation}
At a free surface, the fluid is subject to the constraint that the traction $\vec{T}$ applied at the surface is zero:
\begin{equation}
\label{freesurfacebc}
\vec{T} = (-pI + \tau)\vec{n} = \vec{0}
\end{equation}
In the above, $\vec{n}$ is taken to be the outward normal of the free surface and $I$ represents the identity matrix.

\subsection{A Note on the Simplified Stokes Equations}
\label{simplerstokes}
In situations where the viscosity is spatially constant, some manipulations are often carried out to reduce the contribution of viscosity to a simpler form. Specifically:
\begin{eqnarray}
\nabla \cdot \tau & = & \nabla \cdot \left( \mu(\nabla \vec{u} + (\nabla \vec{u})^T) \right)
\\ & = & \mu \nabla \cdot \nabla \vec{u} + \mu \nabla \cdot (\nabla \vec{u})^T
\\ & = & \mu \nabla \cdot \nabla \vec{u} + \mu \nabla (\nabla \cdot \vec{u})
\\ & = & \mu \nabla \cdot \nabla \vec{u} \label{laplacianviscosity}
\end{eqnarray}
where a simple vector calculus identity and the divergence free property of $\vec{u}$ have been used to eliminate the second term on the right hand side.  This reduces the contribution of viscosity to a simple Laplace operator applied to each component of velocity independently.  However, the natural boundary conditions of this modified problem are quite different from those of the original problem, as discussed by Limache et al.\ \cite{Limache2007,Limache2008}; if care is not taken this can lead to non-physical solutions, particularly for free surface flows.  We therefore prefer the fully general form of the viscous terms, even for constant viscosity \cite{Batty2008}.

\section{Variational Formulations for Time-Dependent Stokes Flow}
A backwards Euler time discretization of the governing equations yields:
\begin{eqnarray}
\frac{\rho}{\Delta t}(\vec{u} - \vec{u}^*) & = & \nabla\cdot\mathbf{\tau} - \nabla p \label{stokesequations1} \\
\nabla \cdot \vec{u} & = & 0 \label{stokesequations2} \\
\tau & = & \mu(\nabla \vec{u} + (\nabla \vec{u})^T) \label{stokesequations3}
\end{eqnarray}
In these expressions $\vec{u}$ is the final velocity field, $\vec{u}^*$ is the input (possibly divergent) velocity field, and $\Delta t$ is the size of the time step.

An equivalent variational formulation is the following:
\begin{equation}
\max_{p,\tau} \min_{\vec{u}} \iiint_{\Omega_L} \frac{\rho}{2}\|\vec{u}-\vec{u}^*\|^2 - \Delta t p \nabla \cdot \vec{u} + \Delta t \tau : \left( \frac{\nabla \vec{u} + (\nabla \vec{u})^T}{2} \right) - \frac{\Delta t}{4 \mu} \lVert \tau \rVert^2_F \label{eq:stokesfreesurface}
\end{equation}
The domain of integration is the liquid (non-air) region, $\Omega_L$.  Calculus of variations can be used to show that the optimality conditions for this problem yield precisely the equations of the original PDE problem above, with the zero traction free surface boundary condition (\ref{freesurfacebc}) enforced as a natural boundary condition.

Similarly, the following variational formulation enforces the Stokes equations inside the domain, but with static solid boundary conditions ($\vec{u} = 0$).
\begin{equation}
\max_{p,\tau} \min_{\vec{u}} \iiint_{\Omega_F} \frac{\rho}{2} \lVert \vec{u}-\vec{u}^* \rVert^2 + \Delta t \vec{u} \cdot \left( \nabla p - \nabla \cdot \tau \right) - \frac{\Delta t}{4 \mu} \lVert \tau \rVert^2_F  \label{eq:stokessolidwall}
\end{equation}
Here the integration is performed over the fluid (non-solid) region, $\Omega_F$. Note that these two variational formulations differ essentially by an integration by parts operation on the middle terms.

\section{Discretization}

With a particular variational formulation in hand, we proceed to directly discretize the required integrals in a manner similar to that proposed by Batty et al.\ \cite{Batty2007,Batty2008}. This results in a discrete optimization problem in which the boundary conditions are enforced naturally and implicitly. This is in contrast with the more common approach of first discretizing the PDE form itself, which necessitates the explicit handling of the potentially difficult boundary conditions outlined earlier.

\begin{figure}
  \centering
  \includegraphics[width=2in]{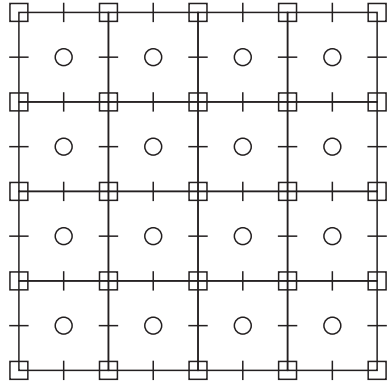}
  \caption{The standard staggered pressure-velocity grid layout in 2D, with stress components added. Circles indicate the locations of pressure and diagonal stress components ($\tau_{xx}$).  Dashes across cell faces indicate horizontal and vertical velocity components.  Small squares at cell corners indicate the locations of off-diagonal stress components ($\tau_{xy}$).}
  \label{fig:macgrid}
\end{figure}

\begin{figure}
  \centering
  \includegraphics[width=2in]{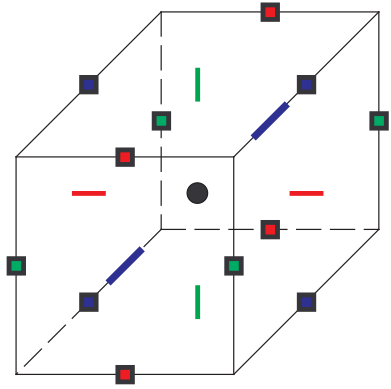}
  \caption{The standard staggered pressure-velocity grid layout in 3D, with stress components added. The black circle indicates the location of pressure and diagonal stress components ($\tau_{xx},\tau_{yy}$). The colored squares on cell edges indicate the locations of off-diagonal stress components ($\tau_{yz}$ is red, $\tau_{xz}$ is green, $\tau_{xy}$ is blue).  Dashes across cell faces indicate velocity components ($u$ is red, $v$ is green, $w$ is blue).}
  \label{fig:macgrid3d}
\end{figure}

We discretize the derivatives using centred finite differences on the classic staggered (MAC) grid, with pressures at cell centres, and velocity components on cell faces \cite{Harlow1965}. To support viscosity we must also place the components of the deviatoric stress tensor on our grid. The most natural way to do this is to locate diagonal components of the stress tensor ($\tau_{xx},\tau_{yy},\tau_{zz}$) at cell centres, and off-diagonal components ($\tau_{xy},\tau_{xz},\tau_{yz}$) on cell edges (nodes in two dimensions).  This arrangement is illustrated in 2D and 3D in Figures \ref{fig:macgrid} and \ref{fig:macgrid3d}, respectively. Straightforward centred differencing can then be used to compute the required derivatives in the correct locations.  This approach was first proposed by Darwish et al.\ \cite{Darwish1992} for two dimensions, and later extended to three dimensions by Mompean and Deville \cite{Mompean1997}. While it has traditionally been applied to non-Newtonian fluids in which the constitutive equations are more complex, we find its simplicity and elegance useful for the purely Newtonian flows we consider.

Note that because the deviatoric stress $\tau$ does not measure compression, we are assured that $\operatorname{Tr}(\tau) = 0$ which in 2D implies that $\tau_{xx} = -\tau_{yy}$.  We can therefore simplify the necessary computations by solving just for $\tau_{xx}$ rather than both quantities; in the final linear system, this will yield equations of the form $\tau_{xx} = \mu \left(\frac{\partial{u}}{\partial{x}} - \frac{\partial{v}}{\partial{y}}\right)$, thereby retaining symmetry.  Similar transformations apply in 3D to eliminate $\tau_{zz} = -(\tau_{xx}+\tau_{yy})$, yielding $\tau_{xx} + \frac{1}{2} \tau_{yy} = \mu \left(\frac{\partial{u}}{\partial{x}} - \frac{\partial{w}}{\partial{z}}\right)$ and $\tau_{yy} + \frac{1}{2}\tau_{xx} = \mu \left(\frac{\partial{v}}{\partial{y}} - \frac{\partial{w}}{\partial{z}}\right)$.  Naturally, the stress tensor will also be symmetric, so that $\tau_{xy} = \tau_{yx}$, $\tau_{xz} = \tau_{zx}$, and $\tau_{yz} = \tau_{zy}$, which further reduces the number of variables to be computed.

\begin{figure}
  \centering
  \includegraphics[width=4.5in]{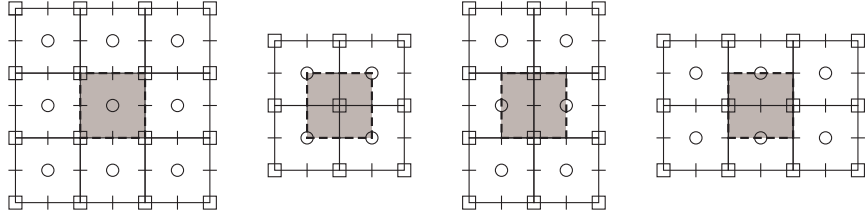}
  \caption{Control volumes around each sample location in 2D. From left to right: pressure and diagonal stress control volume, off-diagonal stress control volume, horizontal velocity control volume, vertical velocity control volume.}
  \label{fig:controlvolumes}
\end{figure}

\begin{figure}
  \centering
  \includegraphics[width=5in]{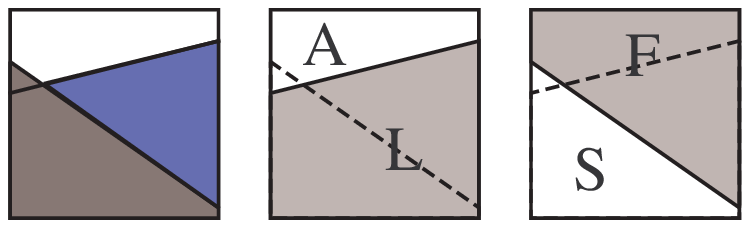}
  \caption{An illustration of volume fraction regions for solid and free surface weighting. Left:  A geometric scenario in which a solid (dark gray) meets a body of liquid (blue) in the presence of air (white). Note the air-liquid interface extrapolated into the solid. Middle:  The volume fraction region for a liquid (ie. non-air) weight, $W_L$, shown in gray, and its complementing air fraction $W_A$ shown in white; the presence of the solid is ignored.  Right:  The volume fraction region for a fluid (ie. non-solid) weight, $W_F$,  shown in gray, and its complementing solid fraction, $W_S$, shown in white; in this case the position of the liquid-air interface is ignored.}
  \label{fig:volumefractions}
\end{figure}

We approximate the integral comprising the variational forms by scaling each term by the fractional volume of material in a cell-sized control volume surrounding the appropriate sample point, and sum over all cells. For example, terms that lie on faces are scaled by the volume of fluid in a square (or cubic) control volume surrounding the face centre. The necessary two-dimensional control volumes are illustrated in Figure \ref{fig:controlvolumes}. (While we could apply a higher order approximation of the integrals, it is this simple piecewise constant approximation that ensures we retain the same stencils as a standard grid-aligned finite difference discretization.) For free surface boundary conditions, we need to estimate volume fractions interior to the liquid (ie. not air), indicated by weights $W_L$.  Later, we will also need the corresponding air fractions, $W_A = I - W_L$ (where $I$ the identity).  Likewise, for solid boundary conditions we estimate the volume fractions $W_F$ of a cell that is inside the fluid (ie. not solid), and its complementary solid fraction, $W_S = I - W_F$. These volume fractions are illustrated in Figure~\ref{fig:volumefractions}, and can, for example, be estimated from a level set representation using the method of Min and Gibou \cite{Min2007}. We will see that this simple piecewise constant approximation of the integrals ensures that we retain simple centred finite difference stencils, while volume fraction weighting handles boundary conditions.


In equation (\ref{eq:stokesfreesurface}), we are integrating over the liquid region $\Omega_L$, so we must use volume fractions with subscript $L$. The first term of the equation consists of velocity data that lies on faces, so we estimate the integral using fractional volumes associated to each velocity face sample.  We indicate this using superscripts on the weights that indicate the type of control volume being considered; in this case the weight used is $W_L^u$. Similarly, the second term consists of pressures and divergences that are conceptually located at cell centres, so we use cell-centred weights, $W_L^p$. Finally, the last two terms are associated with stresses, which lie on cell edges and centres, so we use the fractions $W_L^\tau$.

Applying this framework to (\ref{eq:stokesfreesurface}) leads to the following discrete Stokes problem with free surface boundaries:
\begin{equation}
\max_{p,\tau} \min_{u} \frac{1}{2} (u-u^*)^T P W_L^u (u-u^*) + \Delta t  p^T  W_L^p G^T u + \Delta t  \tau^T  W_L^\tau D u - \frac{\Delta t}{4} \tau^T M^{-1} W_L^\tau \tau
\end{equation}
In this expression $P$ is a diagonal matrix of densities per velocity sample and $M$ is a diagonal matrix of viscosity coefficients per stress sample. The derivatives are discretized with staggered grid finite difference operators: $G$ is the usual discrete gradient, and $D$ is the discrete deformation rate applied to velocity (ie. $Du \approx \frac{1}{2}(\nabla u + (\nabla u)^T)$). Note that the negative transposes of $G$ and $D$ are the corresponding discrete vector and tensor divergence operators, respectively. The various $W$ terms are diagonal matrices consisting of the volume fractions described above.

Since this discrete optimization problem is a quadratic, the optimality conditions yield the following symmetric indefinite linear system:
\begin{equation}
\left(
\begin{array}{ccc}
\frac{1}{\Delta t} P W_L^u & D^T W_L^\tau              & G W_L^p \\
W_L^\tau D                 & -\frac{1}{2}M^{-1} W_L^\tau & 0\\
W_L^p G^T                 & 0                             & 0 \\
\end{array}
\right)
\left(
\begin{array}{c}
u\\
\tau\\
p\\
\end{array}
\right)
=
\left(
\begin{array}{c}
\frac{1}{\Delta t} P W_L^u u^*\\
0\\
0\\
\end{array}
\right) \label{eq:stokesfreesurfacesystem}
\end{equation}

The solid boundary Stokes problem (\ref{eq:stokessolidwall}) can be discretized in much the same manner, except that integrals are computed over the fluid (non-solid) region using volume fractions $W_F$. The discrete form is:
\begin{equation}
\max_{p,\tau} \min_{u} \frac{1}{2} (u-u^*)^T P W_F^u (u-u^*) + \Delta t  u^T W_F^u (Gp + D^T \tau) - \frac{\Delta t}{4} \tau^T M^{-1} W_F^\tau \tau
\end{equation}
The linear system for the solid wall Stokes problem is:
\begin{equation}
\left(
\begin{array}{ccc}
\frac{1}{\Delta t} P W_F^u & W_F^u D^T                  & W_F^u G \\
D W_F^u                    & -\frac{1}{2}M^{-1} W_F^\tau & 0         \\
G^T W_F^u                 & 0                             & 0         \\
\end{array}
\right)
\left(
\begin{array}{c}
u\\
\tau\\
p\\
\end{array}
\right)
=
\left(
\begin{array}{c}
\frac{1}{\Delta t} P W_F^u u^*\\
0\\
0\\
\end{array}
\right)\label{eq:stokessolidsystem}
\end{equation}

Because these two systems have nearly identical forms, and differ only by the weighting matrices $W$, we can straightforwardly combine them to handle both free surfaces and solid boundaries in the same problem:
\begin{equation}
\left(
\begin{array}{ccc}
\frac{1}{\Delta t} P W_F^u W_L^u  & W_F^u D^T W_L^\tau                  & W_F^u G W_L^p \\
W_L^\tau D W_F^u                   & -\frac{1}{2}M^{-1} W_F^\tau W_L^\tau & 0         \\
W_L^p G^T W_F^u                         & 0                                 & 0         \\
\end{array}
\right)
\left(
\begin{array}{c}
u\\
\tau\\
p\\
\end{array}
\right)
=
\left(
\begin{array}{c}
\frac{1}{\Delta t} P W_F^u W_L^u u^*\\
0\\
0\\
\end{array}
\right)\label{eq:stokesmixedsystem}
\end{equation}
By combining the two formulations only at the discrete level, we are able to exploit the natural boundary conditions to handle \emph{both} boundary types together, despite the fact that in each of the two continuous variational formulations only one of the two boundaries can be considered natural.

For this Stokes system, eliminating stress by applying the Schur complement to the centre block yields the symmetric indefinite velocity-pressure system most commonly associated with the Stokes problem. However, as outlined earlier we prefer to solve \emph{symmetric positive-definite} systems, as they are generally more amenable to common black box solvers such as preconditioned conjugate gradient methods, domain decomposition, etc. Conveniently, the upper-left block of the full system is a diagonal matrix which can be trivially inverted.  We therefore perform a Schur complement on this block to eliminate \emph{velocity} and arrive at a sparse symmetric positive-definite system for pressure and stress. This is exactly analogous to classic pressure projection methods for incompressibility, in which eliminating velocity yields a symmetric positive-definite Poisson equation for pressure alone. Of course, it should be emphasized that this technique can only be applied in the time-dependent Stokes case; otherwise the upper-left block is simply zero making a Schur complement impossible.

The symmetric positive-definite form is:
\begin{eqnarray}
\left(
\begin{array}{cc}
A_{11} & A_{12} \\
A_{12}^T & A_{22}
\end{array}
\right)
\left(
\begin{array}{c}
\tau\\
p
\end{array}
\right)
=
\left(
\begin{array}{c}
W_L^\tau D  W_F^u u^*\\
W_L^p G^T  W_F^u u^*\\
\end{array}
\right)
\label{eq:stokesmixedsystemspd}
\end{eqnarray}
where the blocks of the matrix are
\begin{eqnarray*}
A_{11} & = & \frac{1}{2}M^{-1} W_L^\tau W_F^\tau + \Delta t W_L^\tau D P^{-1} {W_L^u}^{-1} W_F^u D^T W_L^\tau \\
A_{12} & = & \Delta t W_L^\tau D P^{-1} {W_L^u}^{-1} W_F^u G W_L^p \\
A_{22} & = & \Delta t W_L^p G^T P^{-1} {W_L^u}^{-1} W_F^u G W_L^p
\end{eqnarray*}

Our results for the Stokes problem consistently indicate first order convergence for all variables in $L^1$, and first order convergence for velocity in $L^\infty$. Stress fails to converge in $L^\infty$ due to noisy errors along the boundaries. Given that stress is computed as the gradient of velocity, it is not entirely surprising that it loses one order of accuracy in $L^\infty$. Fortunately, for most practical applications the behaviour of velocities is of greater importance.

\section{Non-Homogeneous Boundary Conditions}
The preceding formulation for the Stokes problem exploits natural homogeneous boundary conditions to simplify handling of irregular domains on Cartesian grids.  However, many practical situations will call for \emph{non-homogeneous} boundary conditions where the boundary values are non-zero. For example, moving solid boundaries will require non-zero boundary velocities to be enforced, as will inflow and outflow boundaries. Similarly, non-zero pressure boundary conditions have been used to support surface tension (eg. \cite{Enright2003}). To incorporate non-homogeneous boundary values into our framework in a consistent manner, we introduce additional terms to account for the work done by the boundary itself.
\subsection{Prescribed Traction Boundaries}
To add a prescribed traction boundary, we need to account for the work done by the traction at the surface. We do this by adding the following boundary term to (\ref{eq:stokesfreesurface}):
\begin{equation}
\iint_{\partial \Omega_A} \Delta t \vec{u}\cdot(p_{BC}I - \tau_{BC}) \vec{n}
\end{equation}
where $p_{BC}$ and $\tau_{BC}$ are the prescribed pressure and deviatoric stress, respectively, and $\vec{n}$ in this case is the outward normal with respect to the air (non-liquid) region, $\Omega_A$. (Of course, only the resulting normal component, ie. surface traction, will be enforced.)
This surface integral can be converted to a volume integral through integration by parts:
\begin{equation}
\Delta t \iiint_{\Omega_A} p_{BC} \nabla \cdot \vec{u} - \tau_{BC} : \left(\frac{\nabla \vec{u} + (\nabla \vec{u})^T}{2}\right) + \vec{u} \cdot (\nabla p_{BC} - \nabla \cdot \tau_{BC})
\end{equation}
This ensures that all the relevant quantities are at grid locations that are consistent with the functionals considered previously. This also avoids the need to discretize surface integrals which would add further complications and potentially sacrifice the convenience of centered difference stencils.

Using $W_A$ terms to indicate the air fraction of a particular control volume, the discretized form is:
\begin{equation}
\Delta t \left( -p_{BC}^T W_A^p G^T u - \tau_{BC}^T W_A^\tau D u  + u^T W_A^u (G p_{BC} + D^T \tau_{BC})\right)
\end{equation}

The new terms modify the right hand side of the linear system (\ref{eq:stokesfreesurfacesystem}), to become:
\begin{equation}
\left(
\begin{array}{c}
\frac{1}{\Delta t} P W_L^u u^* + G W_A^p p_{BC} - W_A^u Gp_{BC} + D^T W_A^\tau \tau_{BC}  - W_A^u D^T  \tau_{BC}\\
0\\
0\\
\end{array}
\right)
\end{equation}
Although the air region may extend far from the actual liquid surface, we only need to apply modifications to the right hand side for rows of the system in which the matrix has non-zero entries, indicating that there is liquid present. This means that in practice only the interface between the air and liquid regions plays a role.

\subsection{Prescribed Velocity Boundaries}
In the common case of moving solid boundaries with prescribed velocities $\vec{u}_{BC}$, we account for the work done by the solid on the fluid by adding the following term to (\ref{eq:stokessolidwall}):
\begin{equation}
\iint_{\partial \Omega_S} \Delta t \vec{u}_{BC}\cdot(p I - \tau) \vec{n}
\end{equation}
where $\vec{n}$ is the outward normal to the solid region, $\Omega_S$.  In volume integral form we have:
\begin{equation}
\Delta t \iiint_{\Omega_S} p \nabla \cdot \vec{u}_{BC} - \tau : \left(\frac{\nabla \vec{u}_{BC} + (\nabla \vec{u}_{BC})^T}{2}\right) + \vec{u}_{BC} \cdot (\nabla p - \nabla \cdot \tau)
\end{equation}
Labelling solid fractions $W_S$ and discretizing, we arrive at the following term:
\begin{equation}
\Delta t \left( -p^T W_S^p G^T u_{BC} - \tau^T W_S^\tau D u_{BC}  + u_{BC}^T W_S^u (G p + D^T \tau) \right)
\end{equation}
This results in a modification to the right hand side of the linear system (\ref{eq:stokessolidsystem}), to become:
\begin{equation}
\left(
\begin{array}{c}
\frac{1}{\Delta t} P W_F^u u^*\\
-D W_S^uu_{BC} + W_S^\tau D u_{BC}\\
-G^T W_S^u u_{BC} + W_S^p G^T u_{BC} \\
\end{array}
\right)
\end{equation}
These right hand side modifications are also only applied to rows in which the matrix has valid non-zero entries.

\section{Generalization and Further Factorization}
At this point, we highlight the abstract form of the problem studied above, which helps see the discrete nature of the transformation we use and illustrates how to generalize our technique to other problems such as two-way fluid-solid interaction. This is a special case of a weighted linear least-squares problem with block diagonal regularization subject to linear equality constraints, which we state using bold elements to avoid confusion with prior notation:
\begin{equation}\label{genMin}
\min_{\begin{array}{c}
        \mathbf{u} \\
        \mathbf{C}\mathbf{u}+\mathbf{d}=0
      \end{array}}
   \tfrac{1}{2}(\mathbf{u}-\mathbf{u}^*)^T \mathbf{M} (\mathbf{u}-\mathbf{u}^*)
 + \tfrac{1}{2} (\mathbf{b}-\mathbf{A}\mathbf{u})^T \mathbf{W} (\mathbf{b}-\mathbf{A}\mathbf{u}).
\end{equation}
For the Stokes discretization above, the velocity vector $\mathbf{u}$ here corresponds to the vector $u$ of all fluid velocities and $\mathbf{M}$, the block diagonal regularization matrix which we term the mass matrix for the abstract problem, corresponds to the (simply diagonal) product $PW_F^uW_L^u$. The first term in the objective is thus a kinetic energy norm of the difference between the old (or predicted) velocity $\mathbf{u}^*$ and the new velocity $\mathbf{u}$ which we are finding. The constraint matrix $\mathbf{C}$ and constraint vector $\mathbf{d}$ correspond to the negative weighted discrete divergence operator $\Delta t W_L^pG^TW_F^u$ and the zero vector (for the divergence of the Stokes velocity in the interior) modified as necessary by non-homogeneous boundary conditions. Finally, in the weighted least-squares term the matrix $\mathbf{A}$ corresponds to $DW_F^u$ (i.e.\ the weighted measure of deformation rate), the vector $\mathbf{b}$ is zero in the interior of the Stokes problem but modified by non-homogenous solid boundaries as needed, and the weighting matrix $\mathbf{W}$ corresponds to $2\Delta t W_L^\tau M {W_F^\tau}^{-1}$ (i.e.\ the weighted, time-scaled viscous coefficients).

Rearranged into this format, our Stokes discretization looks like:
\begin{equation}
\min_{\begin{array}{c}
        u \\
       \Delta t W_L^pG^TW_F^uu=0
      \end{array}}
   \tfrac{1}{2}(u-u^*)^T PW_F^uW_L^u (u-u^*)
 + \Delta t (DW_F^uu)^T W_L^\tau M {W_F^\tau}^{-1} (DW_F^uu)
\end{equation}
The solution is a balance between staying close to the old velocity (weighted by density) and minimizing the deformation rate (weighted by viscosity and time step) with cell fractions accounting for irregular geometry, subject to incompressibility. This makes evident the connection to Helmholtz's minimum dissipation theorem for Stokes flow \cite{Batchelor1967}.

As before, we can write down the KKT optimality conditions, introducing a Lagrange multiplier $\mathbf{\lambda}$ for the linear constraint:
\begin{equation}\label{generalKKT}
\left(\begin{array}{cc}
\mathbf{M}+\mathbf{A}^T\mathbf{W}\mathbf{A} & \mathbf{C}^T \\
\mathbf{C} & 0
\end{array}\right)
\left(\begin{array}{c}
\mathbf{u}\\
\mathbf{\lambda}
\end{array}\right)
=
\left(\begin{array}{c}
\mathbf{M}\mathbf{u}^*-\mathbf{A}^T\mathbf{W}\mathbf{b} \\
-\mathbf{d}
\end{array}\right).
\end{equation}
Up to a scale factor, the new variable $\mathbf{\lambda}$ corresponds to pressure in the Stokes problem.

This is of course a symmetric indefinite matrix, but our variable change to pressure and viscous stress generalizes here as well to arrive at a sparse SPD matrix. First introduce the weighted least-squares residual
\begin{equation}\label{residual}
\mathbf{r}=\mathbf{W}(\mathbf{b}-\mathbf{A}\mathbf{u}),
\end{equation}
which in the Stokes problem is the weighted and time-scaled viscous stress. Augmenting system (\ref{generalKKT}) with equation (\ref{residual}) multiplied by $\mathbf{W^{-1}}$ we arrive at:
\begin{equation}
\left(\begin{array}{ccc}
\mathbf{M} & -\mathbf{A}^T & \mathbf{C}^T \\
-\mathbf{A} & -\mathbf{W}^{-1} & 0 \\
\mathbf{C} & 0 & 0
\end{array}\right)
\left(\begin{array}{c}
\mathbf{u}\\
\mathbf{r}\\
\mathbf{\lambda}
\end{array}\right)
=
\left(\begin{array}{c}
\mathbf{M}\mathbf{u}^* \\
-\mathbf{b}\\
-\mathbf{d}
\end{array}\right).
\end{equation}
Eliminating velocity with the first equation $\mathbf{u}=\mathbf{u}^*+\mathbf{M}^{-1}\mathbf{A}^T\mathbf{r}-\mathbf{M}^{-1}\mathbf{C}^T\mathbf{\lambda}$ leaves us with the SPD Schur complement problem for the residual and the Lagrange multiplier:
\begin{equation}\label{sparseSPD}
\left(\begin{array}{cc}
\mathbf{W}^{-1}+\mathbf{AM}^{-1}\mathbf{A}^T & -\mathbf{AM}^{-1}\mathbf{C}^T \\
-\mathbf{CM}^{-1}\mathbf{A}^T & \mathbf{CM}^{-1}\mathbf{C}^T
\end{array}\right)
\left(\begin{array}{c}
\mathbf{r}\\
\mathbf{\lambda}
\end{array}\right)
=
\left(\begin{array}{c}
\mathbf{b}-\mathbf{Au}^* \\
\mathbf{d}+\mathbf{Cu}^*
\end{array}\right).
\end{equation}
A key advantage to this system is that, under the assumption that the mass matrix $\mathbf{M}$ and the weight matrix $\mathbf{W}$ are block diagonal or more specifically have sparse inverses, and that $\mathbf{A}$ and $\mathbf{C}$ don't have overly dense columns, this is a sparse matrix itself. (If $\mathbf{W}$ isn't of this form, using $\mathbf{\tilde{r}}=\mathbf{b}-\mathbf{Au}$ can restore sparsity.)

Being sparse and SPD, many black-box linear solvers are available which may not apply to the original indefinite system. In our studies, we used PCG with a parallelized algebraic multiplicative Schwartz overlapping domain-decomposition preconditioner, for example.

We can go further, however, reducing the requirement on $A$ and $C$. In particular, the matrix in equation (\ref{sparseSPD}) can be factored conveniently as:
\begin{multline}
\left(\begin{array}{cc}
\mathbf{W}^{-1}+\mathbf{AM}^{-1}\mathbf{A}^T & -\mathbf{AM}^{-1}\mathbf{C}^T \\
-\mathbf{CM}^{-1}\mathbf{A}^T & \mathbf{CM}^{-1}\mathbf{C}^T
\end{array}\right)
\\
=
\left(\begin{array}{cc}
-\mathbf{A} & \mathbf{I} \\
\mathbf{C} & 0
\end{array}\right)
\left(\begin{array}{cc}
\mathbf{M}^{-1} &  \\
  & \mathbf{W}^{-1}
\end{array}\right)
\left(\begin{array}{cc}
-\mathbf{A}^T & \mathbf{C}^T \\
\mathbf{I} & 0
\end{array}\right).
\end{multline}
These factors are obviously as sparse as the original components. Note also that the right-hand side of equation (\ref{sparseSPD}) can be written as
\begin{equation}
\left(\begin{array}{c}
\mathbf{b}-\mathbf{Au}^* \\
\mathbf{d}+\mathbf{Cu}^*
\end{array}\right)
=
\left(\begin{array}{cc}
-\mathbf{A} & \mathbf{I} \\
\mathbf{C} & 0
\end{array}\right)
\left(\begin{array}{cc}
\mathbf{M}^{-1} &  \\
  & \mathbf{W}^{-1}
\end{array}\right)
\left(\begin{array}{c}
\mathbf{Mu}^* \\
\mathbf{Wb}
\end{array}\right)
+
\left(\begin{array}{c}
0 \\
\mathbf{d}
\end{array}\right).
\end{equation}
In the case of Stokes with homogeneous boundary conditions, $\mathbf{d}=0$, and it is then clear that equation (\ref{sparseSPD}) gives the normal equations for a sparse, weighted, and \textit{unconstrained} least-squares problem:
\begin{equation}
\min_{\mathbf{r},\mathbf{\lambda}} \left\|
\left(\begin{array}{c}
\mathbf{Mu}^* \\
\mathbf{Wb}
\end{array}\right)
-
\left(\begin{array}{cc}
-\mathbf{A}^T & \mathbf{C}^T \\
\mathbf{I} & 0
\end{array}\right)
\left(\begin{array}{c}
\mathbf{r} \\
\mathbf{\lambda}
\end{array}\right)
\right\|^2_{[\mathbf{M}^{-1}, \mathbf{W}^{-1}]}.
\end{equation}
Specialized solvers such as LSQR \cite{Paige1982} may then offer a significant advantage.

The primary advantage of this generalization, and factorization, lies however in the ease of extending the formulation to more general dynamics. For example, Batty et al.\cite{Batty2007} coupled rigid body dynamics with inviscid incompressible flow using a pressure-only subset of equation ($\ref{sparseSPD}$) expressed as an unconstrained least-squares problem. However, for rigid bodies overlapping many fluid grid cells the system suffered a large dense block corresponding to those cells, leading Robinson-Mosher et al.\ to pursue an indefinite form \cite{Robinson-Mosher2008}. In the factored form, however, the rigid body merely corresponds to six fairly dense rows in $\mathbf{A}$ and $\mathbf{C}$, which can be handled much more efficiently in both storage and multiplication with the factored form. Elastic and/or constrained solid dynamics can be phrased in the same minimization form as equation (\ref{genMin}), e.g.\ English and Bridson's isometrically deforming membranes \cite{English2008}, further opening a clear route to efficiently solvable general solid-fluid coupling: the additional degrees of freedom of the solid are appended to $\mathbf{u}$, the solid mass matrix is appended to $\mathbf{M}$, additional solid constraints are appended to $\mathbf{C}$ and $\mathbf{d}$, and elastic potential energy terms are expressed as a possibly non-linear least-squares terms appended to $\mathbf{A}$ and $\mathbf{B}$ (as is quite natural for hyperelastic finite element models, and points out the utility of a Hessian-free Gauss-Newton iteration for solving them). Finally, we note that while Robinson-Mosher et al.\ discuss some of the above manipulations \cite{Robinson-Mosher2010}, they do so for a less general linear algebra problem, and do not discuss the optimization form for either the initial constrained problem nor the final unconstrained least-squares problem.

\section{Null Space Elimination}
\begin{figure}
  \centering
  \includegraphics[width=2in]{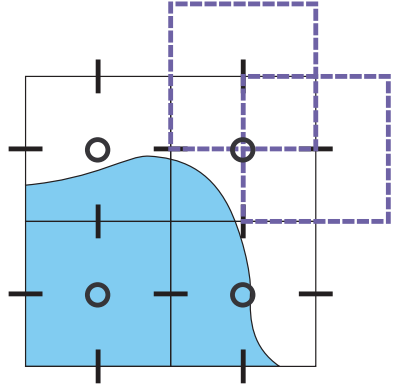}
  \caption{A case in which a null space arises for pressure in a free surface problem. The top right cell contains some liquid, and therefore will have a positive volume weight associated with the divergence constraint on the cell. However, two of its associated velocity face control volumes (indicated by dashed blue squares) contain no liquid and will have zero volume weights.  We identify this pressure sample as invalid, and replace it with the value of the boundary condition (eg. $p = 0$), thereby eliminating the null space.}
  \label{fig:nullspace}
\end{figure}

An issue that arises with our approach is the presence of null spaces due to overlapping volume weights assigned to different terms of the discrete variational problem.  For example, consider the divergence operator for a cell near a free surface boundary; there are volume weights associated with both velocities (face centres) and pressures (cell centres). Cases frequently arise in the discretized system in which a pressure with a non-zero volume weight enforces a divergence constraint on at least one velocity face with zero associated volume. This velocity sample will appear in no other equations because of its zero volume weight, and therefore it may take on an arbitrary value as long as it satisfies the constraint.  An example of such a null space scenario is shown in Figure \ref{fig:nullspace}.

In the final result, only samples with a positive associated volume weight are considered, so the physical solution is not adversely affected.  However, such large spurious null spaces can pose problems when applying standard solvers for sparse linear systems; we would therefore like to eliminate them.

To do so, we identify each variable that enforces a relationship on a sample with zero volume weights.  For example, in the free surface pressure case a non-zero weighted pressure is tagged as invalid if it enforces the divergence-free condition on one or more velocity faces with zero weights.  Likewise, in the solid boundary case if a non-zero weighted velocity sample borders a zero-weighted pressure sample, that velocity sample is tagged as invalid.  This process can likewise be extended to the viscous terms. Once these invalid variables have been identified, they can be straightforwardly eliminated from the linear system, and replaced with the value of the boundary condition, resulting in a modification to the right-hand-side.  This reduced set of equations retains symmetry and has the same solution, but no longer suffers from large null spaces.

There can be one additional null space when the fluid domain is completely enclosed by solid walls. Because only the gradient of pressure affects the final velocities, pressure solutions that differ by a constant are effectively equivalent.  This rank 1 null space doesn't pose substantial problems, so we have not eliminated it; if necessary, it could be removed by arbitrarily fixing one pressure value in the domain.

\section{Convergence Studies}

We have verified that the method computes the exact solution for linear problems even for irregular domains.  This includes the case of hydrostatic fluid with solid (and possibly free surface) boundaries, as well as rigid translations and rotations of liquid bodies with free surface boundaries.  We provide a range of examples below to illustrate the convergence orders achieved by our methods in more difficult scenarios. All of the examples make use of curved boundaries which do not align with the underlying Cartesian grid, and we consider a single time step of the time-dependent problem in question.  The examples test our methods in the presence of free surfaces and both static and moving boundaries. We compute the $L^{\infty}$ and $L^{1}$ errors, where our discrete $L^{1}$ norm is computed as $\|u_h\|_1 = \sum_{i} |u_i| h^d$ for a uniform grid spacing $h = \Delta x$ in $d$ spatial dimensions.

For each case, we transformed the linear system to the sparse symmetric positive-definite form as described above, and solved it with the conjugate gradient method preconditioned with overlapping multiplicative Schwarz domain decomposition. The preconditioner was determined purely algebraically, from a simple graph partition of the sparse matrix; we expect the positive-definite form would allow the use of many other black-box solvers as well. Though we do not consider this iterative solver a core contribution of the paper, Table \ref{solveriterations} presents some representative iteration counts to illustrate its scaling.

\begin{table}
\caption{Solver Behaviour}
\centering
\label{solveriterations}
\begin{tabular}{|c | c | c|}
\hline
 & PCG Iterations & PCG Iterations\\
Grid & Free Surface (\ref{freesurface_section}) & Solid Wall (\ref{solidwall_section})\\
\hline
$16^2$          & 1  & 1 \\
$32^2$          & 1 & 1 \\
$64^2$          & 6 & 4 \\
$128^2$         & 12 & 14 \\
$256^2$         & 27 & 42 \\
$512^2$         & 54 & 91 \\
$1024^2$        & 110 & 195 \\
\hline
\end{tabular}
\end{table}

\subsection{Stokes Flow with Free Surface Boundaries (2D)}
\label{freesurface_section}
\begin{figure}
\centering
\subfigure[]
{
\includegraphics[width=55mm]{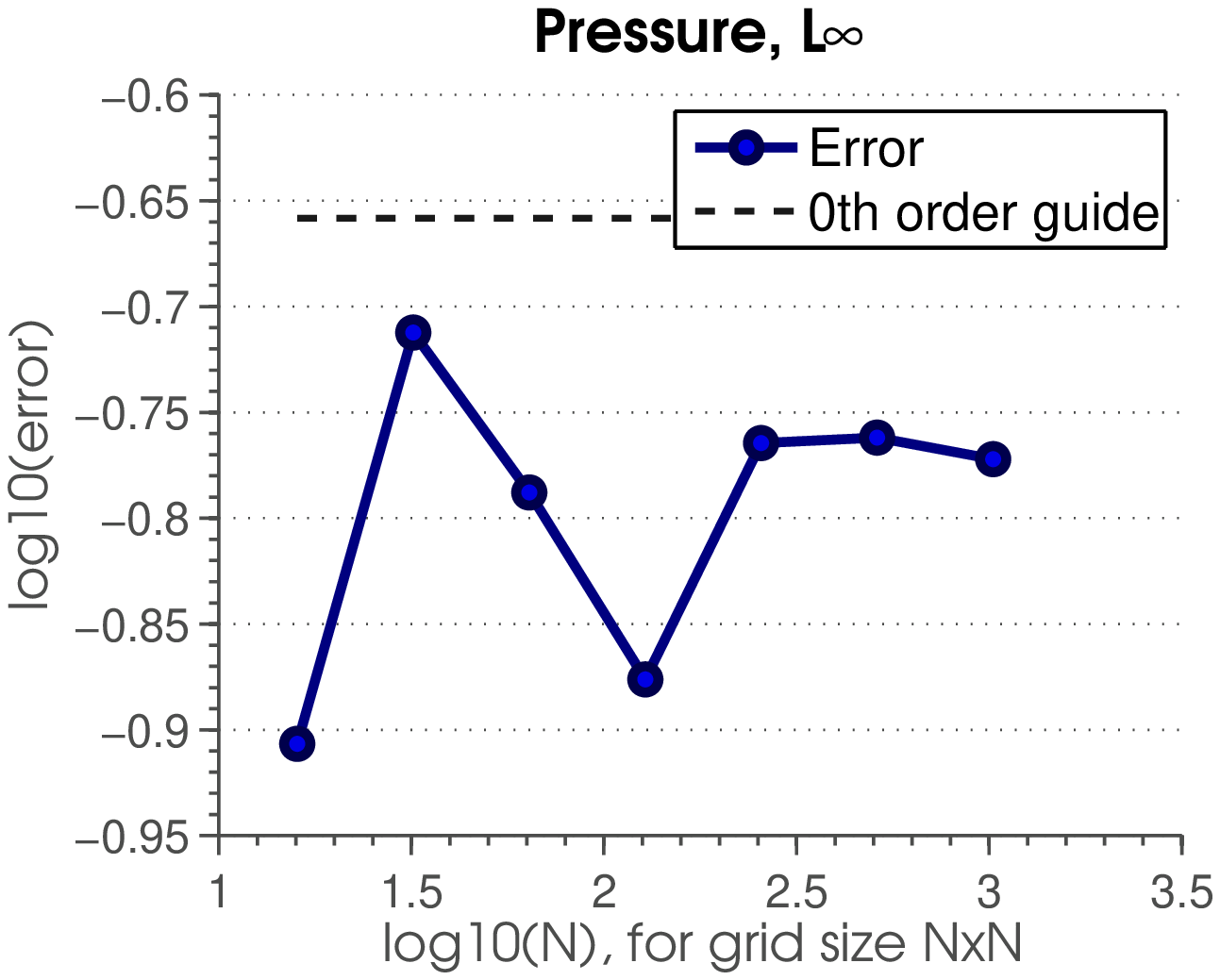}
}
\subfigure[]
{
\includegraphics[width=55mm]{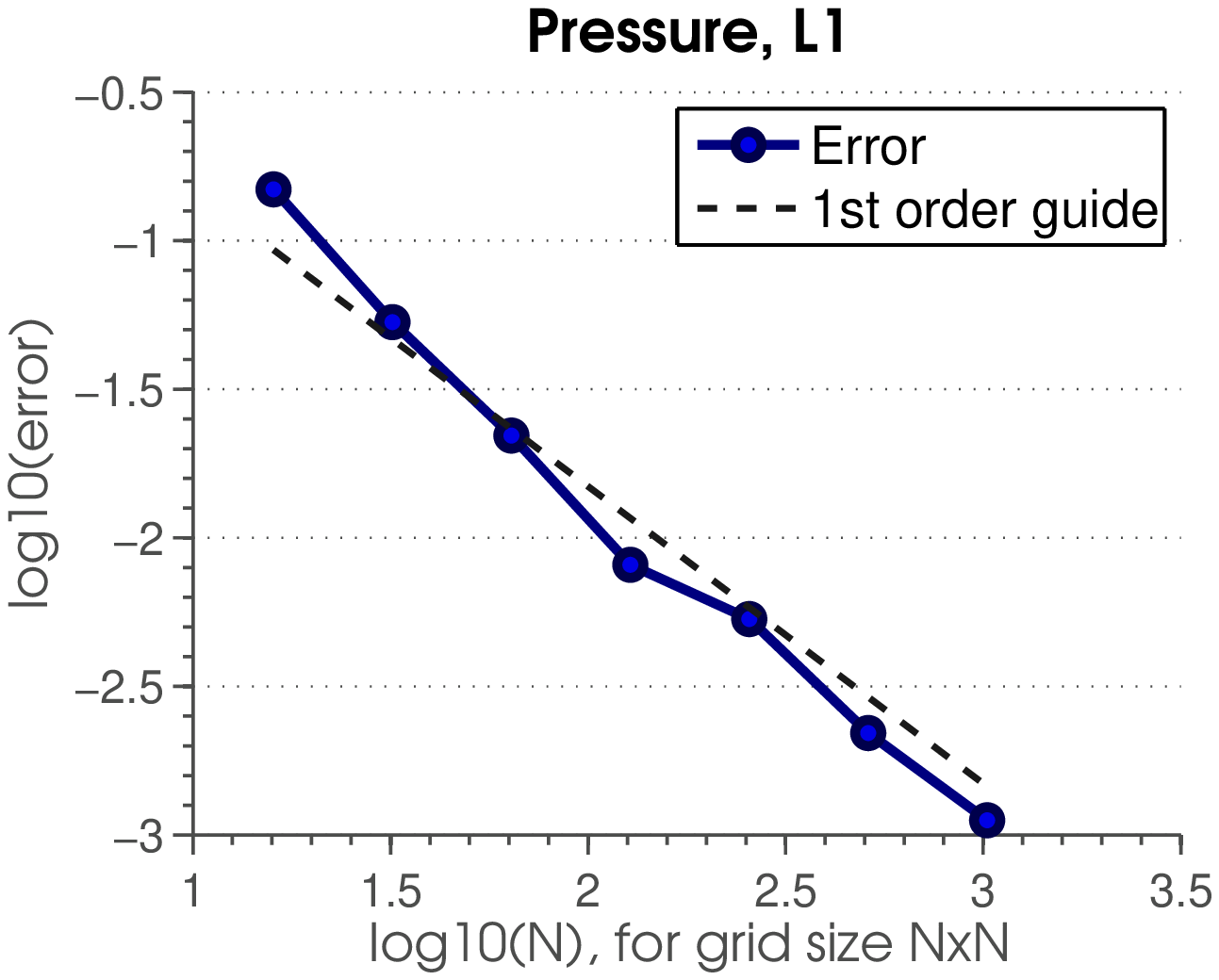}
}
\subfigure[]
{
\includegraphics[width=55mm]{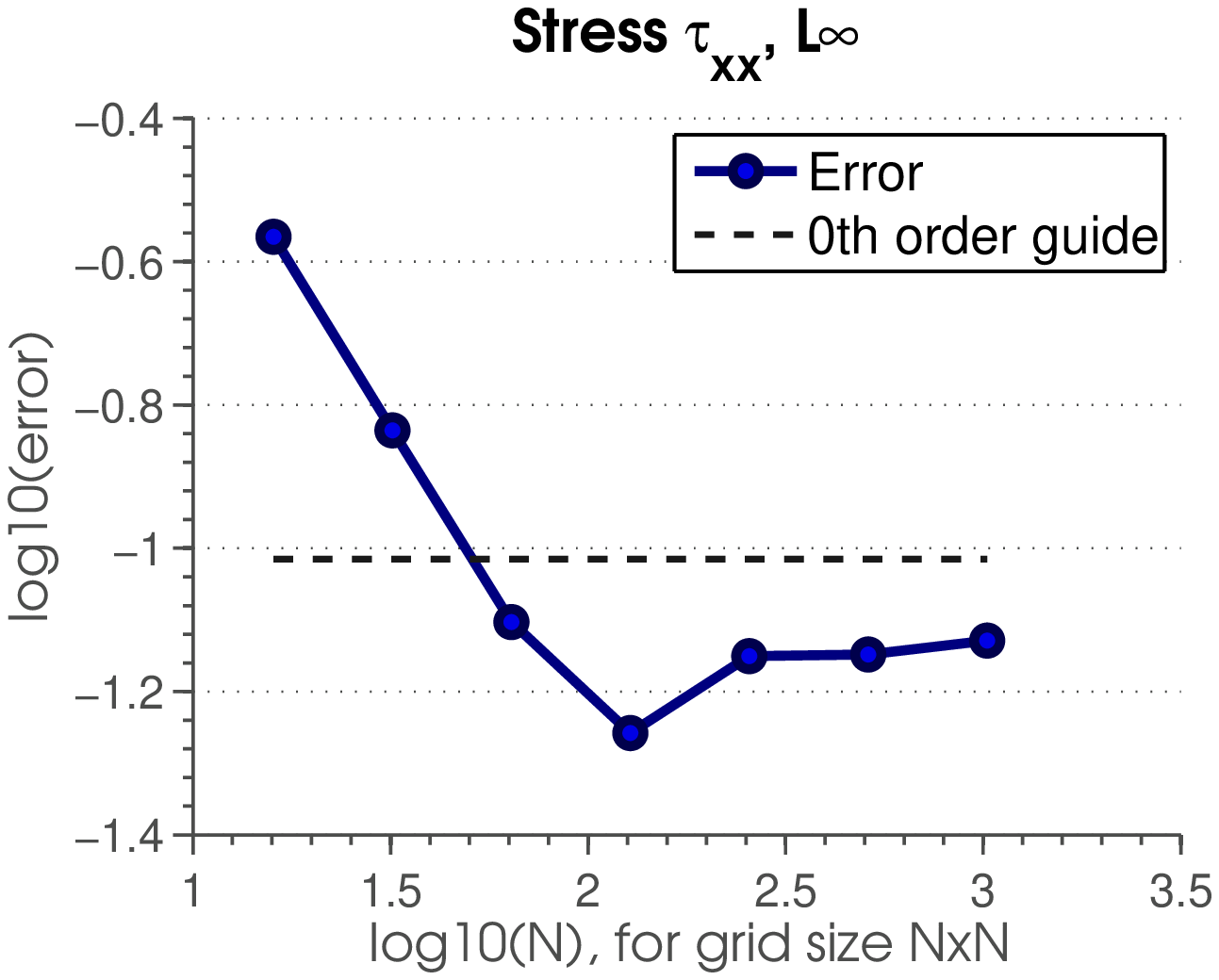}
}
\subfigure[]
{
\includegraphics[width=55mm]{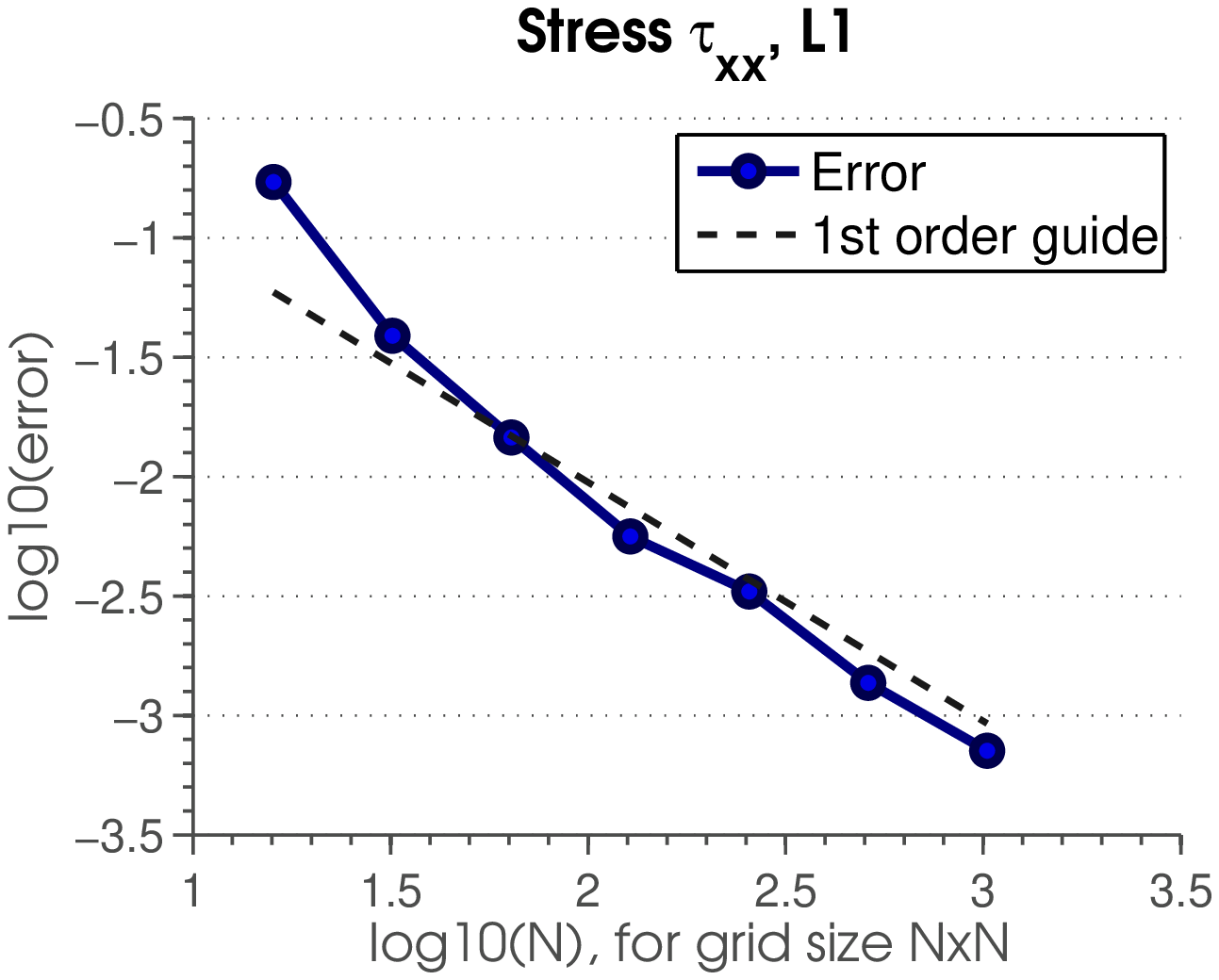}
}
\subfigure[]
{
\includegraphics[width=55mm]{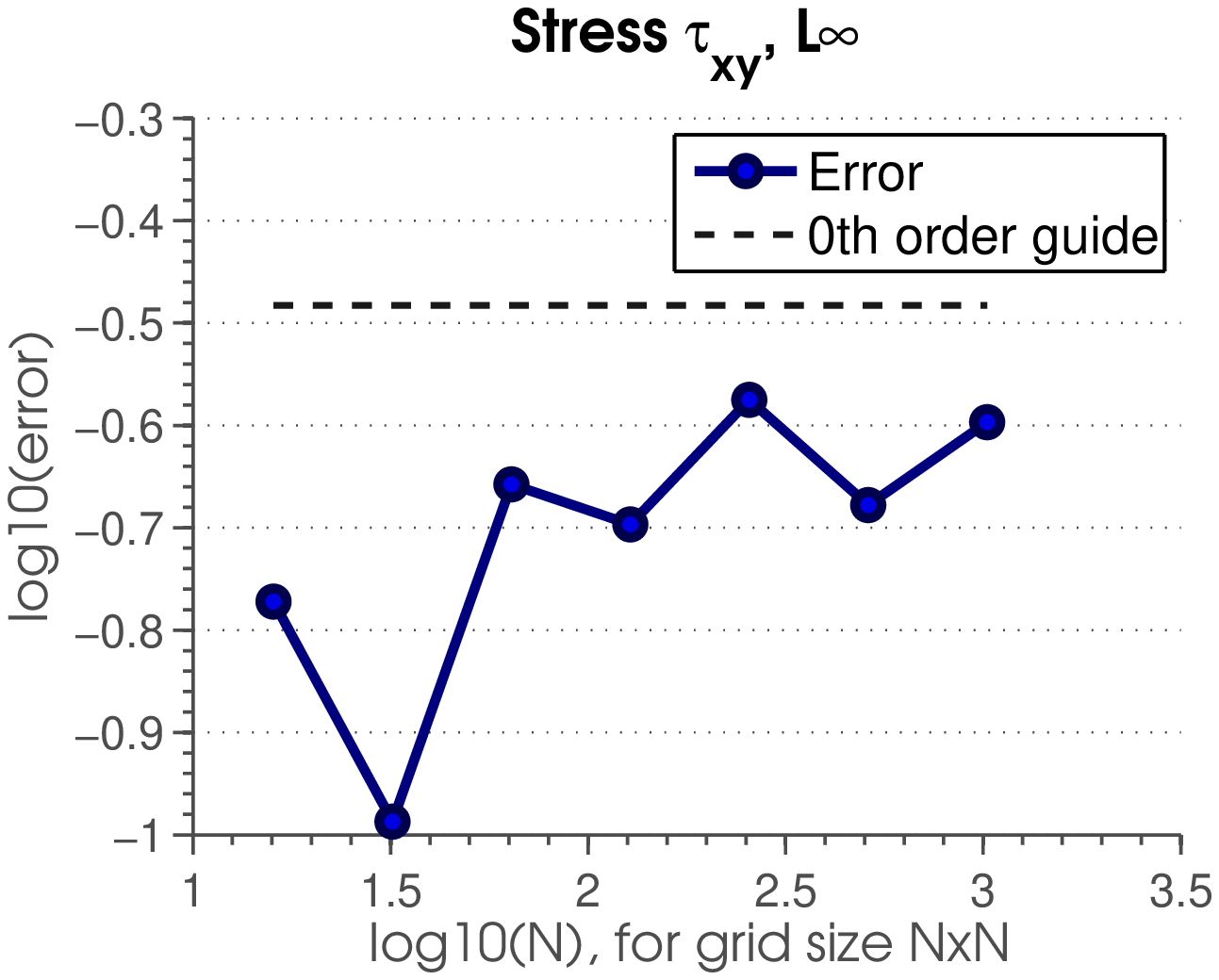}
}
\subfigure[]
{
\includegraphics[width=55mm]{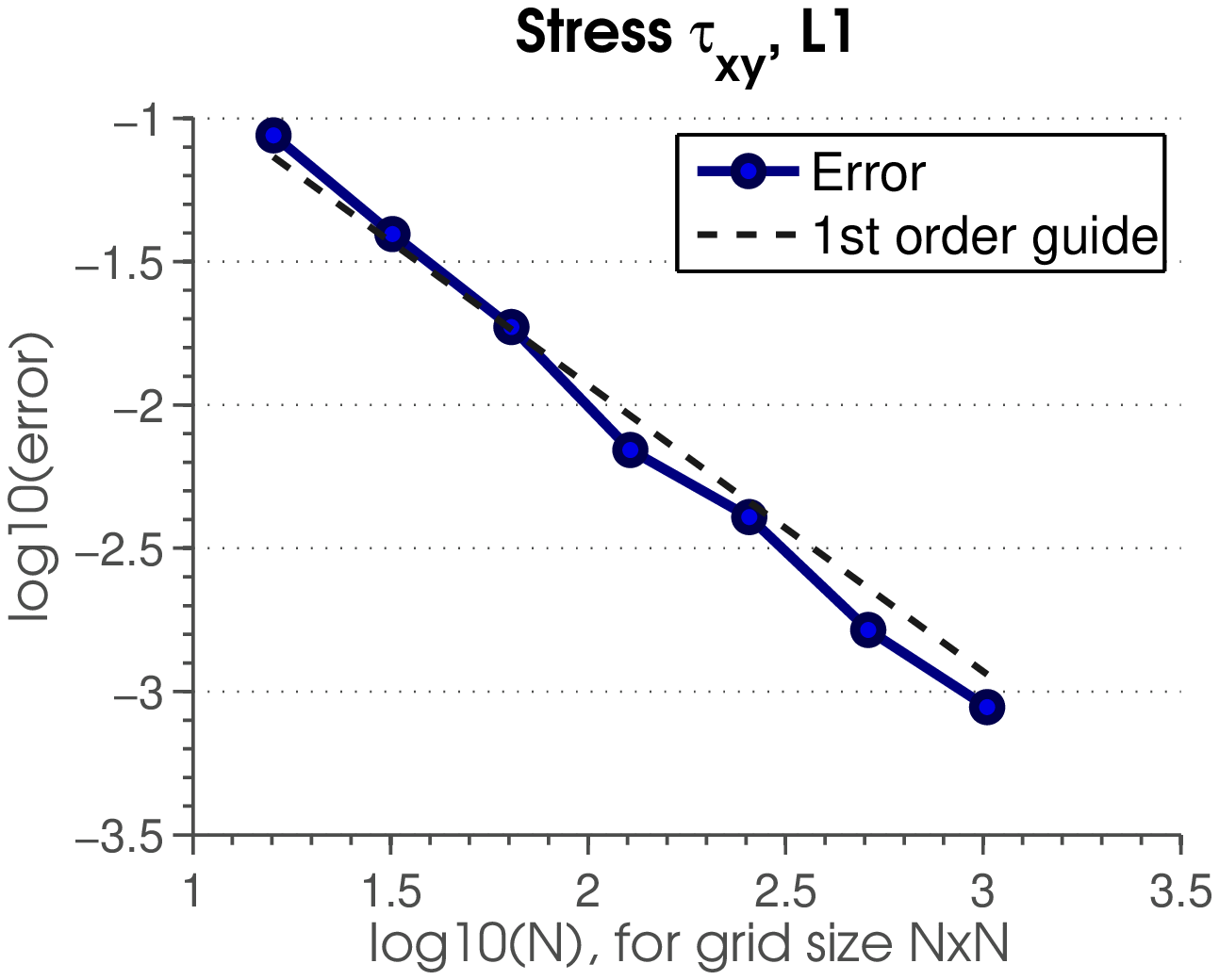}
}
\subfigure[]
{
\includegraphics[width=55mm]{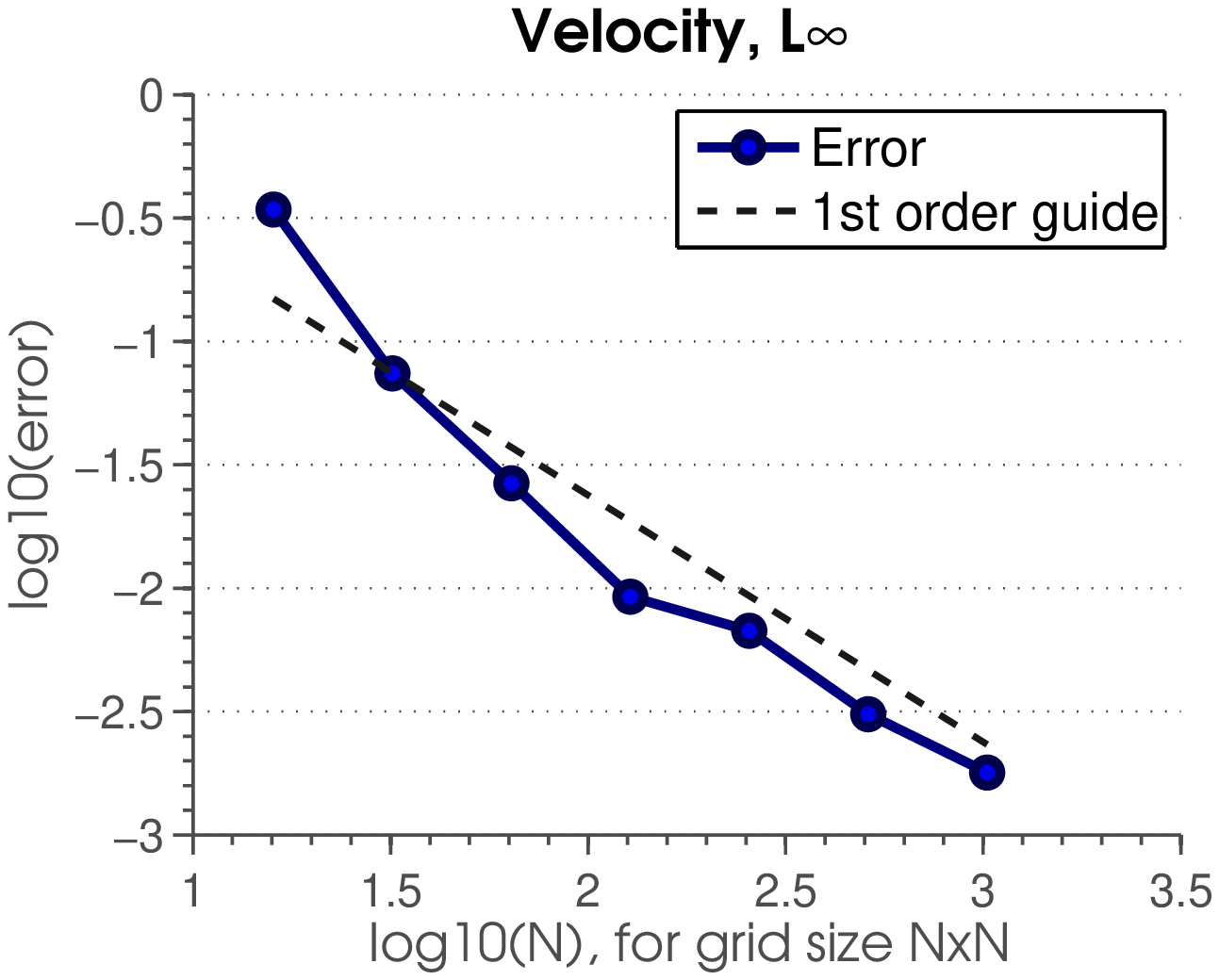}
}
\subfigure[]
{
\includegraphics[width=55mm]{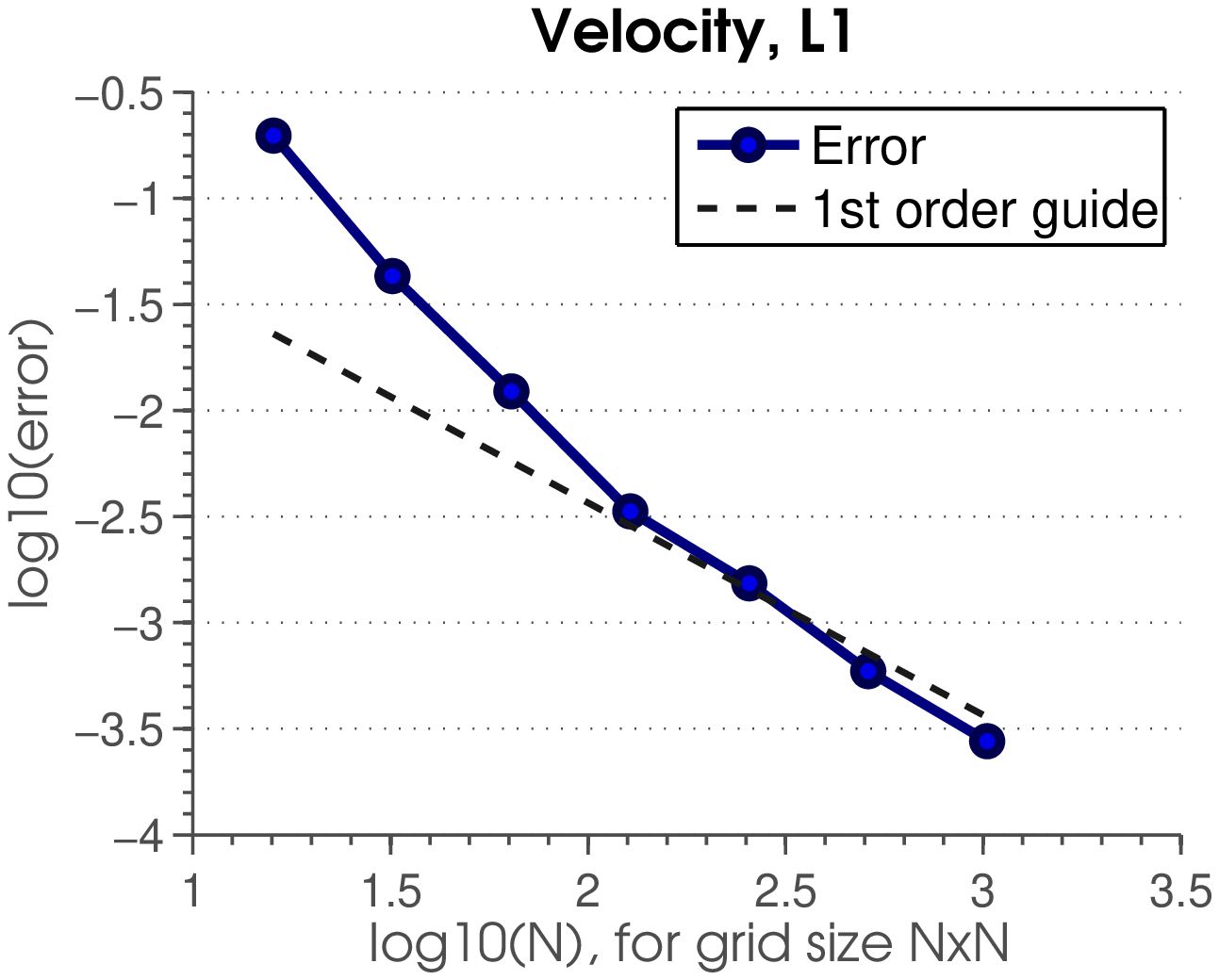}
}
\caption{Convergence graphs for the Stokes problem with free surface boundaries (2D).}
\label{fig:stokes_fs}
\end{figure}

Our free surface Stokes test case is a fluid disk of radius $r=0.75$ centred at the origin, with density $\rho=1$ and viscosity $\mu = 0.1$, computed over a timestep $\Delta t = 1$ .  For simplicity of presentation, we describe the final velocity field in terms of a streamfunction, $\psi$, where the velocity field can be derived as $\vec{u}_{final} = \nabla \times \psi$.  This also guarantees that the velocity field is divergence free.  The streamfunction is:
\begin{equation}
\psi = \frac{128}{81}r^4 \cos(2\theta) \cos(\sqrt 3 \ln r) \left(15 - 30r + 16r^2\right)
\end{equation}
This is a non-trivial velocity field designed to fulfill the free surface zero traction condition at $r=1$, smoothly blended into a zero velocity at the origin ($r=0)$.
The zero traction condition (\ref{freesurfacebc}) enforces a relationship between the surface pressure and the viscous stress resulting from this velocity field. To satisfy this condition, we use the following expression for pressure:
\begin{equation}
p = \frac{512 \sqrt 3}{81} r^2 \mu \sin(2 \theta) \sin(\sqrt 3 \ln r)(15 - 30r + 16r^2)
\end{equation}
The pressure in this expression will be non-zero at the interface; any method to solve this problem will need to correctly handle the coupling between pressure and viscous stresses. From this information, the expressions for the input velocity and final stresses can be derived using equations (\ref{stokesequations1})-(\ref{stokesequations3}).  We used a computer algebra system for this purpose. The convergence results are shown in Table \ref{stokesfreesurfacechart} and Figure \ref{fig:stokes_fs}.

\begin{table}
\caption{Convergence of Stokes with free surface (2D)}
\centering
\label{stokesfreesurfacechart}
\begin{tabular}{|c | c c | c c |}
\hline
Grid & $\|p - p^h\|_{\infty}$ & Order  & $\|p - p^h\|_{1}$ & Order \\
\hline
$16^2$          & 1.2428E-001   &               & 1.4886E-001   & \\
$32^2$          & 1.9439E-001   & -0.65         & 5.3215E-002   & 1.48 \\
$64^2$          & 1.6330E-001   & 0.25          & 2.2135E-002   & 1.27 \\
$128^2$         & 1.3291E-001   & 0.30          & 8.1145E-003   & 1.45 \\
$256^2$         & 1.7242E-001   & -0.38         & 5.3320E-003   & 0.61 \\
$512^2$         & 1.7261E-001   & -0.00         & 2.1992E-003   & 1.28 \\
$1024^2$        & 1.6894E-001   & 0.03          & 1.1239E-003   & 0.97 \\
\hline
Grid & $\|\tau_{xx} - \tau_{xx}^h\|_{\infty}$ & Order  & $\|\tau_{xx} - \tau_{xx}^h\|_{1}$ & Order \\
\hline
$16^2$          & 2.7154E-001   &               & 1.7116E-001   & \\
$32^2$          & 1.4558E-001   & 0.90          & 3.8890E-002   & 2.14 \\
$64^2$          & 7.8877E-002   & 0.88          & 1.4642E-002   & 1.41 \\
$128^2$         & 5.5237E-002   & 0.51          & 5.6140E-003   & 1.38 \\
$256^2$         & 7.0687E-002   & -0.36         & 3.2991E-003   & 0.77 \\
$512^2$         & 7.1097E-002   & -0.01         & 1.3680E-003   & 1.27 \\
$1024^2$        & 7.4305E-002   & -0.06         & 7.1219E-004   & 0.94 \\
\hline
Grid & $\|\tau_{xy} - \tau_{xy}^h\|_{\infty}$ & Order  & $\|\tau_{xy} - \tau_{xy}^h\|_{1}$ & Order \\
\hline
$16^2$          & 1.6864E-001   &               & 8.7332E-002   & \\
$32^2$          & 1.0266E-001   & 0.72          & 3.9547E-002   & 1.14 \\
$64^2$          & 2.1950E-001   & -1.10         & 1.8694E-002   & 1.08 \\
$128^2$         & 2.0094E-001   & 0.13          & 6.9612E-003   & 1.43 \\
$256^2$         & 2.6596E-001   & -0.40         & 4.0559E-003   & 0.78 \\
$512^2$         & 2.0961E-001   & 0.34          & 1.6447E-003   & 1.30 \\
$1024^2$        & 2.5255E-001   & -0.27         & 8.8242E-004   & 0.90 \\
\hline
Grid & $\|u - u^h\|_{\infty}$ & Order  & $\|u - u^h\|_{1}$ & Order \\
\hline
$16^2$          & 3.4171E-001   &               & 1.9727E-001   & \\
$32^2$          & 7.4246E-002   & 2.20          & 4.3033E-002   & 2.20 \\
$64^2$          & 2.6593E-002   & 1.48          & 1.2321E-002   & 1.80 \\
$128^2$         & 9.2292E-003   & 1.53          & 3.3497E-003   & 1.88 \\
$256^2$         & 6.7182E-003   & 0.46          & 1.5327E-003   & 1.13 \\
$512^2$         & 3.0843E-003   & 1.12          & 5.9129E-004   & 1.37 \\
$1024^2$        & 1.7877E-003   & 0.79          & 2.7579E-004   & 1.10 \\
\hline
\end{tabular}
\end{table}

\subsection{Stokes Flow with Solid Wall Boundaries (2D)}
\label{solidwall_section}
\begin{figure}
\centering
\subfigure[]
{
\includegraphics[width=55mm]{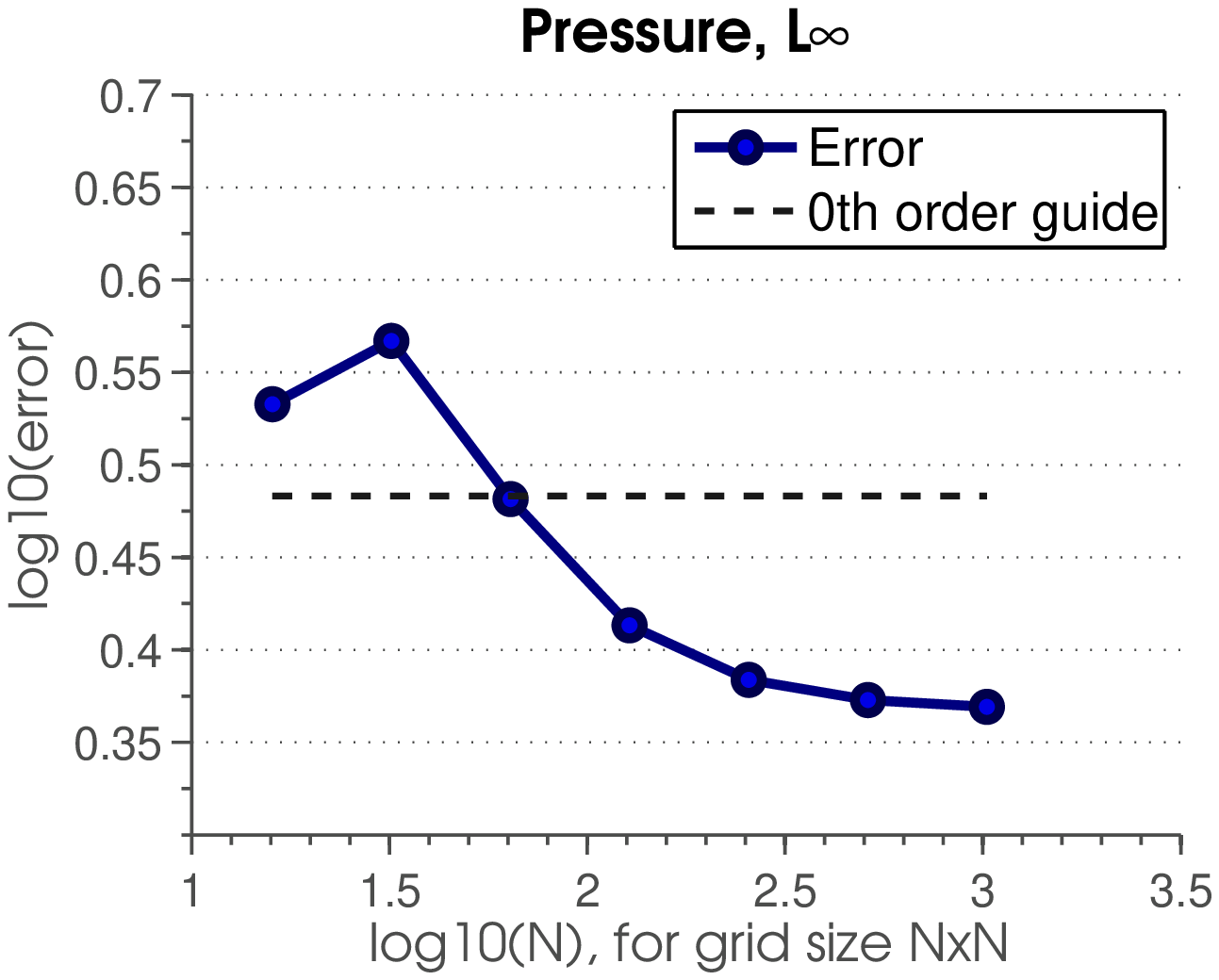}
}
\subfigure[]
{
\includegraphics[width=55mm]{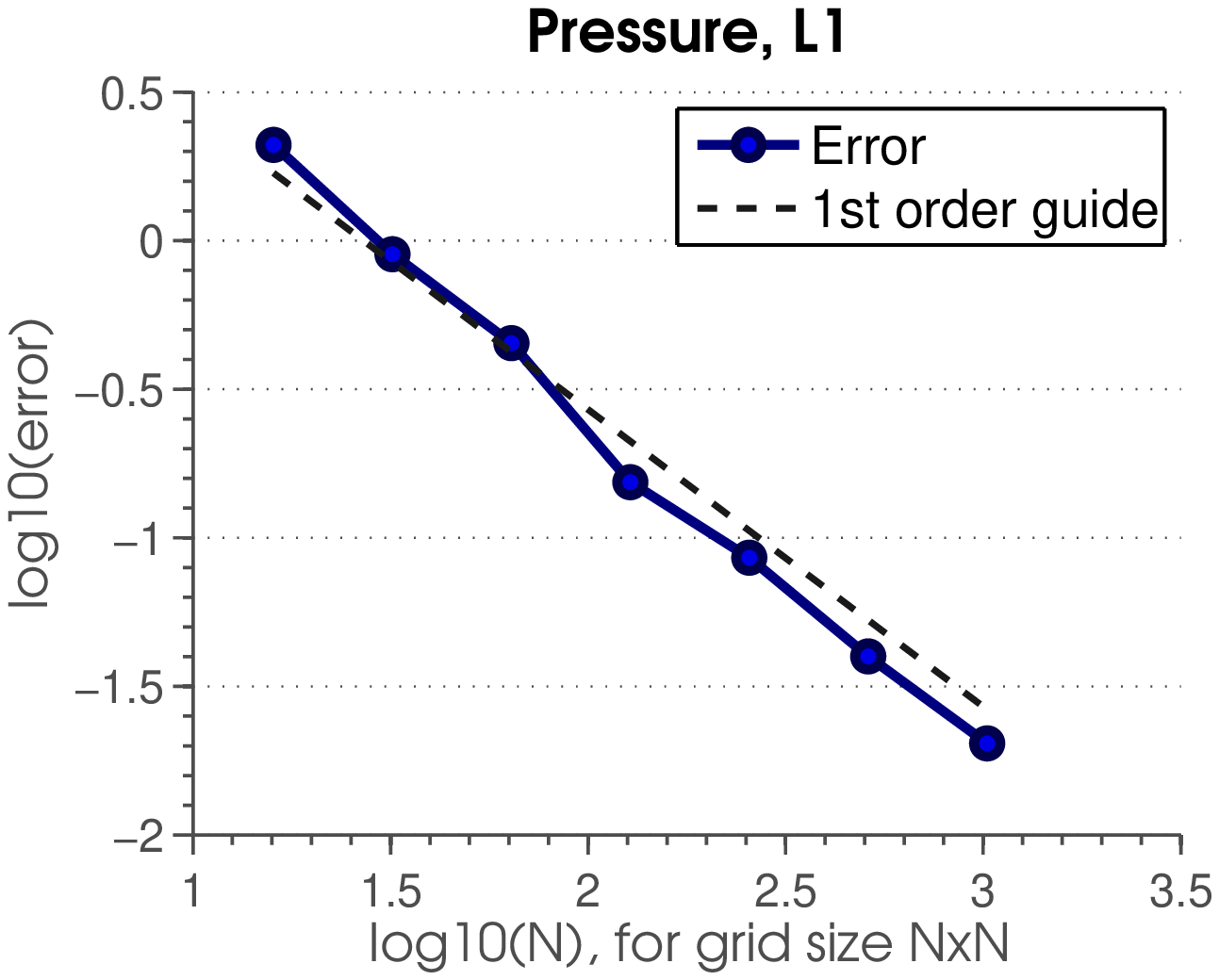}
}
\subfigure[]
{
\includegraphics[width=55mm]{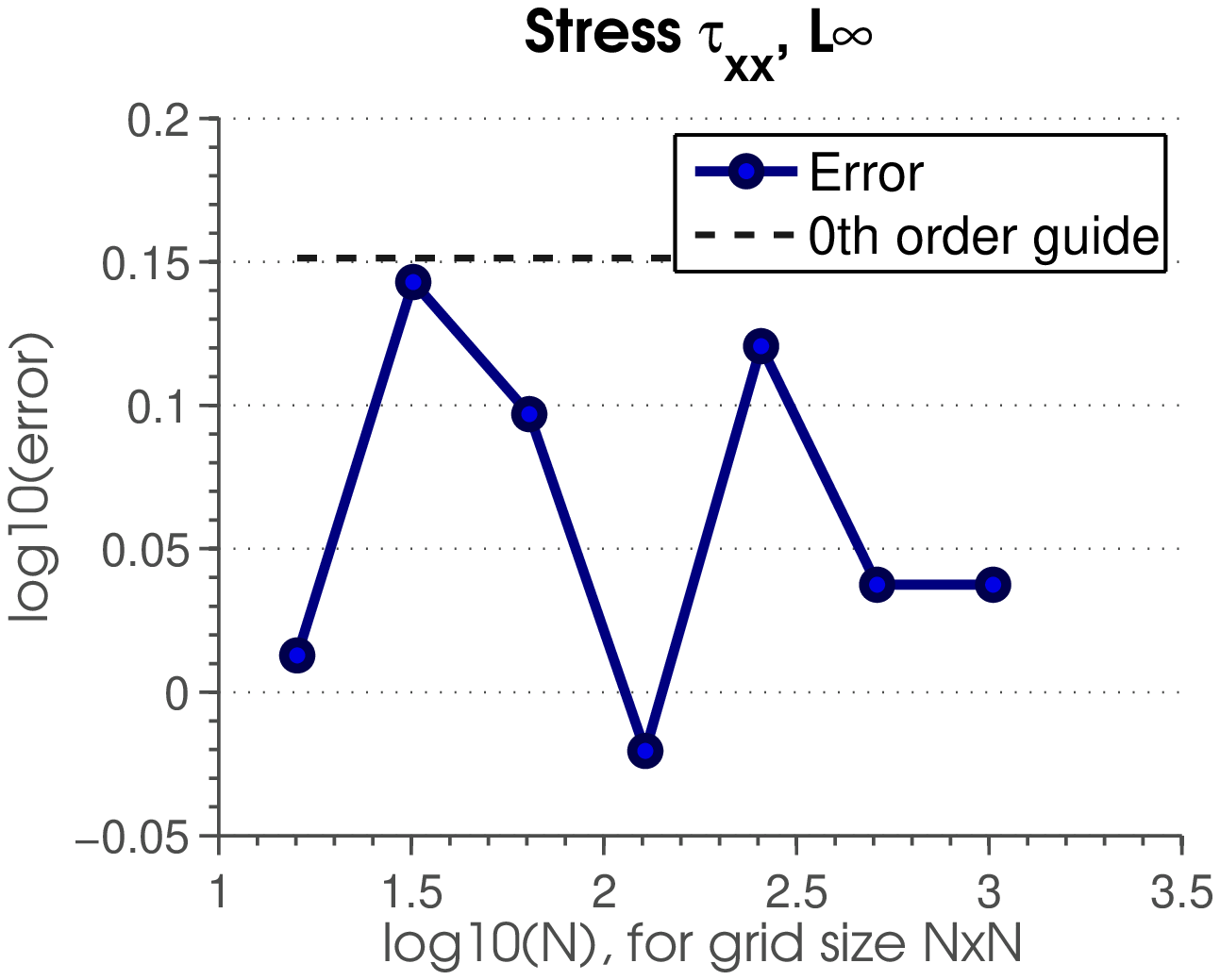}
}
\subfigure[]
{
\includegraphics[width=55mm]{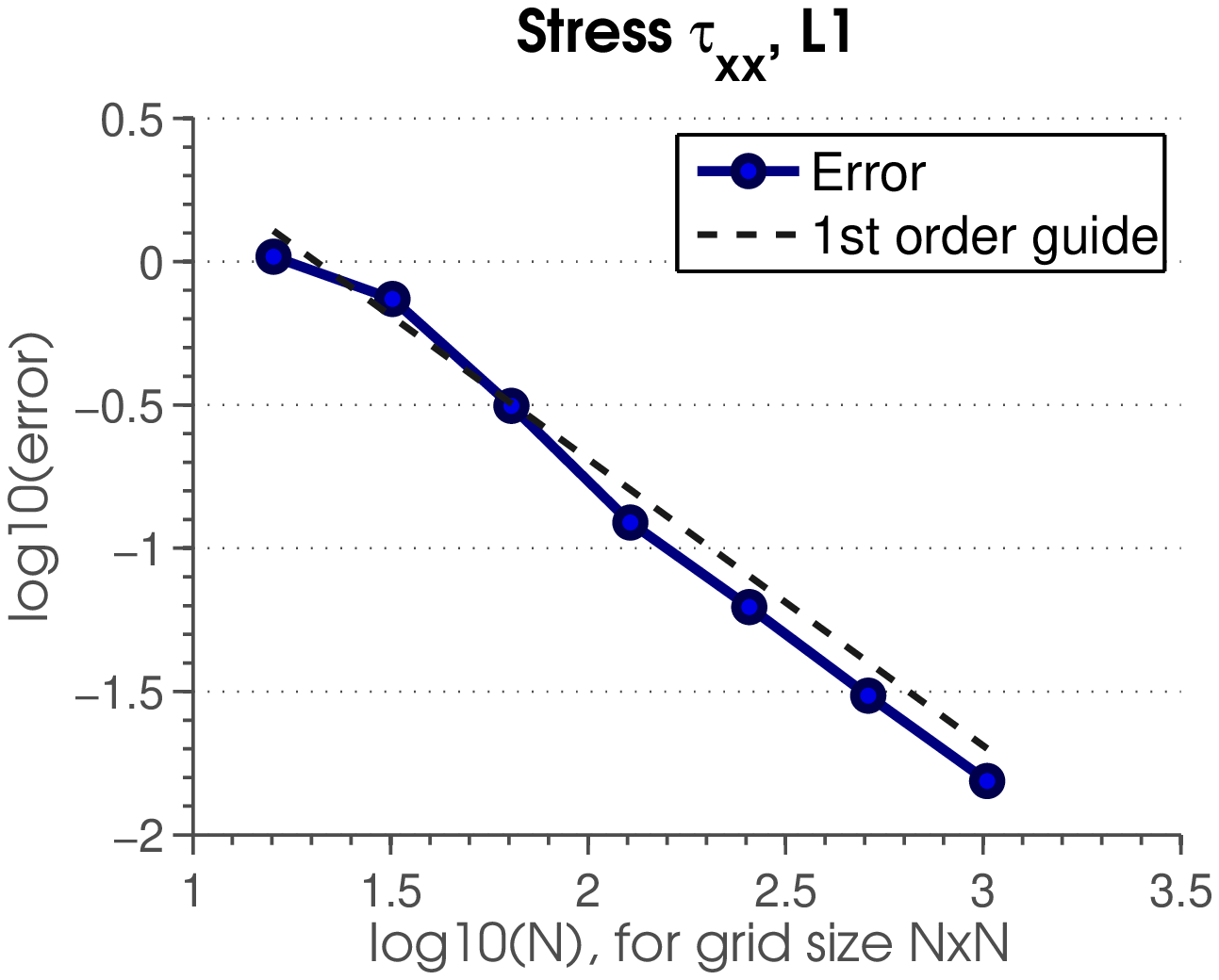}
}
\subfigure[]
{
\includegraphics[width=55mm]{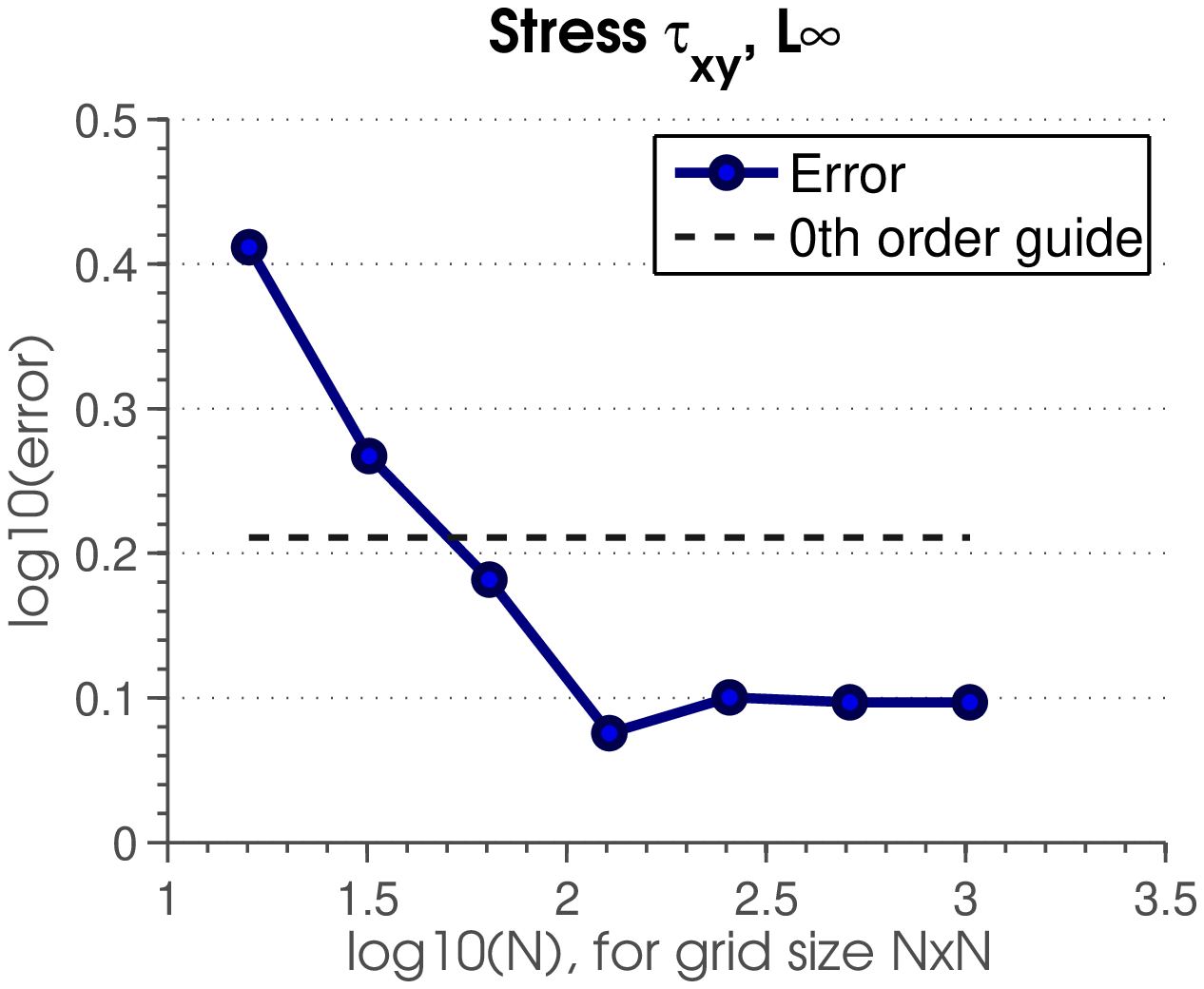}
}
\subfigure[]
{
\includegraphics[width=55mm]{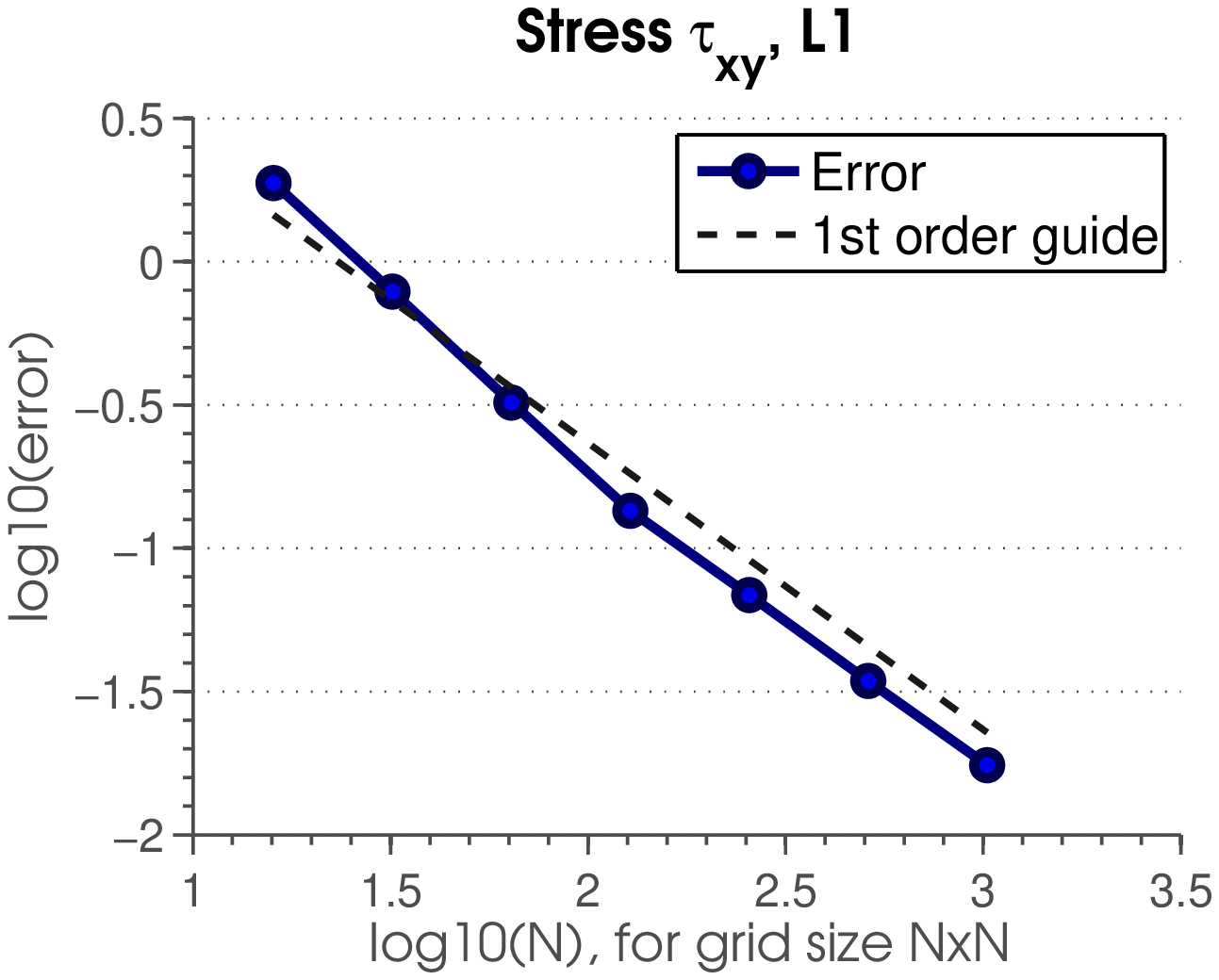}
}
\subfigure[]
{
\includegraphics[width=55mm]{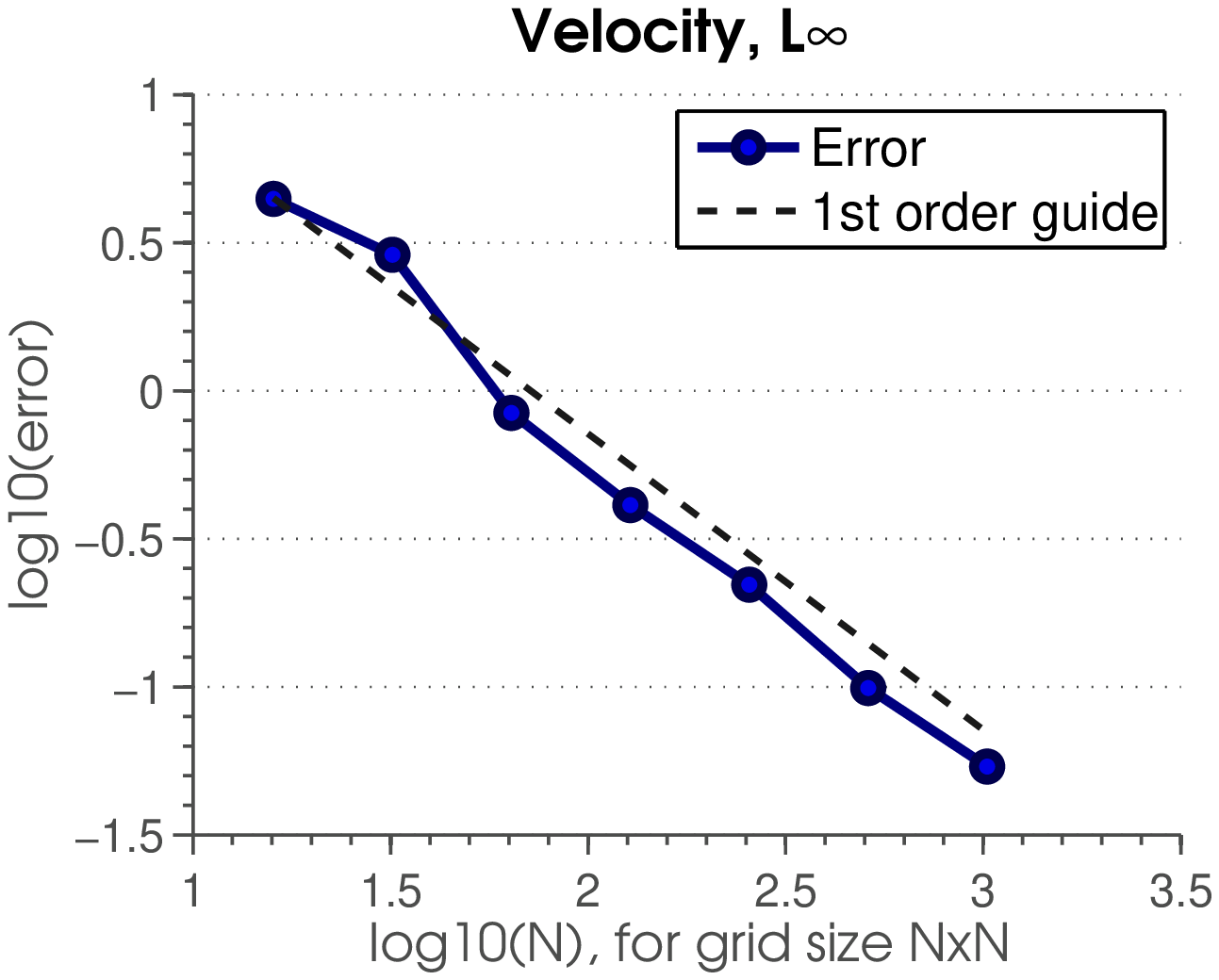}
}
\subfigure[]
{
\includegraphics[width=55mm]{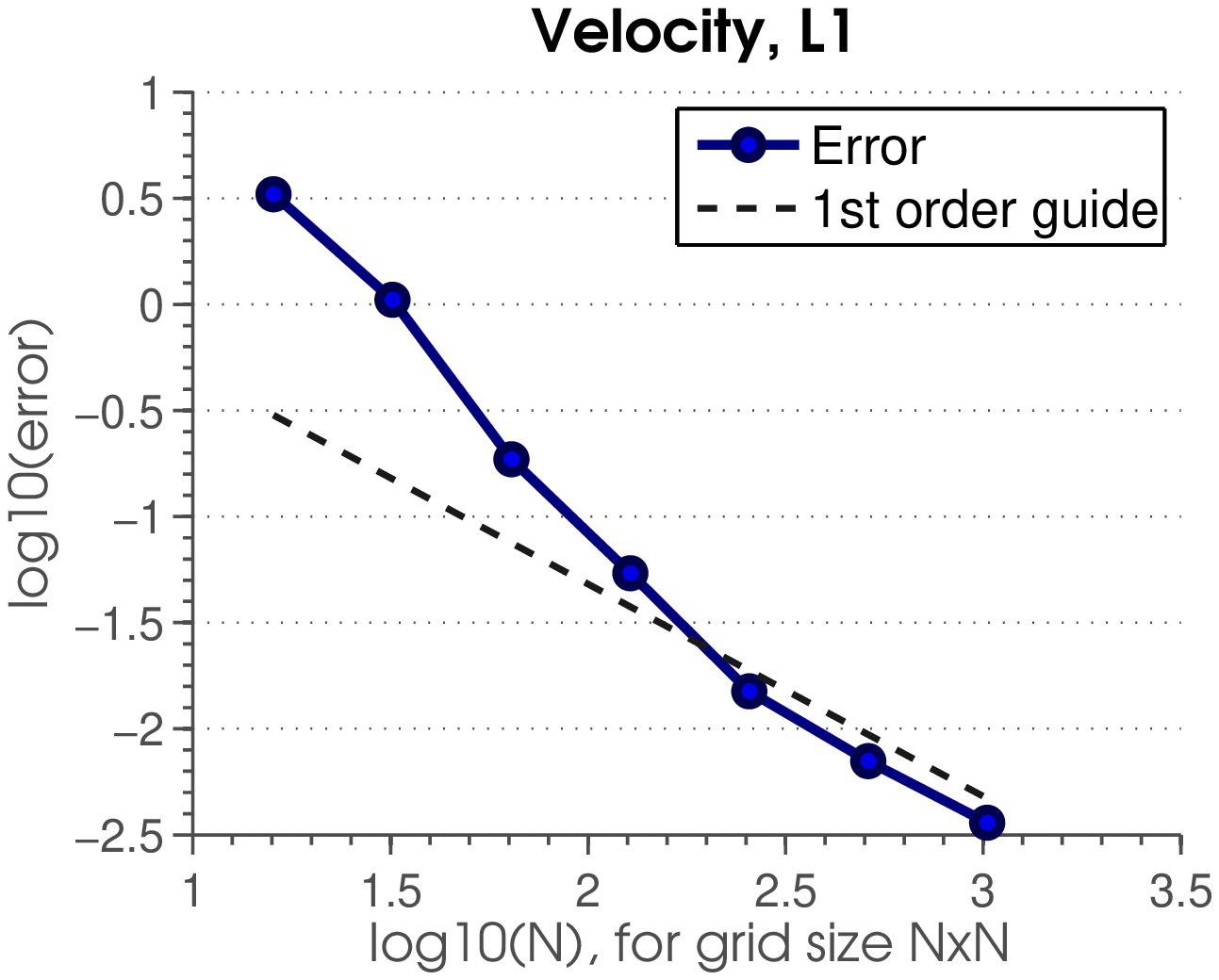}
}
\caption{Convergence graphs for the Stokes problem with solid wall boundaries (2D).}
\label{fig:stokes_solid}
\end{figure}

Our Stokes solid boundary test case is an annulus centred at the origin with inner radius $r=0.5$, outer radius $r=1$, density $\rho=1$, and viscosity $\mu = 0.1$, computed over a timestep $\Delta t = 1$. Inner and outer boundaries are static solids.  We will again use a streamfunction $\psi$ to dictate our velocity field and ensure it is divergence free:
\begin{equation}
\psi = 256r^4-768r^3+832r^2-384r+64
\end{equation}
For pressure, we use:
\begin{equation}
p = r^2 \cos(\theta) \sin(\theta)
\end{equation}
Equations (\ref{stokesequations1})-(\ref{stokesequations3}) can be used to derive the input velocities and final stresses.  The convergence results are shown in Table \ref{stokessolidwallchart} and Figure \ref{fig:stokes_solid}.

\begin{table}
\caption{Convergence of Stokes with solid walls (2D)}
\centering
\label{stokessolidwallchart}
\begin{tabular}{|c | c c | c c |}
\hline
Grid & $\|p - p^h\|_{\infty}$ & Order  & $\|p - p^h\|_{1}$ & Order \\
\hline
$16^2$          & 3.4118E+000    &              & 2.1013E+000    & \\
$32^2$          & 3.6933E+000    & -0.11        & 9.0009E-001    & 1.22 \\
$64^2$          & 3.0338E+000    & 0.28          & 4.5162E-001    & 0.99 \\
$128^2$        & 2.5942E+000    & 0.23          & 1.5373E-001    & 1.55 \\
$256^2$        & 2.4222E+000    & 0.10          & 8.5695E-002    & 0.84 \\
$512^2$        & 2.3573E+000    & 0.04          & 3.9879E-002    & 1.10 \\
$1024^2$        & 2.3381E+000    & 0.01          & 2.0266E-002    & 0.98 \\
\hline
Grid & $\|\tau_{xx} - \tau_{xx}^h\|_{\infty}$ & Order  & $\|\tau_{xx} - \tau_{xx}^h\|_{1}$ & Order \\
\hline
$16^2$          & 1.0293E+000    &              & 1.0421E+000    & \\
$32^2$          & 1.3935E+000    & -0.44        & 7.4091E-001    & 0.49 \\
$64^2$          & 1.2483E+000    & 0.16          & 3.1356E-001    & 1.24 \\
$128^2$        & 9.5438E-001    & 0.39          & 1.2345E-001    & 1.34 \\
$256^2$        & 1.3160E+000    & -0.46        & 6.2448E-002    & 0.98 \\
$512^2$        & 1.0903E+000    & 0.27          & 3.0554E-002    & 1.03 \\
$1024^2$        & 1.0943E+000    & -0.01        & 1.5407E-002    & 0.99 \\
\hline
Grid & $\|\tau_{xy} - \tau_{xy}^h\|_{\infty}$ & Order  & $\|\tau_{xy} - \tau_{xy}^h\|_{1}$ & Order \\
\hline
$16^2$          & 2.5836E+000    &              & 1.8844E+000    & \\
$32^2$          & 1.8510E+000    & 0.48          & 7.8641E-001    & 1.26 \\
$64^2$          & 1.5152E+000    & 0.29          & 3.2160E-001    & 1.29 \\
$128^2$        & 1.1931E+000    & 0.34          & 1.3456E-001    & 1.26 \\
$256^2$        & 1.2560E+000    & -0.07        & 6.8684E-002    & 0.97 \\
$512^2$        & 1.2546E+000    & 0.00          & 3.4391E-002    & 1.00 \\
$1024^2$        & 1.2466E+000    & 0.01          & 1.7487E-002    & 0.98 \\
\hline
Grid & $\|u - u^h\|_{\infty}$ & Order  & $\|u - u^h\|_{1}$ & Order \\
\hline
$16^2$          & 4.4449E+000    &              & 3.2953E+000    & \\
$32^2$          & 2.8765E+000    & 0.63          & 1.0500E+000    & 1.65 \\
$64^2$          & 8.4194E-001    & 1.77          & 1.8614E-001    & 2.50 \\
$128^2$        & 4.1100E-001    & 1.03          & 5.3975E-002    & 1.79 \\
$256^2$        & 2.2147E-001    & 0.89          & 1.4992E-002    & 1.85 \\
$512^2$        & 9.8967E-002    & 1.16          & 7.0519E-003    & 1.09 \\
$1024^2$        & 5.3807E-002    & 0.88          & 3.6056E-003    & 0.97 \\
\hline
\end{tabular}
\end{table}

\subsection{Stokes Flow with Both Solid Wall and Free Surface Boundaries (2D)}

\begin{figure}
\centering
\subfigure[]
{
\includegraphics[width=55mm]{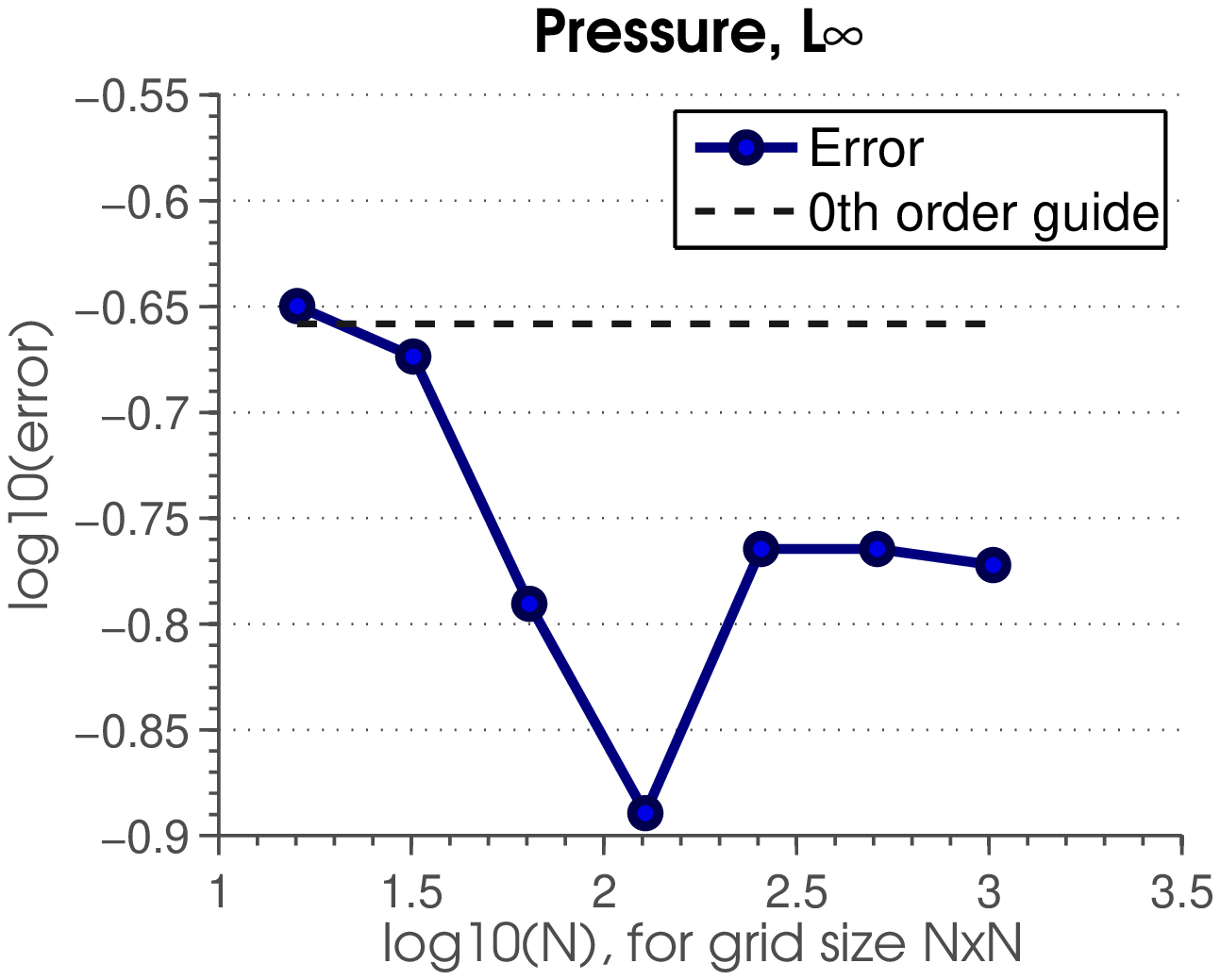}
}
\subfigure[]
{
\includegraphics[width=55mm]{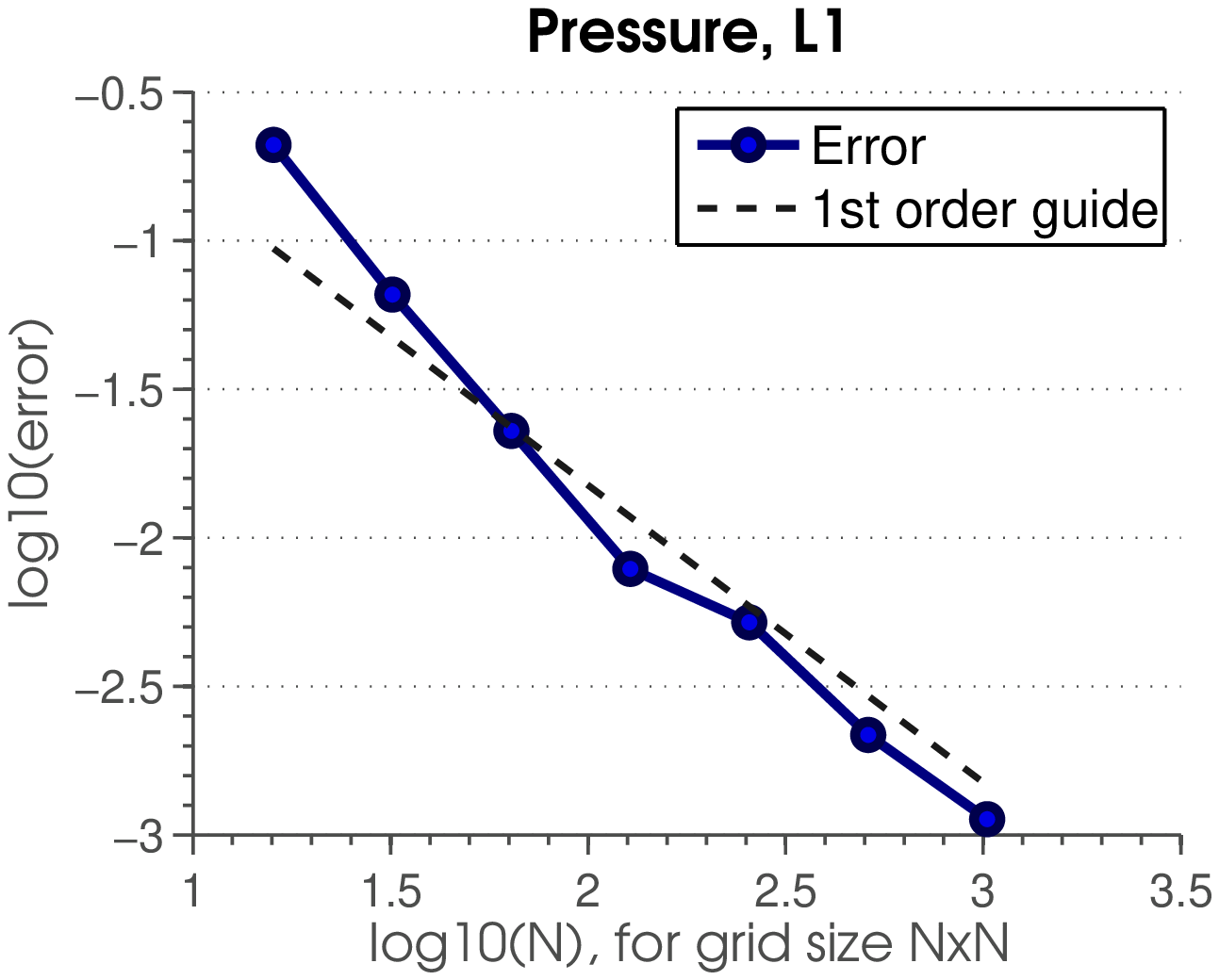}
}
\subfigure[]
{
\includegraphics[width=55mm]{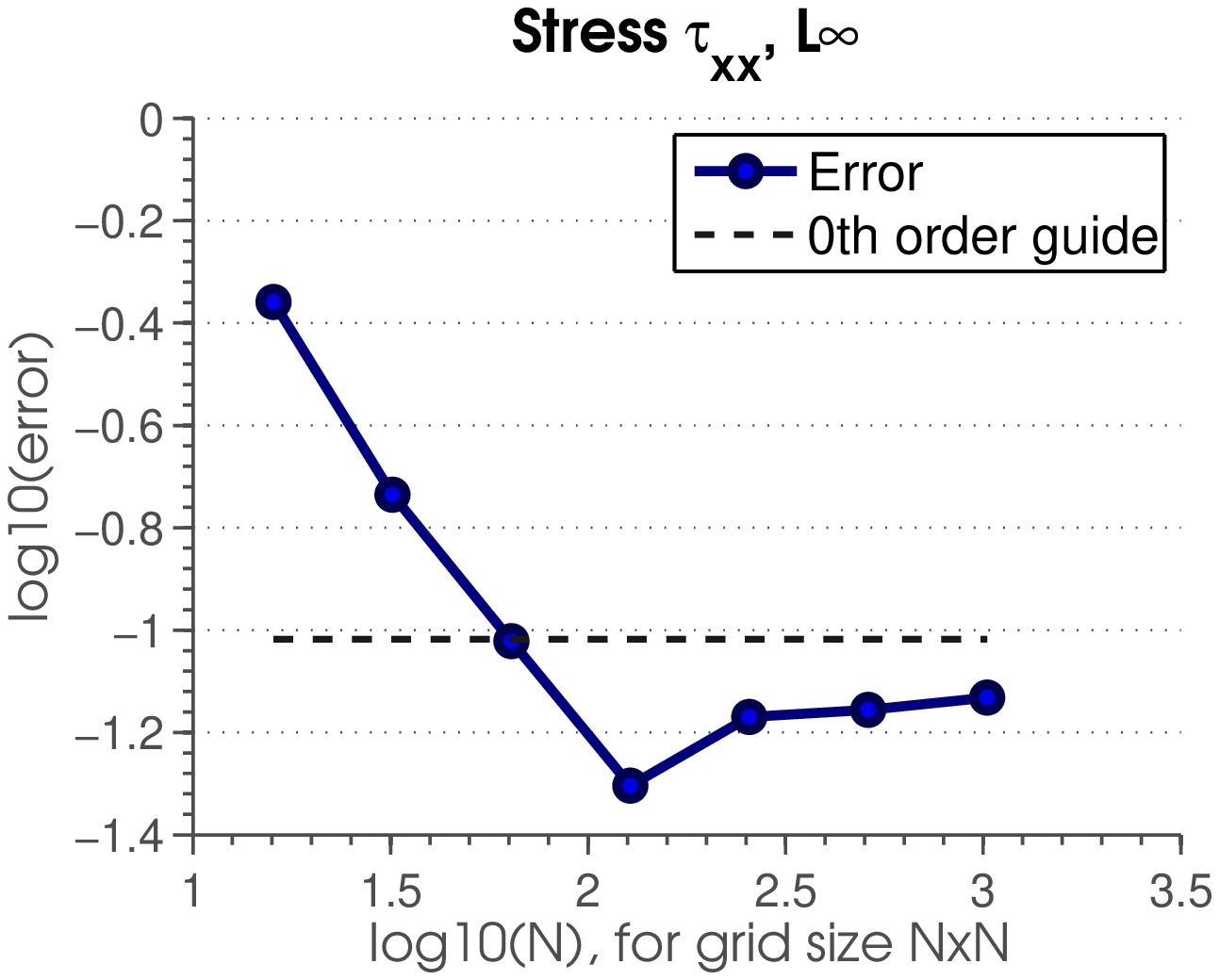}
}
\subfigure[]
{
\includegraphics[width=55mm]{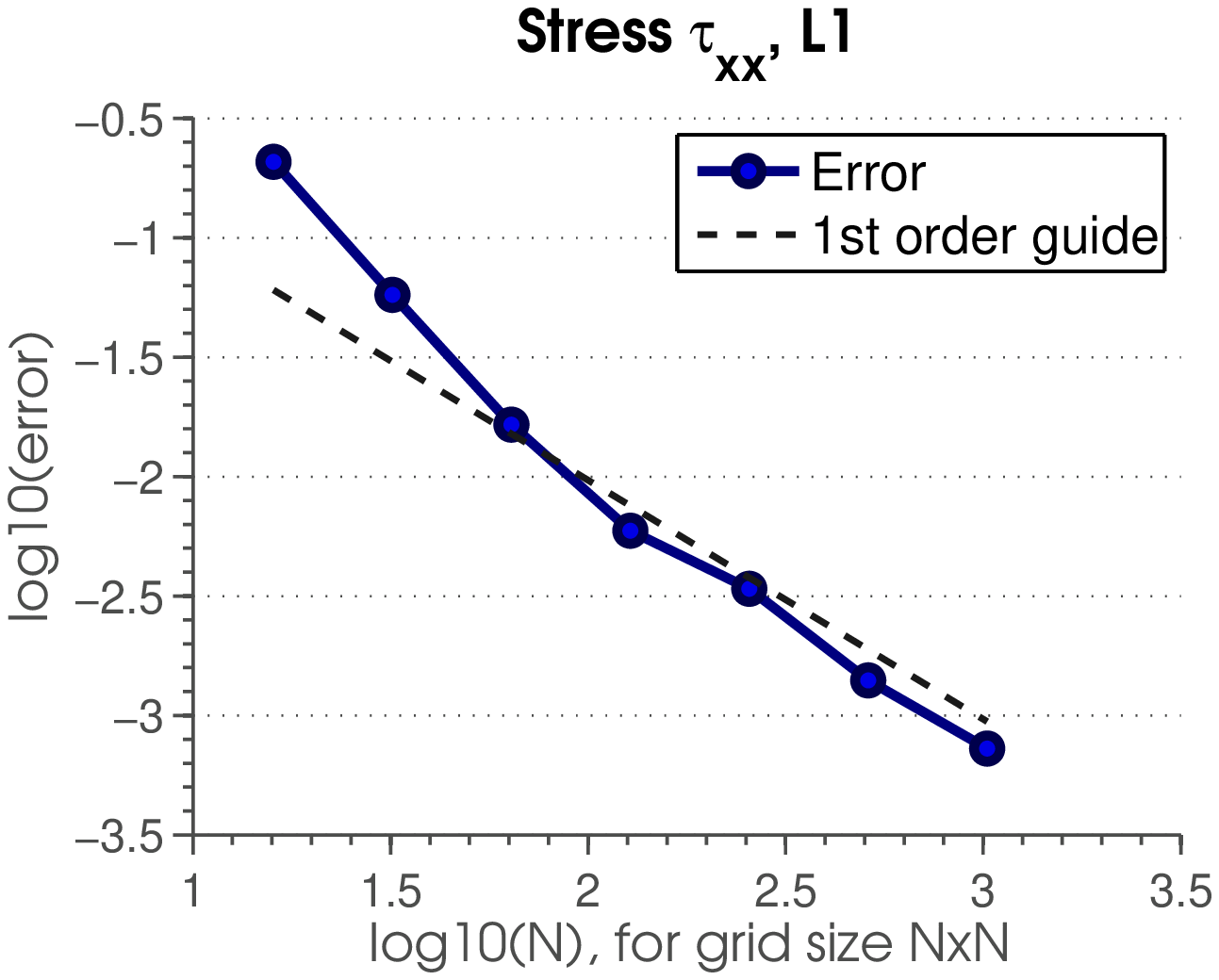}
}
\subfigure[]
{
\includegraphics[width=55mm]{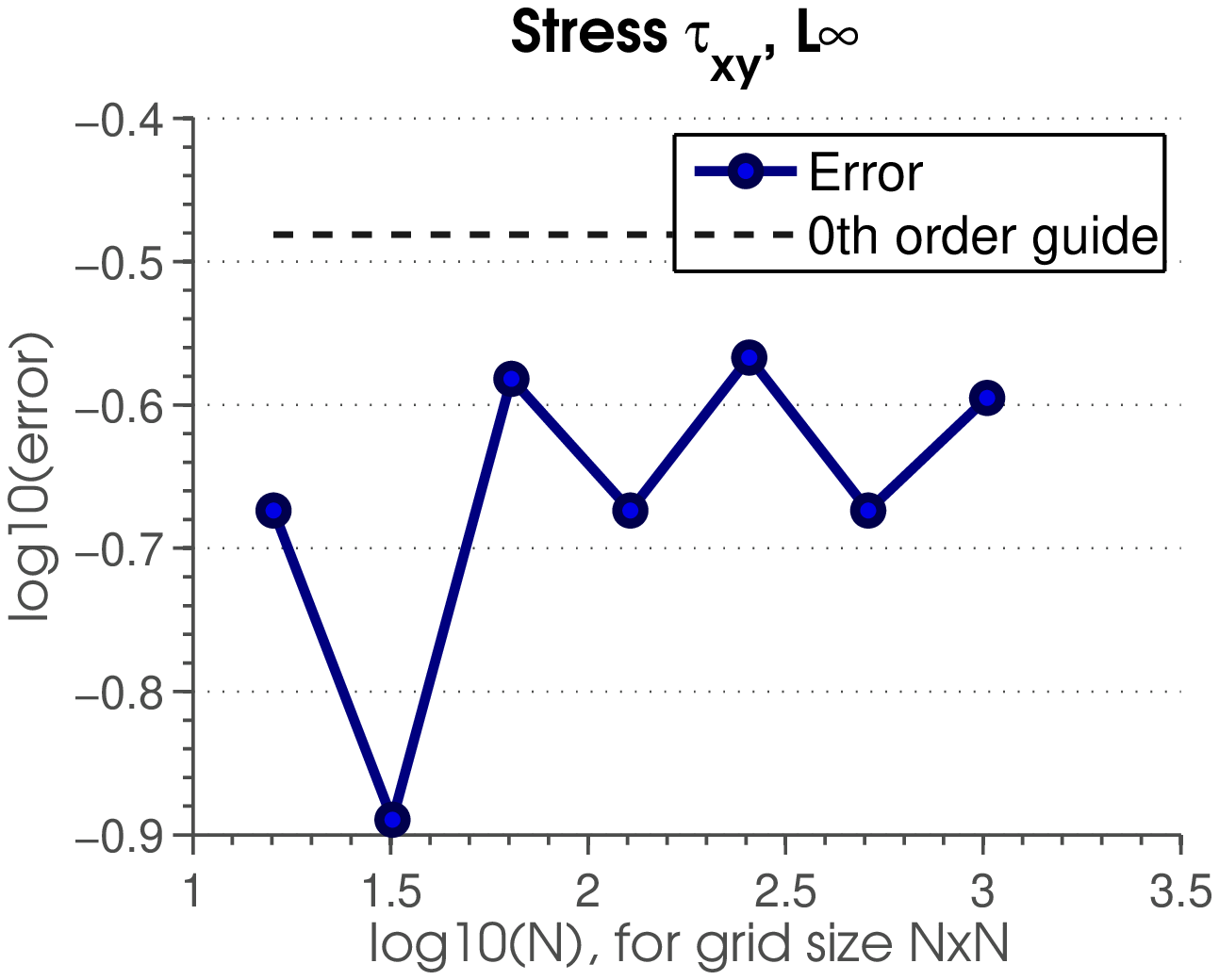}
}
\subfigure[]
{
\includegraphics[width=55mm]{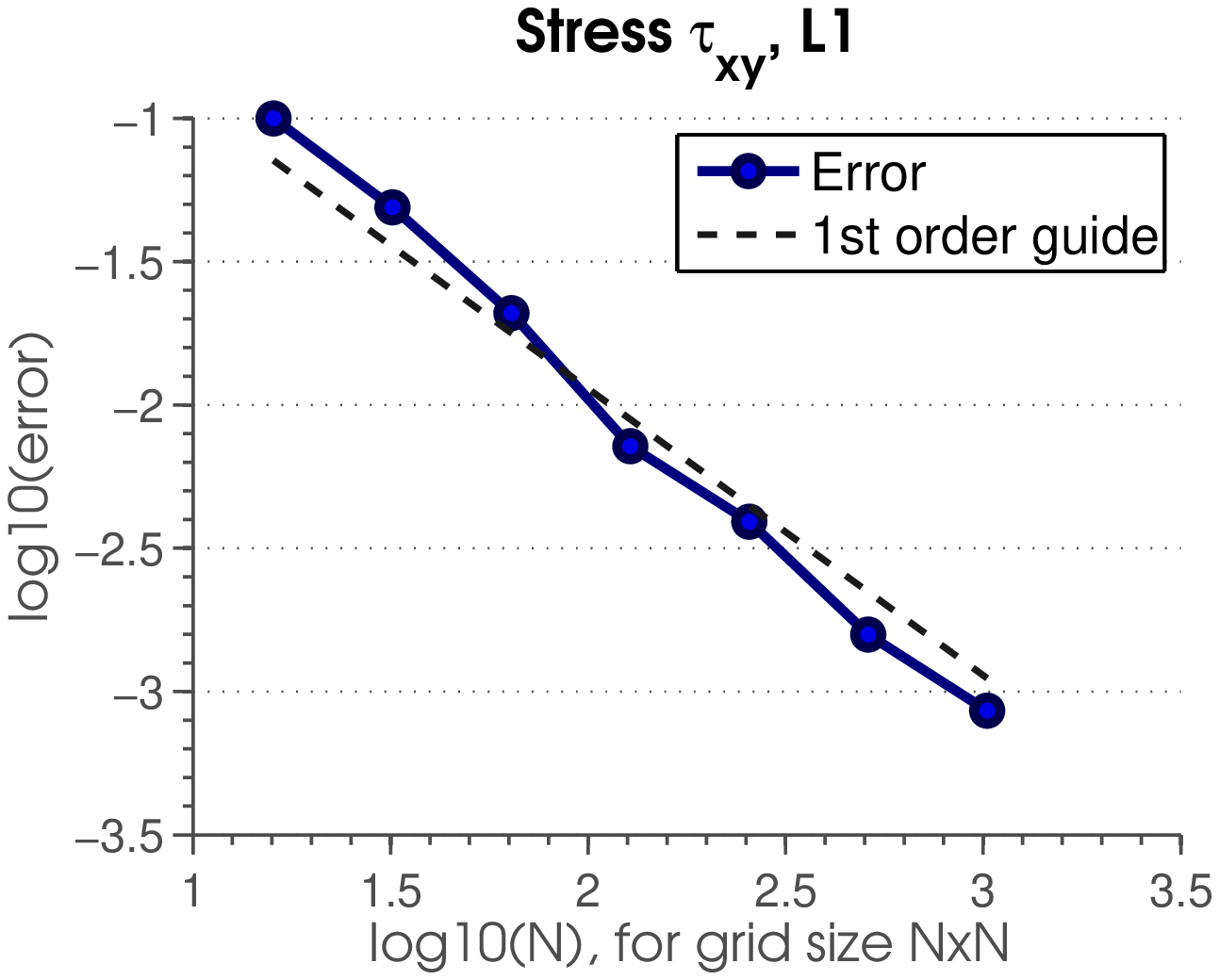}
}
\subfigure[]
{
\includegraphics[width=55mm]{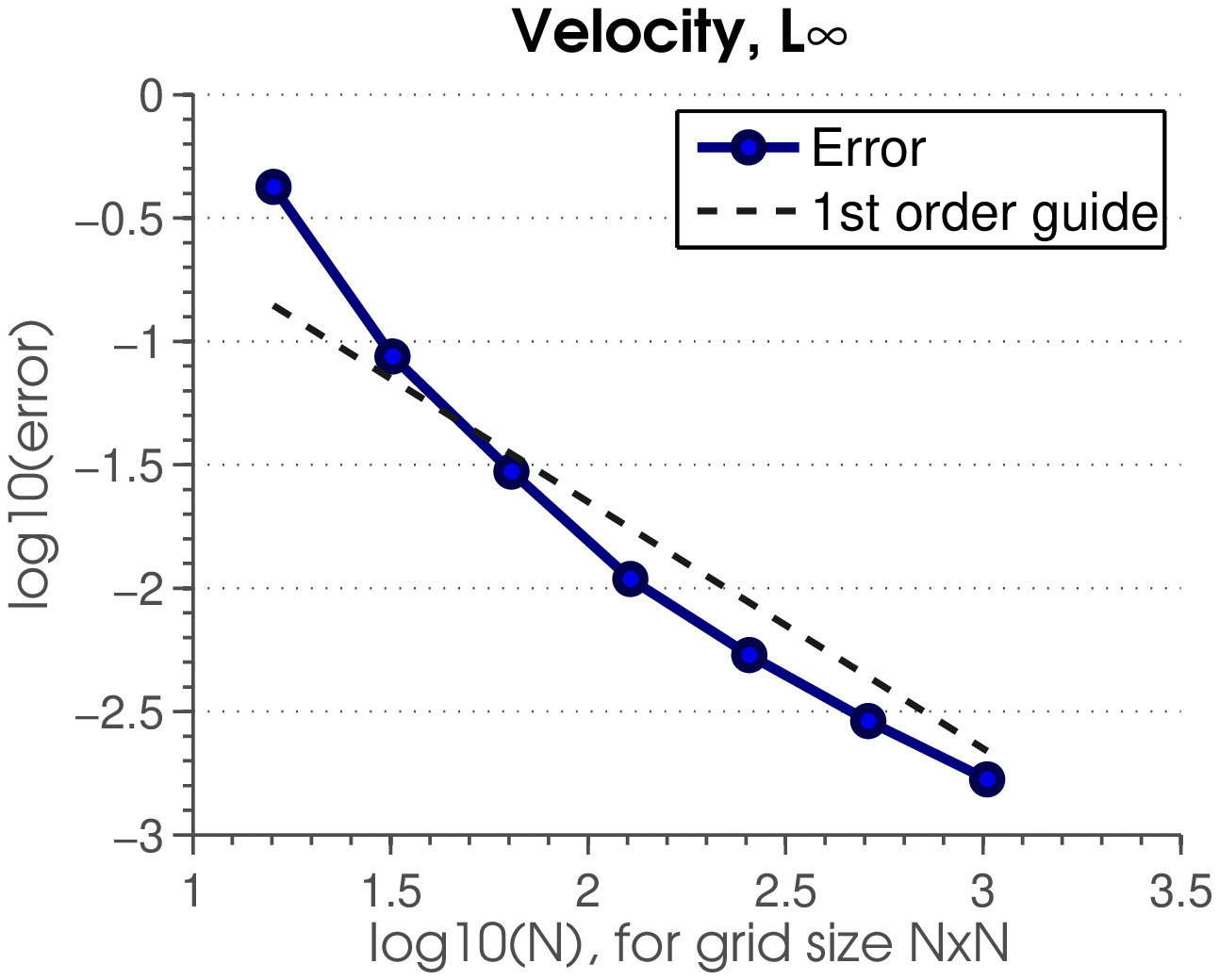}
}
\subfigure[]
{
\includegraphics[width=55mm]{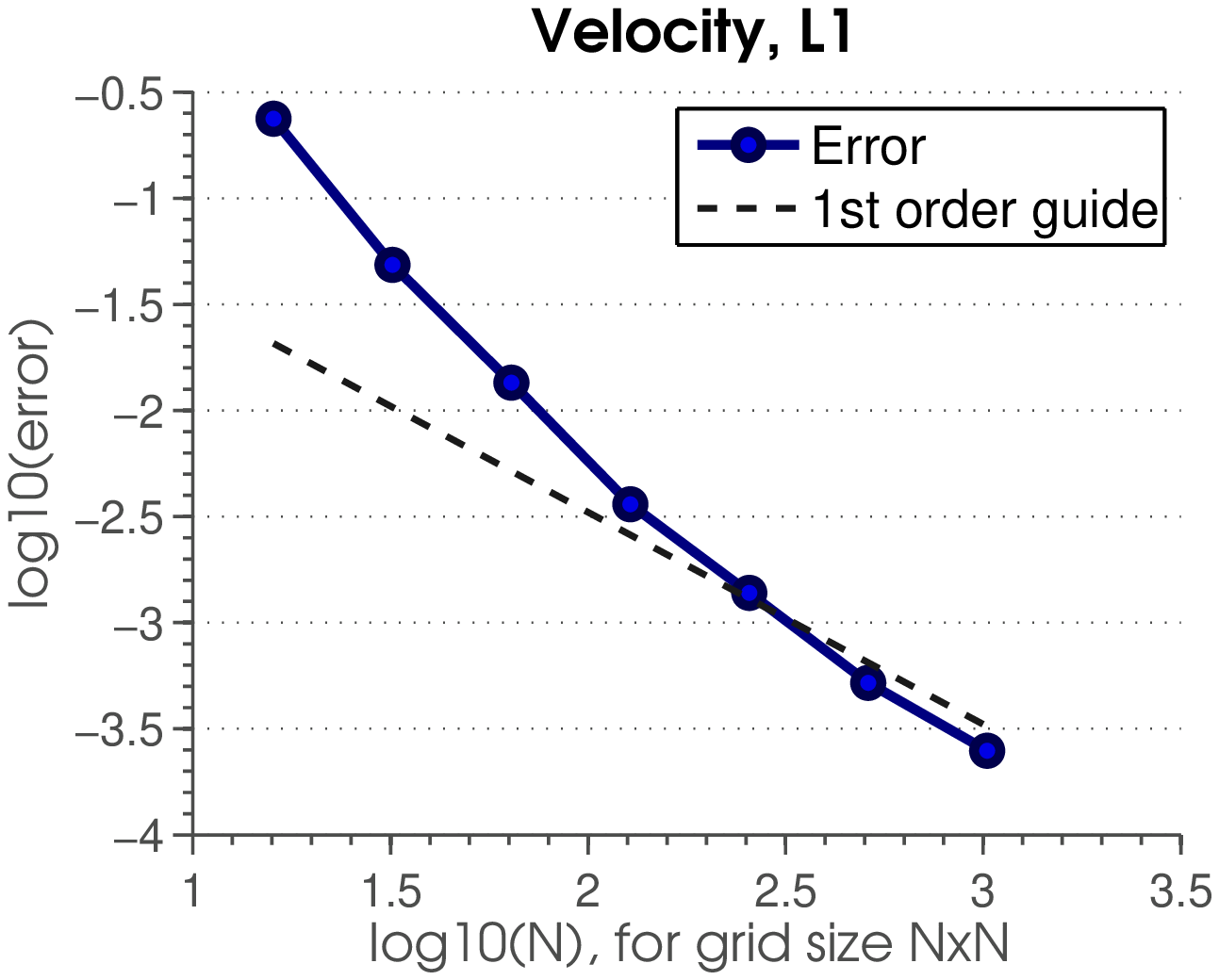}
}
\caption{Convergence graphs for the Stokes problem with both a free surface and a solid wall boundary (2D).}
\label{fig:stokes_both}
\end{figure}

To test a scenario featuring both solid and free surface boundaries, we solve for fluid motion in an annulus centred at the origin with inner radius $r = 0.1$, outer radius $r=0.75$, density $\rho = 1$, and viscosity $\mu = 0.1$, over a timestep $\Delta t = 1$.  The outer boundary is a free surface, and the inner boundary is a static solid.  We will again use a streamfunction $\psi$ to dictate our velocity field and ensure it is divergence free:
\begin{equation}
\psi = \frac{80000}{371293} \cos(2\theta) r^4 \cos(\sqrt 3 \ln r)(169-390r+240r^2)
\end{equation}
For pressure, we use:
\begin{equation}
p = \frac{320000}{371293}\sqrt 3 \mu r^2 \sin(2 \theta) \sin(\sqrt 3 \ln r)(169-390r+240r^2)
\end{equation}
Equations (\ref{stokesequations1})-(\ref{stokesequations3}) can be used to derive the input velocities and final stresses.  The convergence results are shown in Table \ref{stokesmixedchart} and Figure \ref{fig:stokes_both}.
\begin{table}
\caption{Convergence of Stokes with solid walls and a free surface (2D)}
\centering
\label{stokesmixedchart}
\begin{tabular}{|c | c c | c c |}
\hline
Grid & $\|p - p^h\|_{\infty}$ & Order  & $\|p - p^h\|_{1}$ & Order \\
\hline
$16^2$          & 2.2404E-001   &               & 2.1048E-001   & \\
$32^2$          & 2.1216E-001   & 0.08          & 6.5787E-002   & 1.68 \\
$64^2$          & 1.6222E-001   & 0.39          & 2.2932E-002   & 1.52 \\
$128^2$         & 1.2874E-001   & 0.33          & 7.8514E-003   & 1.55 \\
$256^2$         & 1.7168E-001   & -0.42         & 5.1902E-003   & 0.60 \\
$512^2$         & 1.7212E-001   & -0.00         & 2.1723E-003   & 1.26 \\
$1024^2$        & 1.6865E-001   & 0.03          & 1.1268E-003   & 0.95 \\
\hline
Grid & $\|\tau_{xx} - \tau_{xx}^h\|_{\infty}$ & Order  & $\|\tau_{xx} - \tau_{xx}^h\|_{1}$ & Order \\
\hline
$16^2$          & 4.3812E-001   &               & 2.0762E-001   & \\
$32^2$          & 1.8386E-001   & 1.25          & 5.7557E-002   & 1.85 \\
$64^2$          & 9.5150E-002   & 0.95          & 1.6510E-002   & 1.80 \\
$128^2$         & 4.9719E-002   & 0.94          & 5.9357E-003   & 1.48 \\
$256^2$         & 6.7733E-002   & -0.45         & 3.3888E-003   & 0.81 \\
$512^2$         & 6.9916E-002   & -0.05         & 1.3958E-003   & 1.28 \\
$1024^2$        & 7.3807E-002   & -0.08         & 7.2576E-004   & 0.94 \\
\hline
Grid & $\|\tau_{xy} - \tau_{xy}^h\|_{\infty}$ & Order  & $\|\tau_{xy} - \tau_{xy}^h\|_{1}$ & Order \\
\hline
$16^2$          & 2.1174E-001   &               & 1.0047E-001   & \\
$32^2$          & 1.2921E-001   & 0.71          & 4.8916E-002   & 1.04 \\
$64^2$          & 2.6177E-001   & -1.02         & 2.0869E-002   & 1.23 \\
$128^2$         & 2.1197E-001   & 0.30          & 7.1786E-003   & 1.54 \\
$256^2$         & 2.7070E-001   & -0.35         & 3.9133E-003   & 0.88 \\
$512^2$         & 2.1190E-001   & 0.35          & 1.5764E-003   & 1.31 \\
$1024^2$        & 2.5361E-001   & -0.26         & 8.5715E-004   & 0.88 \\
\hline
Grid & $\|u - u^h\|_{\infty}$ & Order  & $\|u - u^h\|_{1}$ & Order \\
\hline
$16^2$          & 4.2290E-001   &               & 2.3654E-001   & \\
$32^2$          & 8.6887E-002   & 2.28          & 4.8516E-002   & 2.29 \\
$64^2$          & 2.9712E-002   & 1.55          & 1.3486E-002   & 1.85 \\
$128^2$         & 1.0943E-002   & 1.44          & 3.6100E-003   & 1.90 \\
$256^2$         & 5.3552E-003   & 1.03          & 1.3797E-003   & 1.39 \\
$512^2$         & 2.8869E-003   & 0.89          & 5.2047E-004   & 1.41 \\
$1024^2$        & 1.6769E-003   & 0.78          & 2.4870E-004   & 1.07 \\
\hline
\end{tabular}
\end{table}

\subsection{Stokes Flow with Prescribed Velocity Solid Boundaries (2D)}
\begin{figure}
\centering
\subfigure[]
{
\includegraphics[width=55mm]{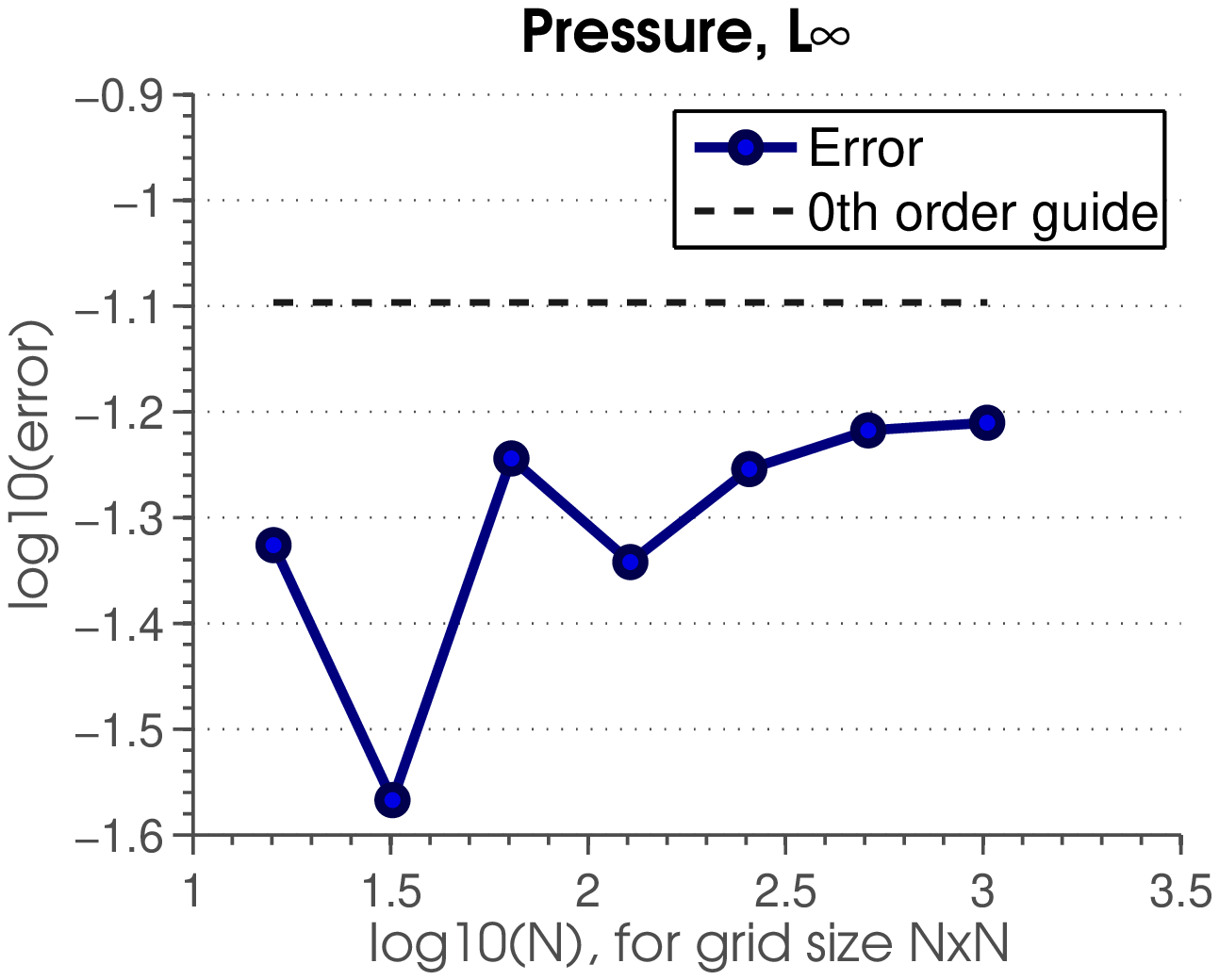}
}
\subfigure[]
{
\includegraphics[width=55mm]{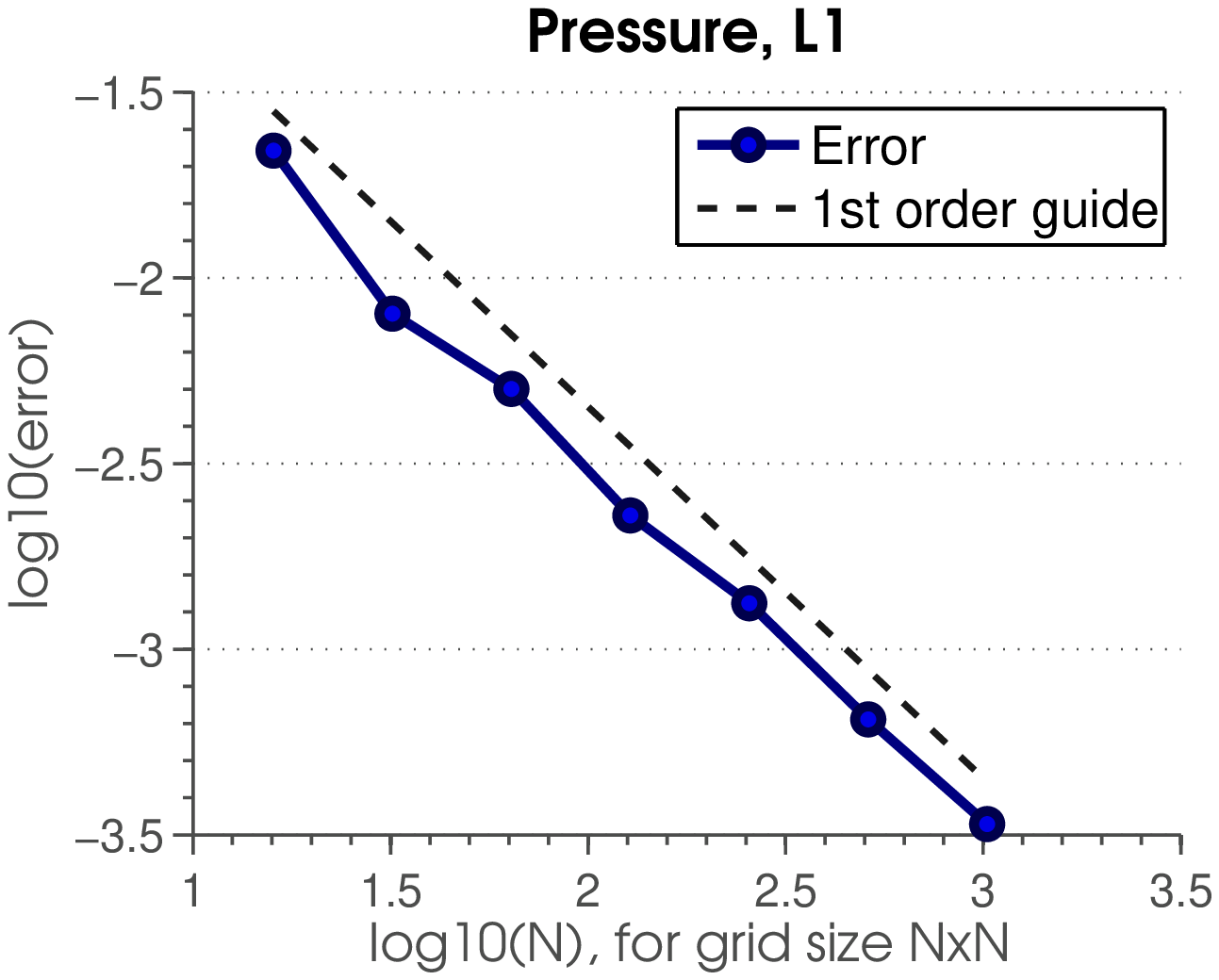}
}
\subfigure[]
{
\includegraphics[width=55mm]{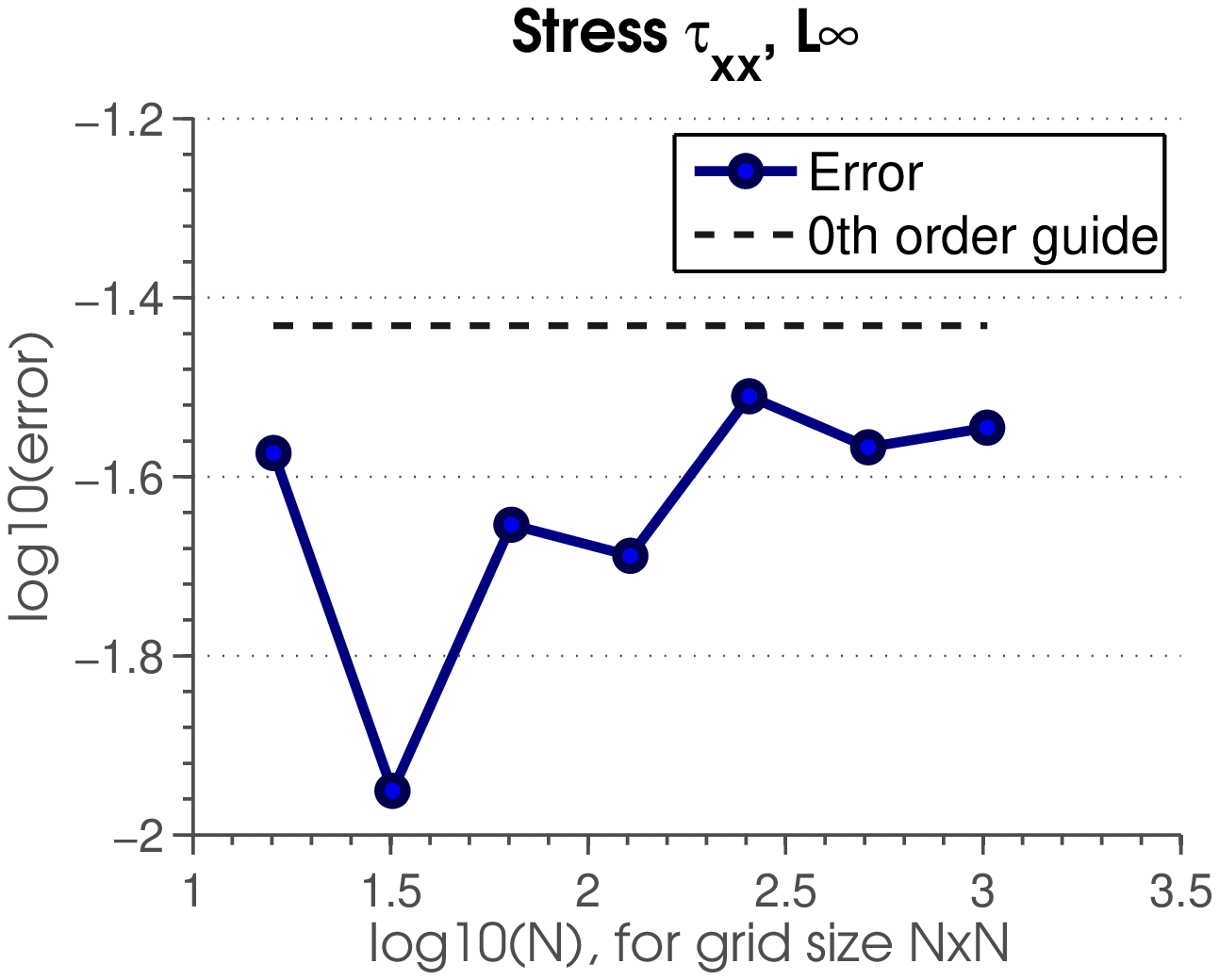}
}
\subfigure[]
{
\includegraphics[width=55mm]{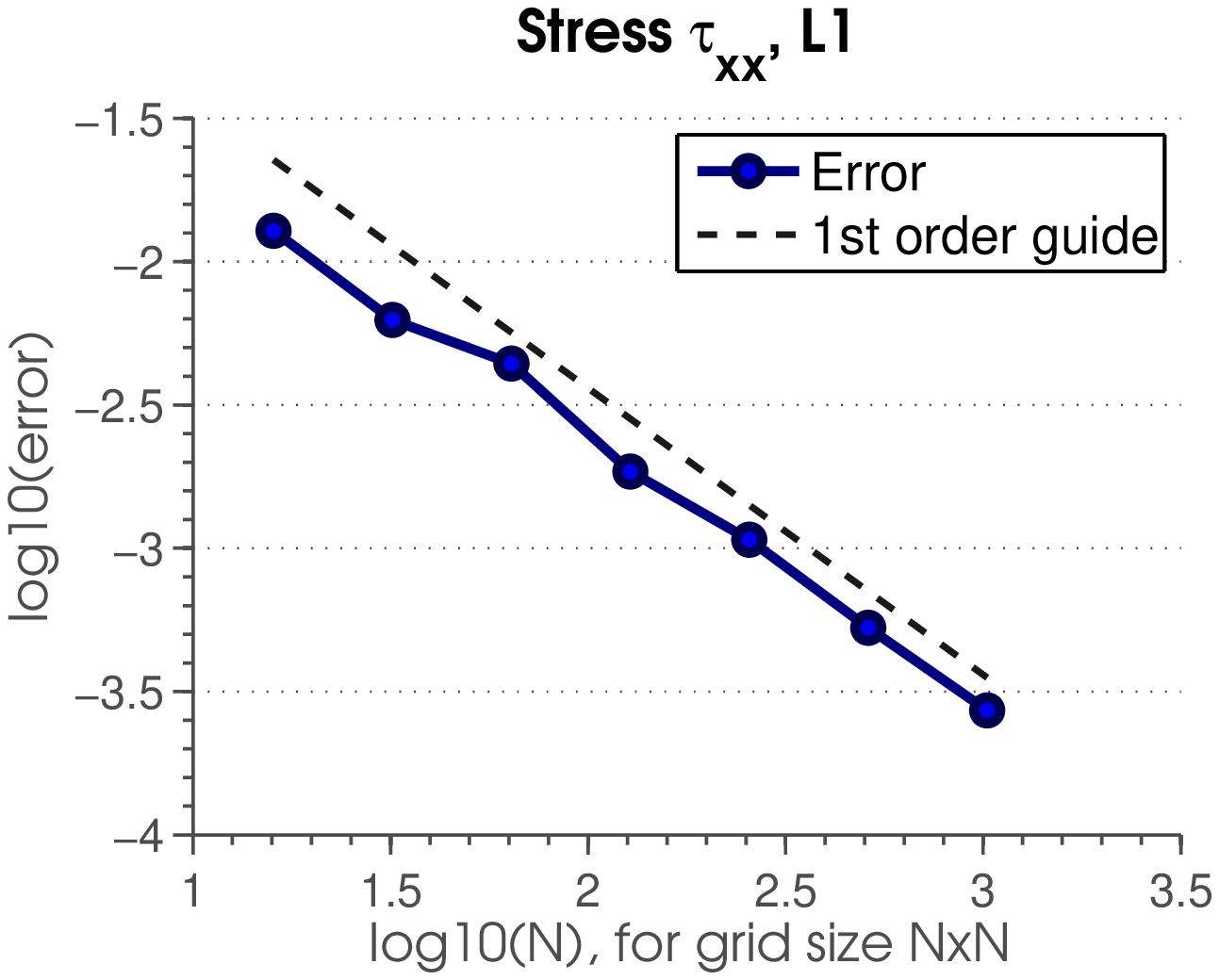}
}
\subfigure[]
{
\includegraphics[width=55mm]{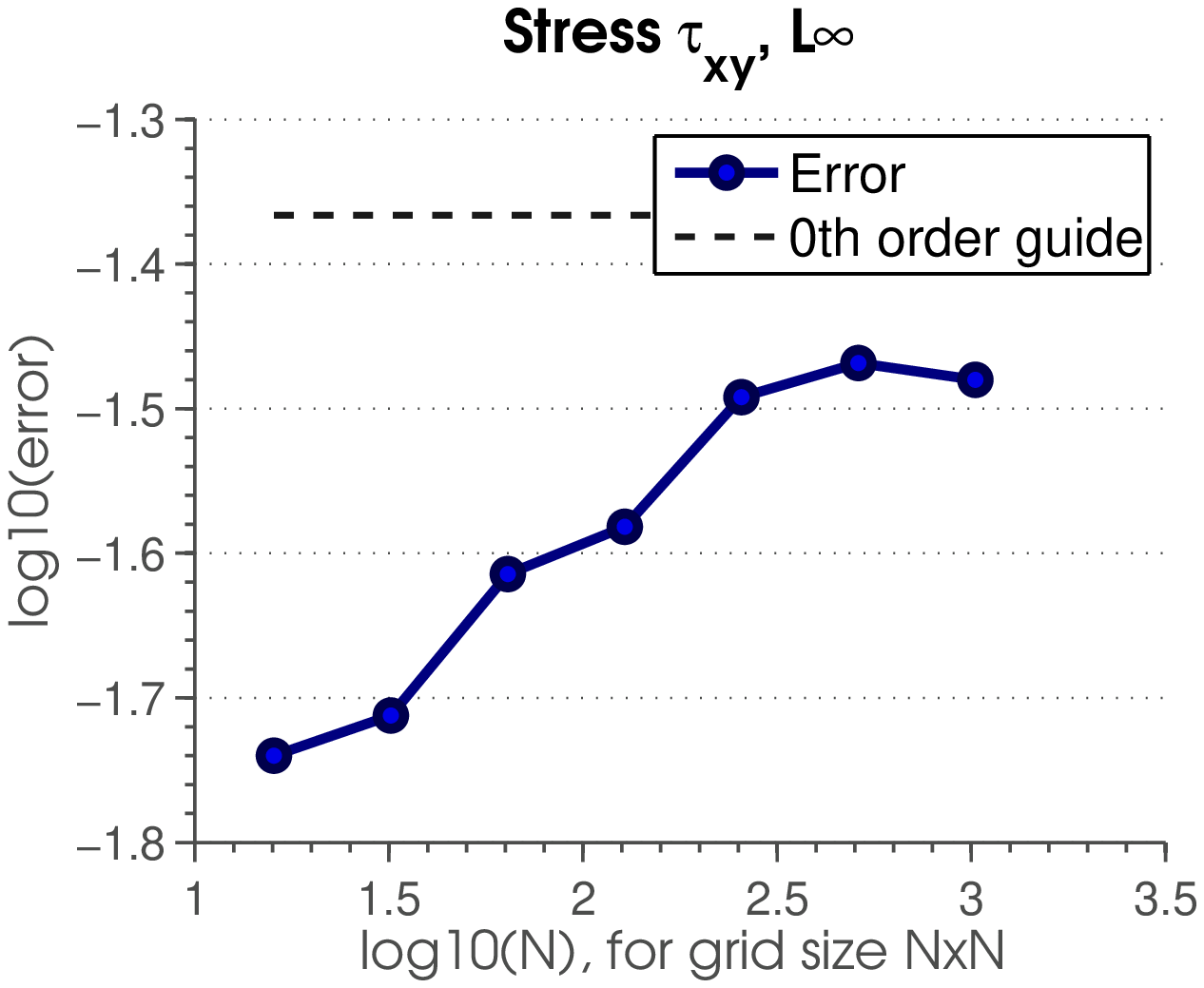}
}
\subfigure[]
{
\includegraphics[width=55mm]{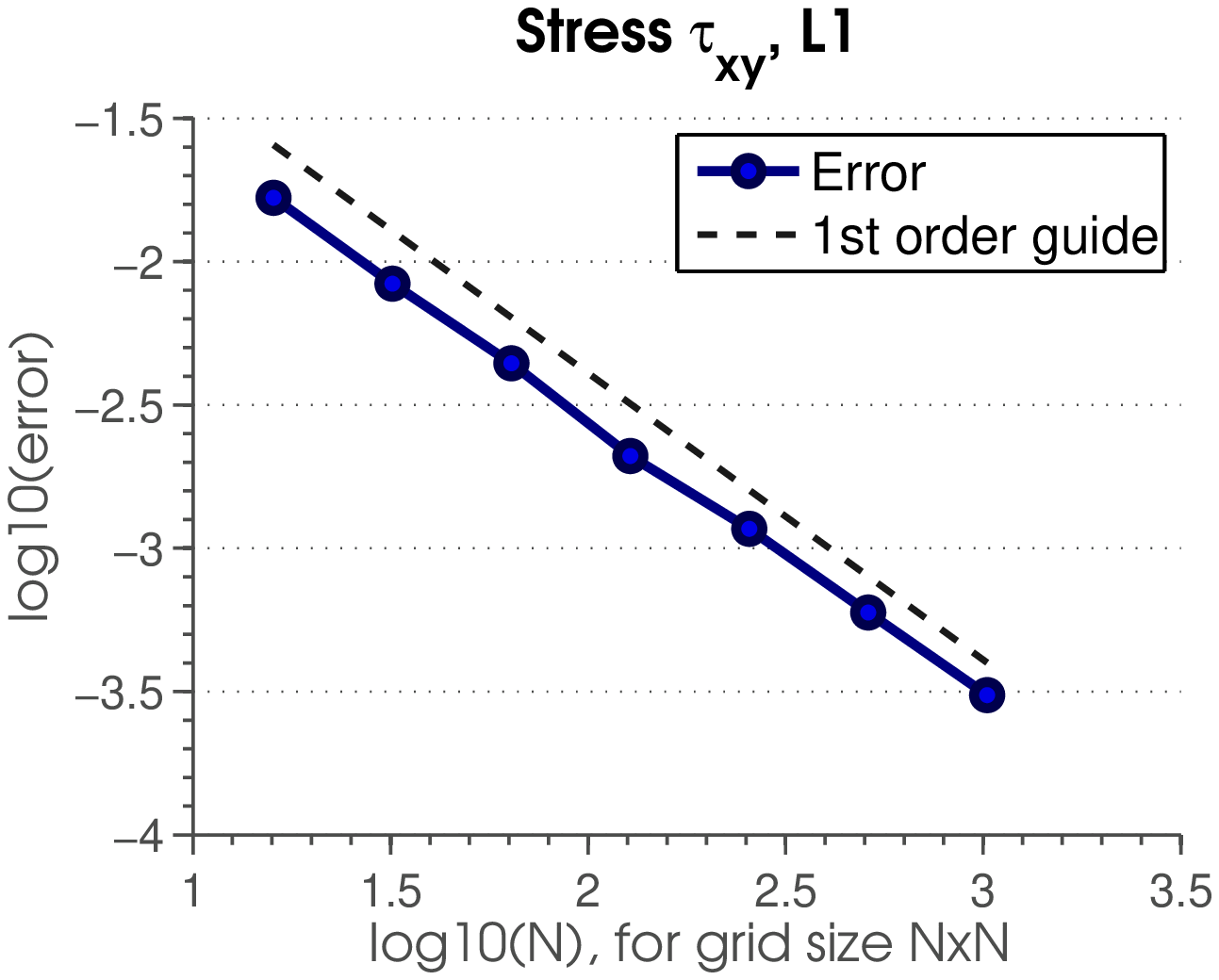}
}
\subfigure[]
{
\includegraphics[width=55mm]{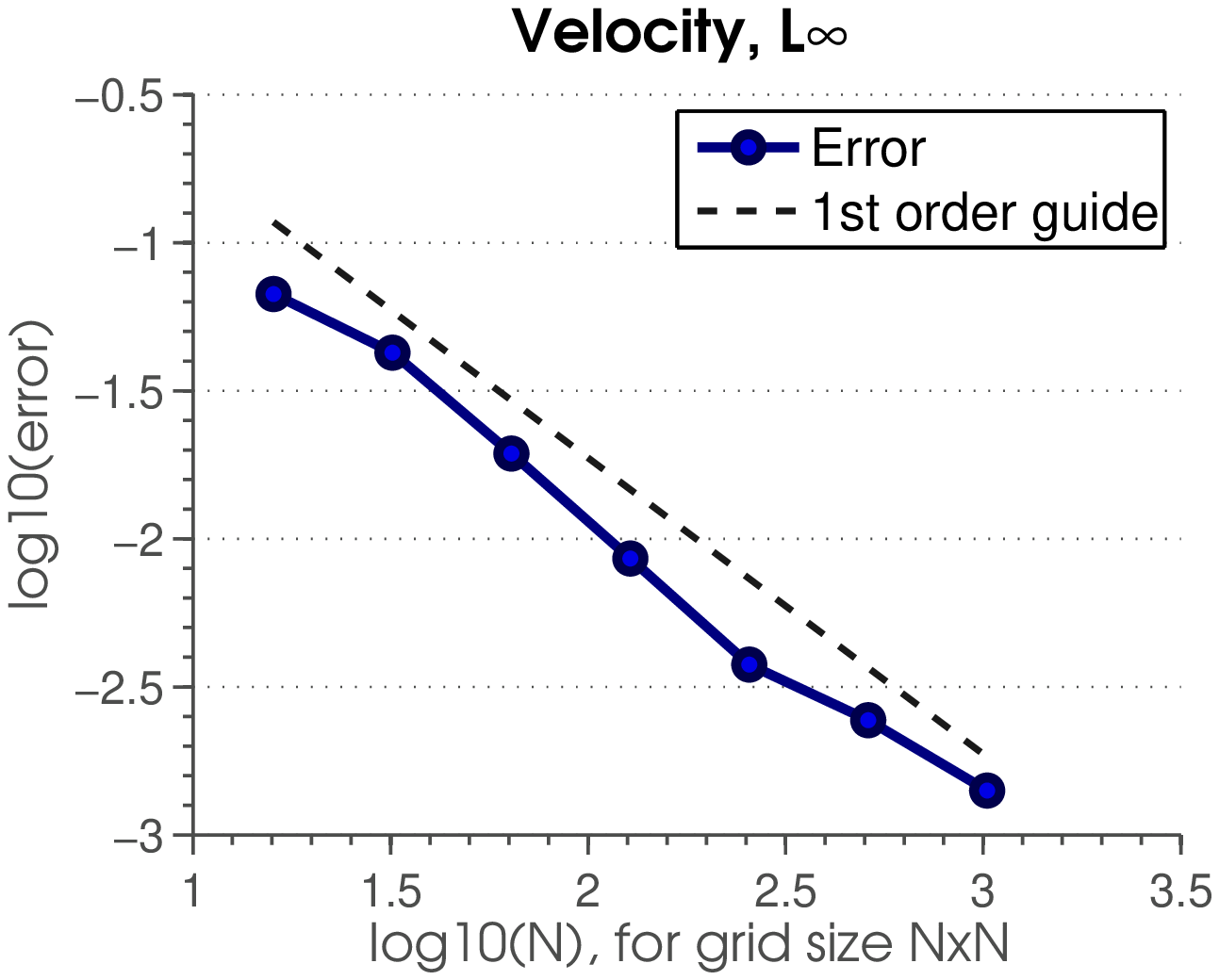}
}
\subfigure[]
{
\includegraphics[width=55mm]{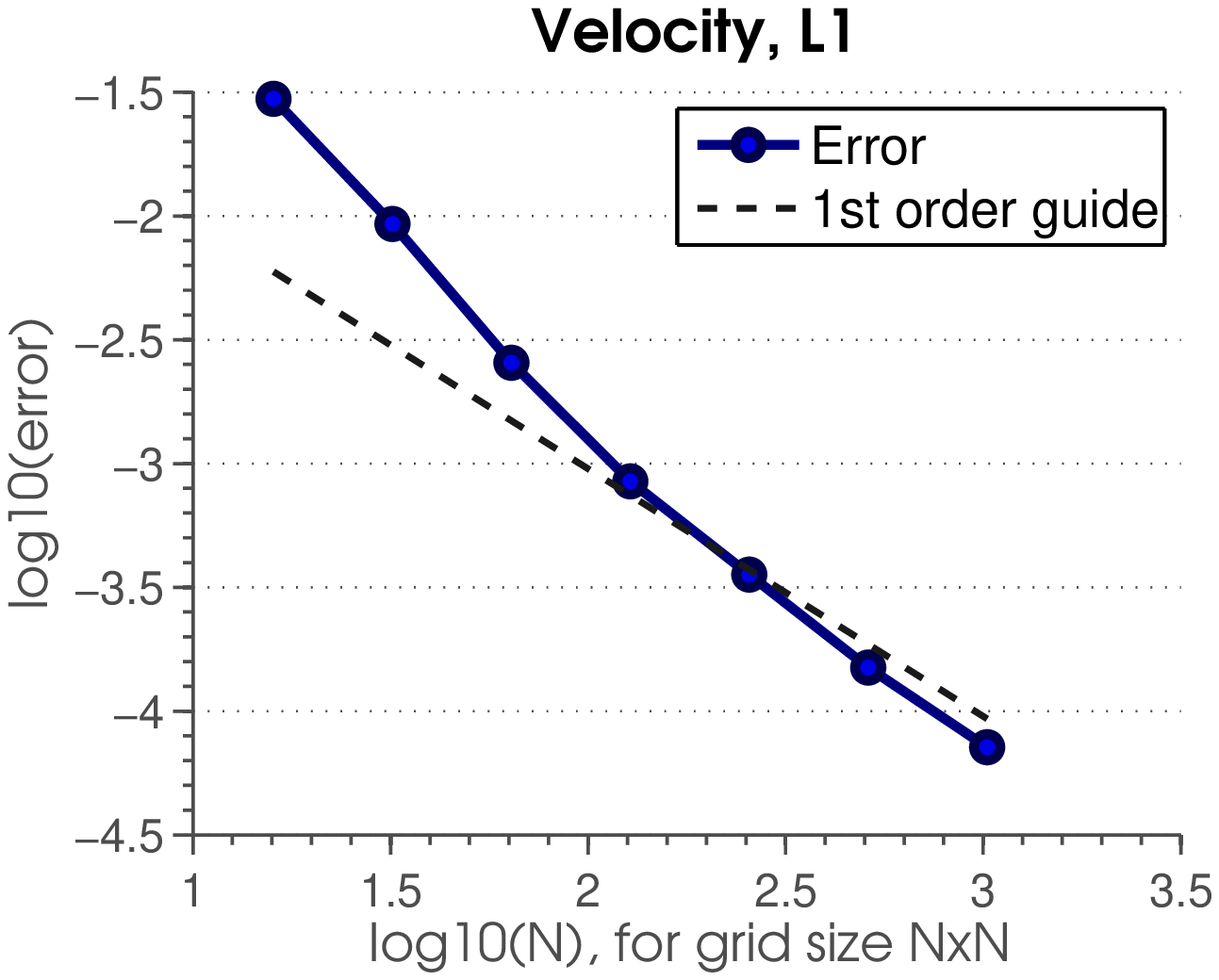}
}
\caption{Convergence graphs for the Stokes problem with a moving solid wall boundary (2D).}
\label{fig:stokes_movingsolid}
\end{figure}

To test solid boundaries with prescribed (non-zero) boundary velocities, we solve for fluid motion in an annulus centred at the origin, with inner radius $r=0.5$, outer radius $r=1$, density $\rho=1$, and viscosity $\mu = 0.1$, over a timestep $\Delta t = 1$. The outer boundary is static, while the inner boundary rotates rigidly with a clockwise angular velocity $\omega = 2$.  For the final velocity we use the streamfunction $\psi$:
\begin{equation}
\psi = r^4-3r^3+\frac{9}{4}r^2+\frac{1}{2}r+\frac{1}{4}
\end{equation}
For pressure, we use:
\begin{equation}
p = r^2 \cos(\theta) \sin(\theta)
\end{equation}
As in the preceding examples, stresses and input velocities can be computed from equations (\ref{stokesequations1})-(\ref{stokesequations3}).
Convergence results are shown in Table \ref{stokesmovingsolidchart} and Figure \ref{fig:stokes_movingsolid}.

\begin{table}
\caption{Convergence of Stokes with prescribed velocity solid walls (2D)}
\centering
\label{stokesmovingsolidchart}
\begin{tabular}{|c | c c | c c |}
\hline
Grid & $\|p - p^h\|_{\infty}$ & Order  & $\|p - p^h\|_{1}$ & Order \\
\hline
$16^2$          & 4.7210E-002    &              & 2.2007E-002    & \\
$32^2$          & 2.7144E-002    & 0.80          & 7.9999E-003    & 1.46 \\
$64^2$          & 5.7009E-002    & -1.07        & 5.0198E-003    & 0.67 \\
$128^2$        & 4.5540E-002    & 0.32          & 2.2860E-003    & 1.13 \\
$256^2$        & 5.5697E-002    & -0.29        & 1.3316E-003    & 0.78 \\
$512^2$        & 6.0562E-002    & -0.12        & 6.4669E-004    & 1.04 \\
$1024^2$        & 6.1643E-002    & -0.03        & 3.3755E-004    & 0.94 \\
\hline
Grid & $\|\tau_{xx} - \tau_{xx}^h\|_{\infty}$ & Order  & $\|\tau_{xx} - \tau_{xx}^h\|_{1}$ & Order \\
\hline
$16^2$          & 2.6730E-002    &              & 1.2788E-002    & \\
$32^2$          & 1.1167E-002    & 1.26          & 6.2639E-003    & 1.03 \\
$64^2$          & 2.2172E-002    & -0.99        & 4.4126E-003    & 0.51 \\
$128^2$        & 2.0462E-002    & 0.12          & 1.8498E-003    & 1.25 \\
$256^2$        & 3.0896E-002    & -0.59        & 1.0722E-003    & 0.79 \\
$512^2$        & 2.7147E-002    & 0.19          & 5.2769E-004    & 1.02 \\
$1024^2$        & 2.8528E-002    & -0.07        & 2.7240E-004    & 0.95 \\
\hline
Grid & $\|\tau_{xy} - \tau_{xy}^h\|_{\infty}$ & Order  & $\|\tau_{xy} - \tau_{xy}^h\|_{1}$ & Order \\
\hline
$16^2$          & 1.8177E-002    &              & 1.6688E-002    & \\
$32^2$          & 1.9415E-002    & -0.10        & 8.3726E-003    & 1.00 \\
$64^2$          & 2.4280E-002    & -0.32        & 4.4206E-003    & 0.92 \\
$128^2$        & 2.6202E-002    & -0.11        & 2.0984E-003    & 1.07 \\
$256^2$        & 3.2232E-002    & -0.30        & 1.1682E-003    & 0.85 \\
$512^2$        & 3.3983E-002    & -0.08        & 5.9705E-004    & 0.97 \\
$1024^2$        & 3.3056E-002    & 0.04          & 3.0660E-004    & 0.96 \\
\hline
Grid & $\|u - u^h\|_{\infty}$ & Order  & $\|u - u^h\|_{1}$ & Order \\
\hline
$16^2$          & 6.6980E-002    &              & 2.9655E-002    & \\
$32^2$          & 4.2451E-002    & 0.66          & 9.2806E-003    & 1.68 \\
$64^2$          & 1.9403E-002    & 1.13          & 2.5526E-003    & 1.86 \\
$128^2$        & 8.5696E-003    & 1.18          & 8.4643E-004    & 1.59 \\
$256^2$        & 3.7606E-003    & 1.19          & 3.5600E-004    & 1.25 \\
$512^2$        & 2.4402E-003    & 0.62          & 1.4966E-004    & 1.25 \\
$1024^2$        & 1.4085E-003    & 0.79          & 7.1476E-005    & 1.07 \\
\hline
\end{tabular}
\end{table}

\subsection{Three Dimensional Stokes Flow}
To test our method in three dimensions where analytical solutions are substantially more difficult to derive, we created a numerical solution at resolution $155^3$, and tested convergence towards this solution.  The test case consists of a sphere of liquid centred at the origin with a free surface at $r=1$, containing a nested static solid sphere of radius $r=0.25$.  Density was set to $\rho=1$ and viscosity to $\mu=0.1$.  We compute a time step of length $\Delta t = 1$ starting from an input velocity field:
 \begin{equation}
\vec{u}_{input} = (0.5 + x\sin(yz), \cos(0.25y) + 0.5xyz, x + 0.5yz^2 + \sqrt{2+y^2})
\end{equation}
Convergence results are shown in Table \ref{stokes3dchart}. Interestingly, this numerical experiment suggests first order convergence in $L^\infty$ even for stress variables, an improvement over the results in 2D. We cannot yet explain why the move to 3D would bring greater accuracy, but several similar tests showed the same pattern.

\begin{table}
\caption{Convergence of 3D Stokes with solid walls and a free surface}
\centering
\label{stokes3dchart}
\begin{tabular}{|c | c c | c c |}
\hline
Grid & $\|p - p^h\|_{\infty}$ & Order  & $\|p - p^h\|_{1}$ & Order \\
\hline
$12^2$          & 1.3912E-002    &               & 1.3312E-002    & \\
$24^2$          & 7.5008E-003    & 0.89          & 6.3879E-003    & 1.06 \\
$48^2$          & 4.9163E-003    & 0.61          & 3.5866E-003    & 0.83 \\
\hline
Grid & $\|\tau_{xx} - \tau_{xx}^h\|_{\infty}$ & Order  & $\|\tau_{xx} - \tau_{xx}^h\|_{1}$ & Order \\
\hline
$12^2$          & 1.7478E-002    &               & 2.0401E-002    & \\
$24^2$          & 1.0965E-002    & 0.67          & 1.1213E-002    & 0.86 \\
$48^2$          & 5.8661E-003    & 0.90          & 5.6263E-003    & 0.99 \\
\hline
Grid & $\|\tau_{yy} - \tau_{yy}^h\|_{\infty}$ & Order  & $\|\tau_{yy} - \tau_{yy}^h\|_{1}$ & Order \\
\hline
$12^2$          & 2.3604E-002    &               & 2.9434E-002    & \\
$24^2$          & 1.3516E-002    & 0.80          & 1.5741E-002    & 0.90 \\
$48^2$          & 6.8033E-003    & 0.99          & 7.7348E-003    & 1.03 \\
\hline
Grid & $\|\tau_{xy} - \tau_{xy}^h\|_{\infty}$ & Order  & $\|\tau_{xy} - \tau_{xy}^h\|_{1}$ & Order \\
\hline
$12^2$          & 6.4309E-003    &               & 7.9982E-003    & \\
$24^2$          & 5.1895E-003    & 0.31          & 4.4602E-003    & 0.84 \\
$48^2$          & 3.1232E-003    & 0.73          & 2.0025E-003    & 1.16 \\
\hline
Grid & $\|\tau_{xz} - \tau_{xz}^h\|_{\infty}$ & Order  & $\|\tau_{xz} - \tau_{xz}^h\|_{1}$ & Order \\
\hline
$12^2$          & 5.5459E-002    &               & 1.3162E-001    & \\
$24^2$          & 3.0026E-002    & 0.89          & 7.6418E-002    & 0.78 \\
$48^2$          & 1.3128E-002    & 1.19          & 3.6082E-002    & 1.08 \\
\hline
Grid & $\|\tau_{yz} - \tau_{yz}^h\|_{\infty}$ & Order  & $\|\tau_{yz} - \tau_{yz}^h\|_{1}$ & Order \\
\hline
$12^2$          & 1.6390E-002    &               & 2.8919E-002    & \\
$24^2$          & 1.0325E-002    & 0.67          & 1.7670E-002    & 0.71 \\
$48^2$          & 5.2896E-003    & 0.96          & 8.7857E-003    & 1.01 \\
\hline
Grid & $\|u - u^h\|_{\infty}$ & Order  & $\|u - u^h\|_{1}$ & Order \\
\hline
$12^2$          & 1.9163E-001    &               & 2.5741E-001    & \\
$24^2$          & 1.2664E-001    & 0.60          & 1.4041E-001    & 0.87 \\
$48^2$          & 7.1250E-002    & 0.83          & 6.3736E-002    & 1.14 \\
\hline
Grid & $\|v - v^h\|_{\infty}$ & Order  & $\|v - v^h\|_{1}$ & Order \\
\hline
$12^2$          & 6.2893E-002    &               & 5.4328E-002    & \\
$24^2$          & 4.7289E-002    & 0.41          & 3.6334E-002    & 0.58 \\
$48^2$          & 2.9547E-002    & 0.68          & 2.0160E-002    & 0.85 \\
\hline
Grid & $\|w - w^h\|_{\infty}$ & Order  & $\|w - w^h\|_{1}$ & Order \\
\hline
$12^2$          & 1.8633E-001    &               & 2.7837E-001    & \\
$24^2$          & 1.2850E-001    & 0.54          & 1.5601E-001    & 0.84 \\
$48^2$          & 7.4279E-002    & 0.79          & 7.1820E-002    & 1.12 \\
\hline
\end{tabular}
\end{table}
\subsection{Discussion}
Table \ref{convergencesummary} summarizes the approximate orders of convergence suggested by our experiments in two dimensions. As noted earlier, we achieve essentially first order convergence, with the exception of pressure and stress in $L^{\infty}$. We now proceed to make some general comments.

 A rigourous convergence theory to describe the method is beyond the scope of the current work.  However, in the absence of boundaries, our methods are equivalent to a straightforward discretization of the original PDE form with second order centred finite differences on a staggered grid, and can therefore expect to achieve uniform second order convergence. Near boundaries this clearly does not hold, and the effective quadrature is likely only first order due to the use of piecewise constant approximations of integral terms. This is in line with our results in $L^{\infty}$ for velocity. Solution gradients can generally be expected to be one order less accurate than the solution itself, and this is also evident in the $O(1)$ errors in the $L^{\infty}$ results for fluid pressure and stress. Interestingly, Chen et al. recently showed that the immersed boundary method for Stokes flow likewise exhibits $O(1)$ errors in pressure in $L^{\infty}$ \cite{Chen2011}.

The virtual node method of Bedrossian et al.\ \cite{Bedrossian2010} for the Poisson equation uses a variational formulation similar to ours, but makes use of piecewise bilinear Cartesian elements near the boundary to estimate the relevant integrals, at the cost of denser stencils for boundary cells. Their results indicate second order convergence which is consistent with the fact that our use of piecewise constant estimates yields first order convergence. This also suggests that applying bilinear elements near boundaries may be effective in raising the convergence order of our method for Stokes flow, while maintaining the benefits of sparsity and positive-definiteness.

Ng et al.\ \cite{Ng2009} pointed out that in the final discretized form, their method for the Poisson equation with Neumann solid boundaries is identical to that of Batty et al.\ \cite{Batty2007} if the face volume weights suggested by the variational perspective are replaced with face area (finite volume) weights. The latter choice leads to an increase in $L^{\infty}$ accuracy for pressure from first to second order and velocity from zeroth to first order. The related ghost fluid method for the Poisson equation with Dirichlet boundaries \cite{Gibou2002} likewise exhibits second order in pressure and first order in velocity, with a different choice of weights. While this hints that alternative diagonal weighting matrices $W$ might raise the order of accuracy of the current method, we have found that directly introducing ghost fluid or finite volume weights in this setting breaks the symmetry of the linear system. Nonetheless, a deeper exploration of the connections between our method and ghost fluid/immersed interface methods, finite volume methods, and finite element methods might provide the key to an improved weighting scheme.

\begin{table}
\caption{Convergence Behaviour in 2D}
\centering
\label{convergencesummary}
\begin{tabular}{|c | c | c |}
\hline
Variable & $L^{\infty}$ & $L^1$ \\
\hline
Pressure        & 0    &  1\\
Deviatoric Stress          & 0    &  1\\
Velocity        & 1    &  1\\
\hline
\end{tabular}
\end{table}

\section{Application to Viscous Jet Buckling}
One particularly fascinating phenomenon exhibited by highly viscous liquids is jet buckling.  When a falling liquid column of sufficient viscosity impacts a solid surface, it will fold or coil over on itself rather than spreading out smoothly.  Relatively few researchers have looked at simulating Newtonian viscous buckling, despite its prevalence in many common liquids such as honey.  To the best of our knowledge, the GENSMAC code of Tom\'{e}, McKee and co-authors is the only prior finite difference scheme to do so in three dimensions \cite{Tome1994,Tome1999,Tome2004,Oishi2008}. However, as noted earlier this approach requires a case-by-case analysis of discrete surface orientations and its implicit formulation entails solving a large non-symmetric linear system. Jet buckling has also been addressed in a finite element setting \cite{Bonito2006} and with an SPH approach \cite{Rafiee2007}.

With this problem in mind, we incorporate our Stokes solver into a simple two-stage fractional step Navier-Stokes routine (similar to that presented in section 2 of the paper by Ng et al.\ \cite{Ng2009}).  First, starting with a velocity field $\vec{u}^n$ at time $t^n$, we compute advection and body forces to produce an intermediate velocity field $\vec{u}^*$:
\begin{equation}
\rho \left( \frac{\vec{u}^* - \vec{u}^n}{\Delta t} - \vec{u}^n \cdot (\nabla \vec{u}^n)\right) = \vec{F}
\end{equation}
We account for advection terms with a first order semi-Lagrangian scheme using bilinear interpolation of velocities. We then simply add any external body forces $\vec{F}$ (gravity in our examples).
From this intermediate velocity, we then simultaneously incorporate viscous forces and project the velocity field to be divergence free using our Stokes solver, to arrive at time $t^{n+1} = t^n + \Delta t$:
\begin{eqnarray}
\rho \frac{(\vec{u}^{n+1} - \vec{u}^*)}{\Delta t} & = & \nabla\cdot\mathbf{\tau}^{n+1} - \nabla p^{n+1}\\
\nabla \cdot \vec{u} & = & 0 \\
\tau^{n+1} & = & \mu(\nabla \vec{u}^{n+1} + (\nabla \vec{u}^{n+1})^T)
\end{eqnarray}
with the appropriate free surface and solid boundary conditions applied.  Tracking of the liquid surface position is performed using a basic semi-Lagrangian level set method (eg. \cite{Enright2005}).

\subsection{Two Dimensional Jet Buckling}
\begin{figure}
\centering
\subfigure[]
{
\includegraphics[width=40mm, clip = true, trim=62mm 27mm 77mm 27mm]{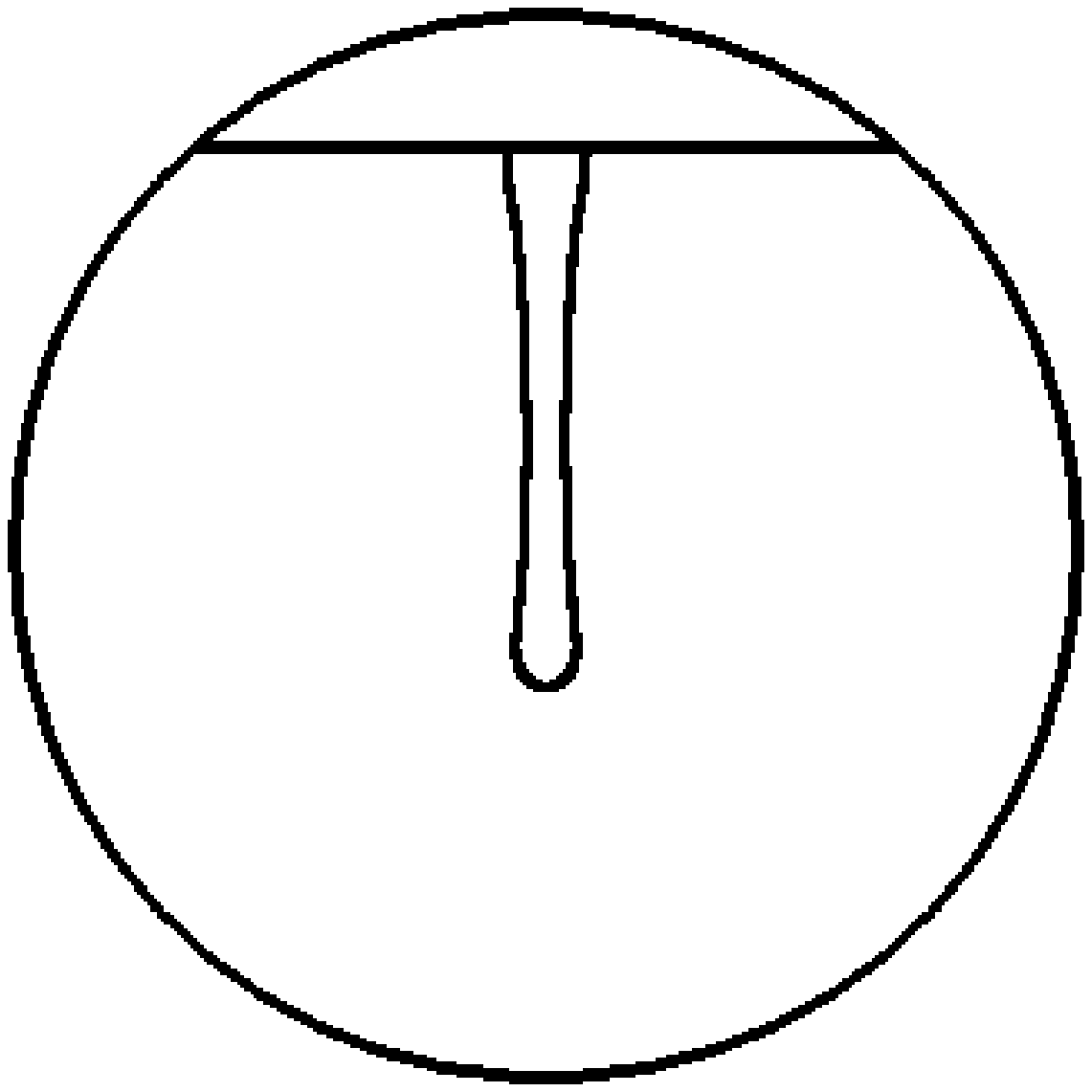}
}
\subfigure[]
{
\includegraphics[width=40mm, clip = true, trim=62mm 27mm 77mm 27mm]{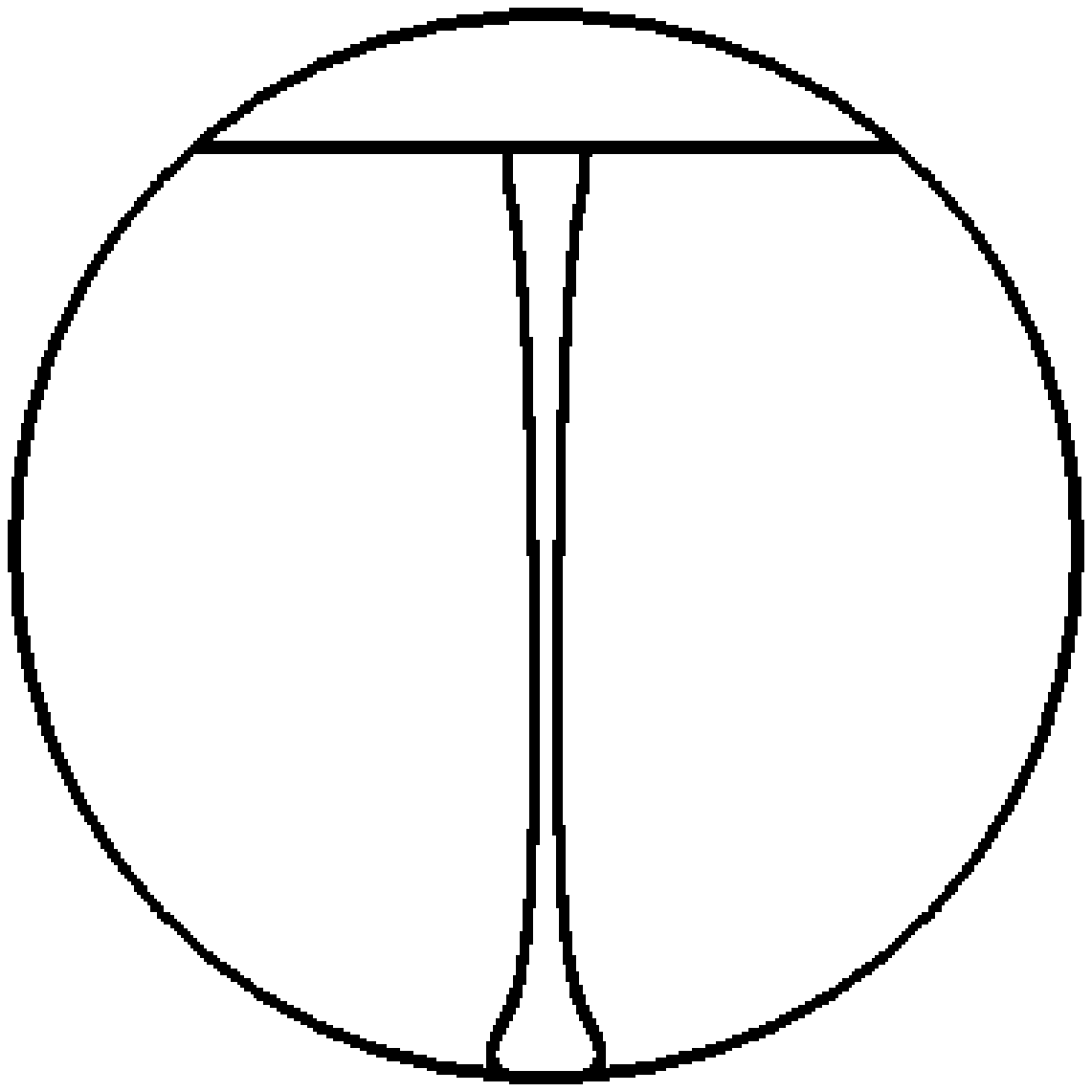}
}
\subfigure[]
{
\includegraphics[width=40mm, clip = true, trim=62mm 27mm 77mm 27mm]{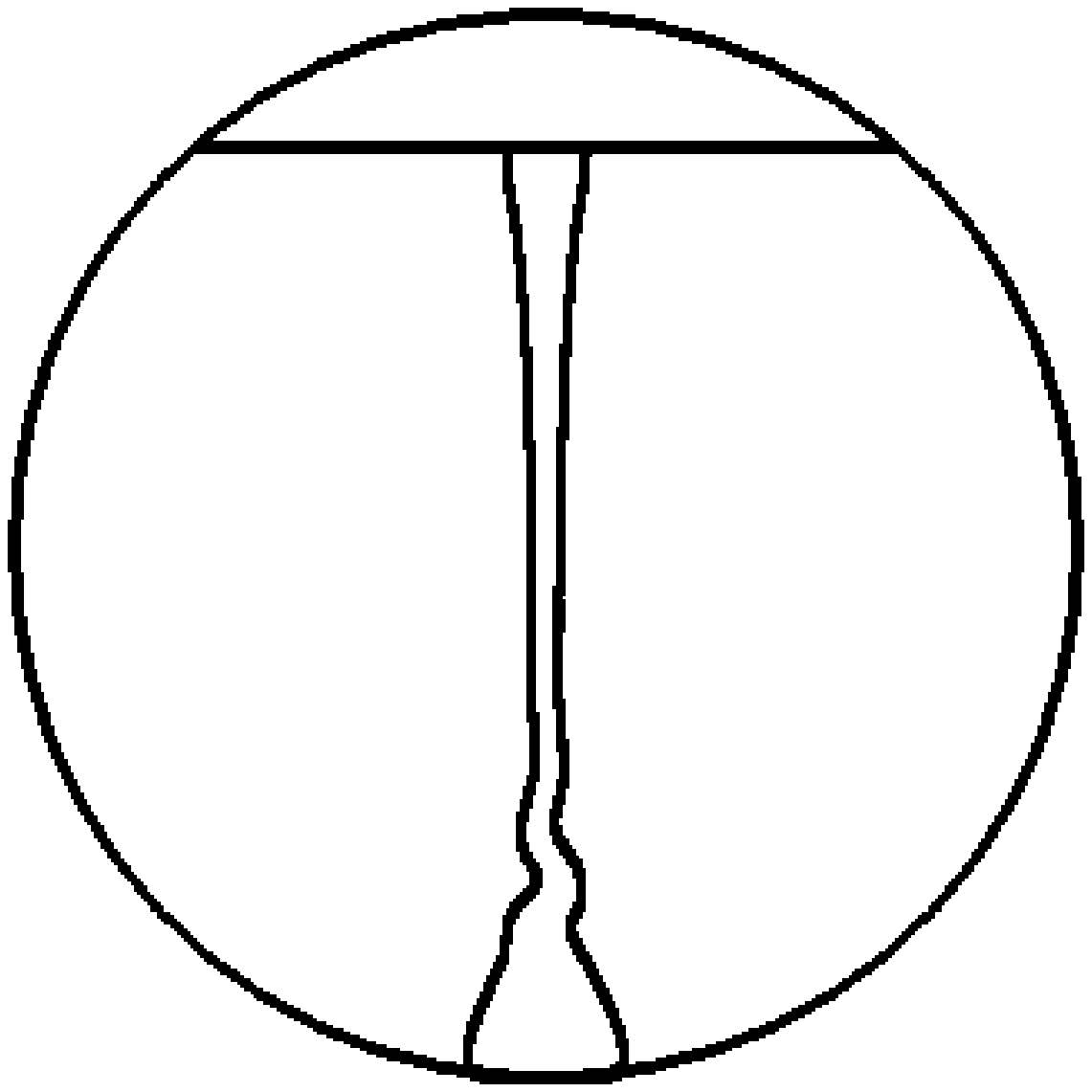}
}
\subfigure[]
{
\includegraphics[width=40mm, clip = true, trim=62mm 27mm 77mm 27mm]{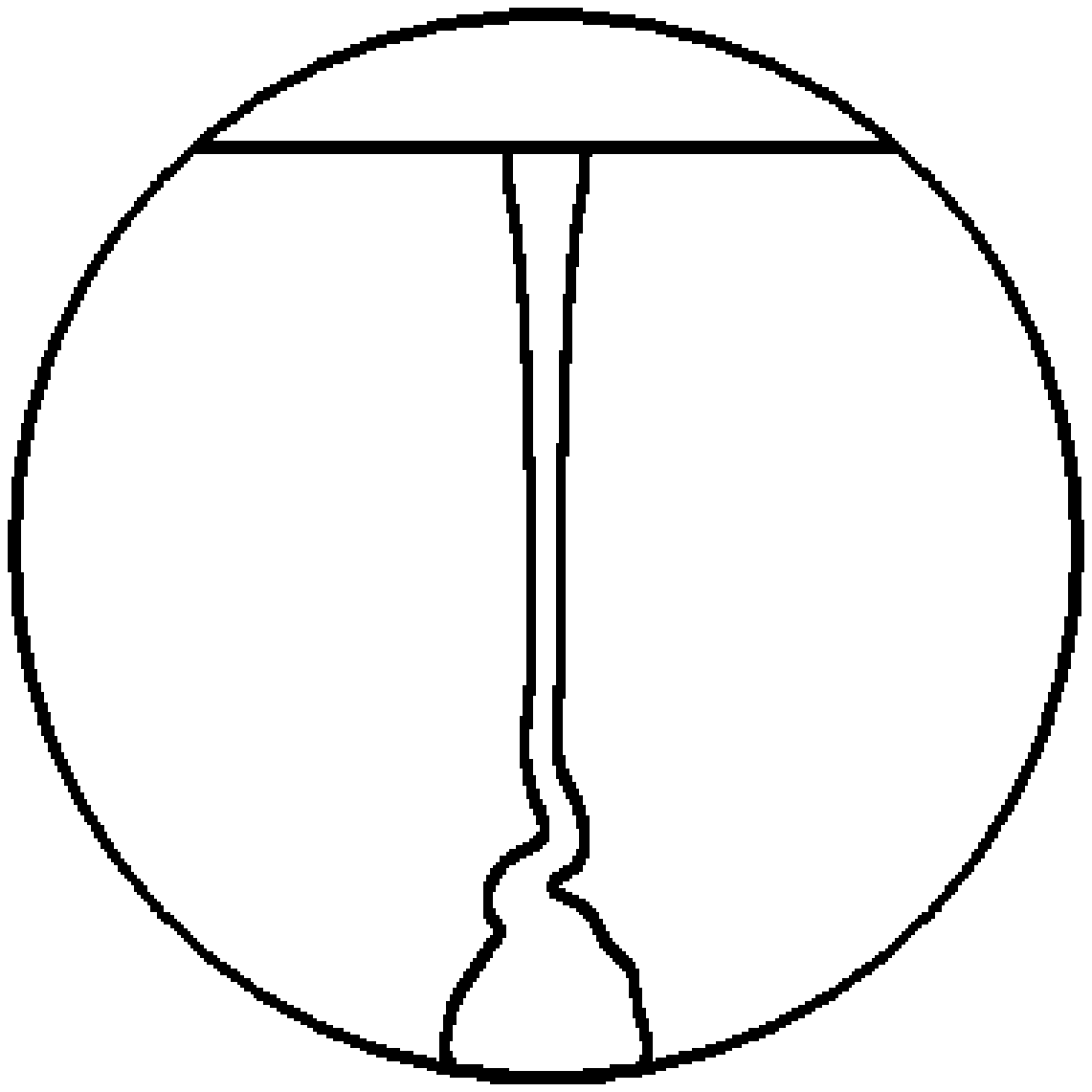}
}
\subfigure[]
{
\includegraphics[width=40mm, clip = true, trim=62mm 27mm 77mm 27mm]{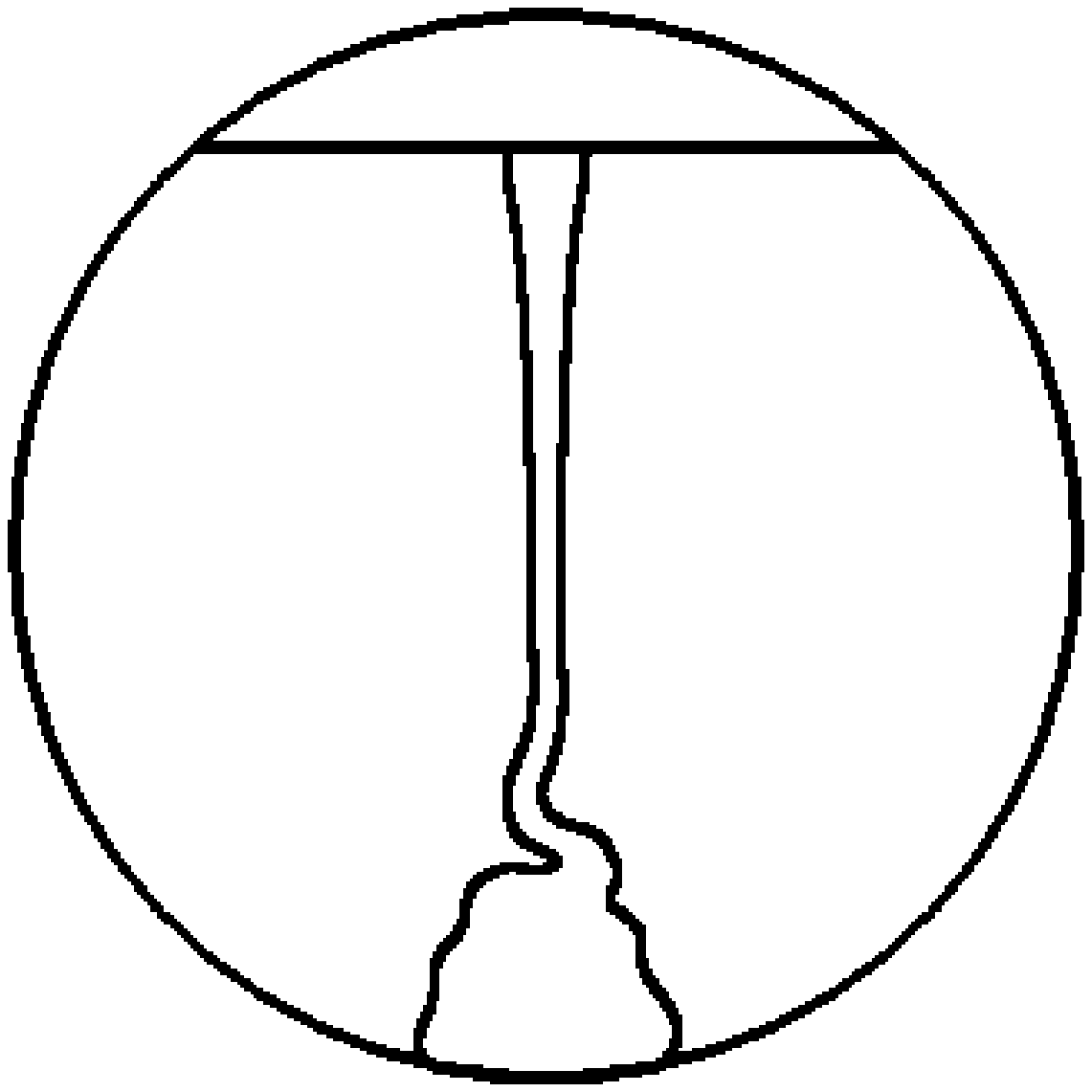}
}
\subfigure[]
{
\includegraphics[width=40mm, clip = true, trim=62mm 27mm 77mm 27mm]{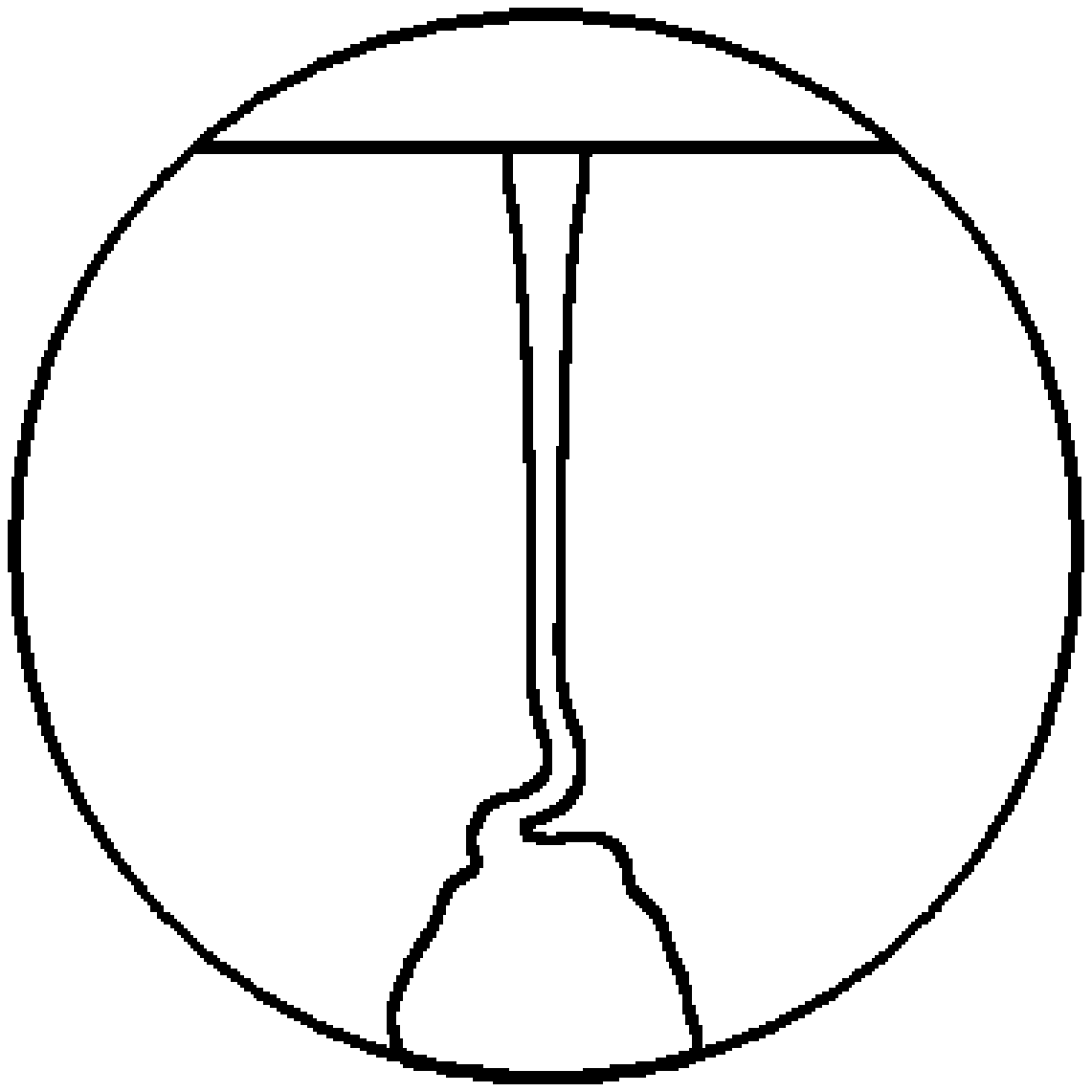}
}
\subfigure[]
{
\includegraphics[width=40mm, clip = true, trim=62mm 27mm 77mm 27mm]{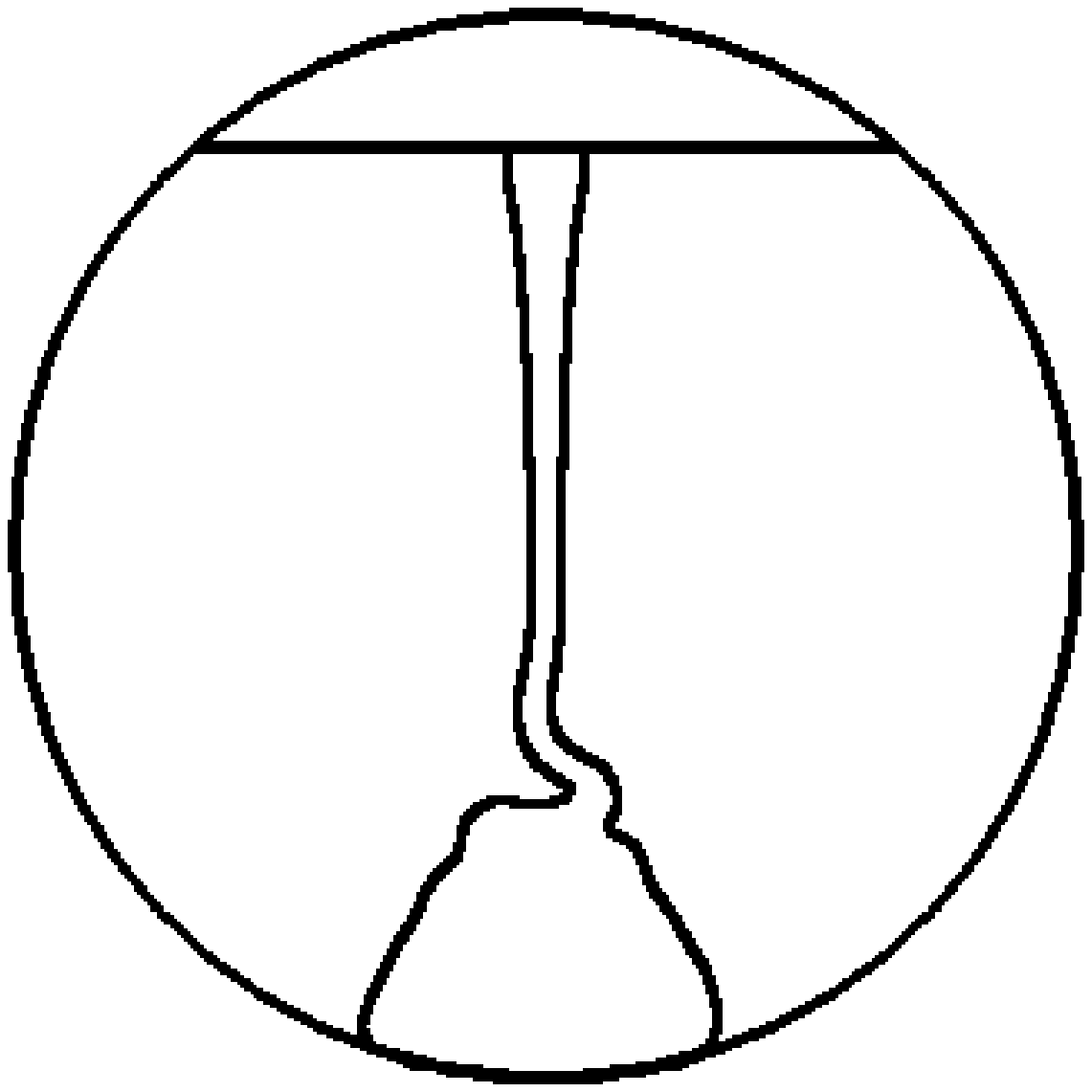}
}
\subfigure[]
{
\includegraphics[width=40mm, clip = true, trim=62mm 27mm 77mm 27mm]{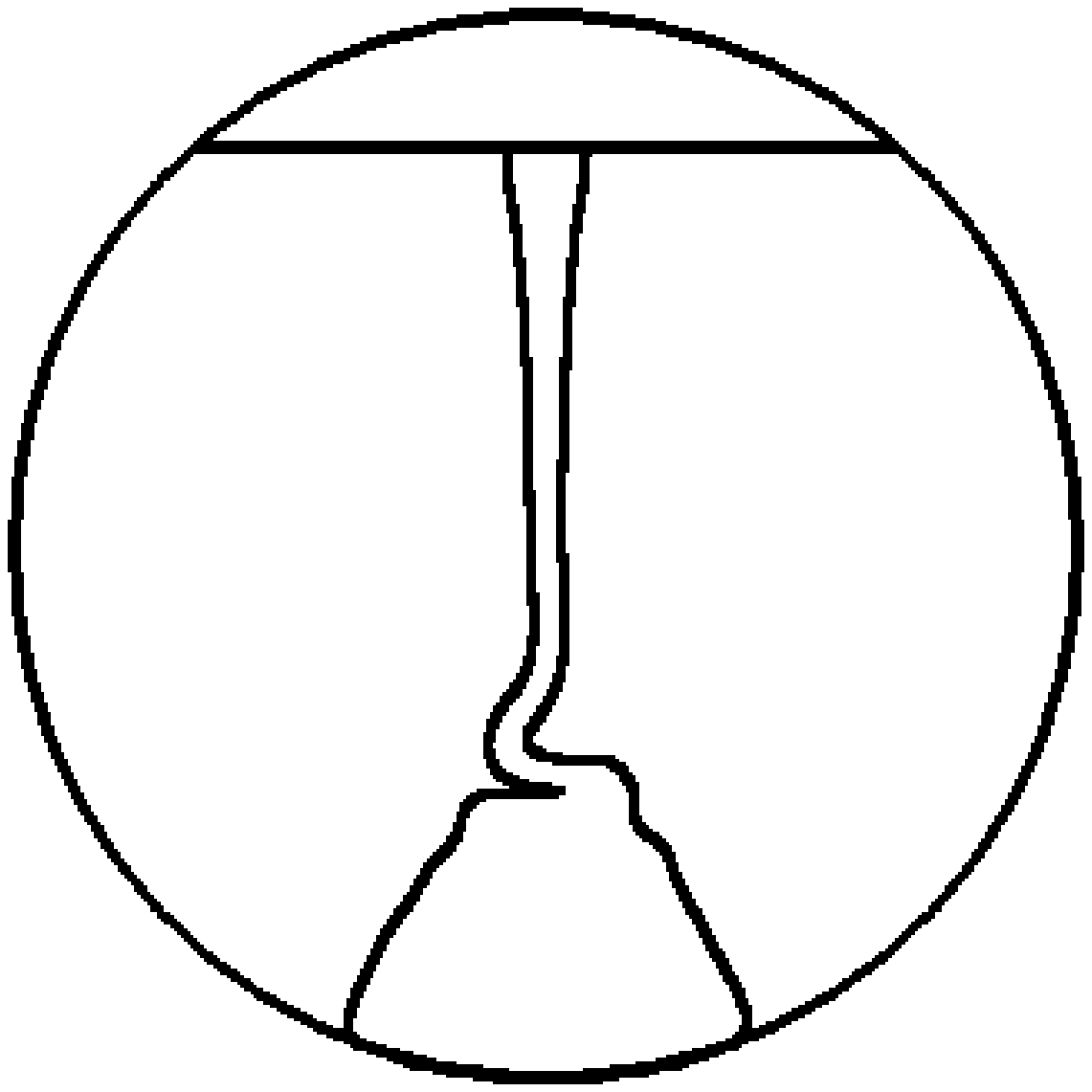}
}
\subfigure[]
{
\includegraphics[width=40mm, clip = true, trim=62mm 27mm 77mm 27mm]{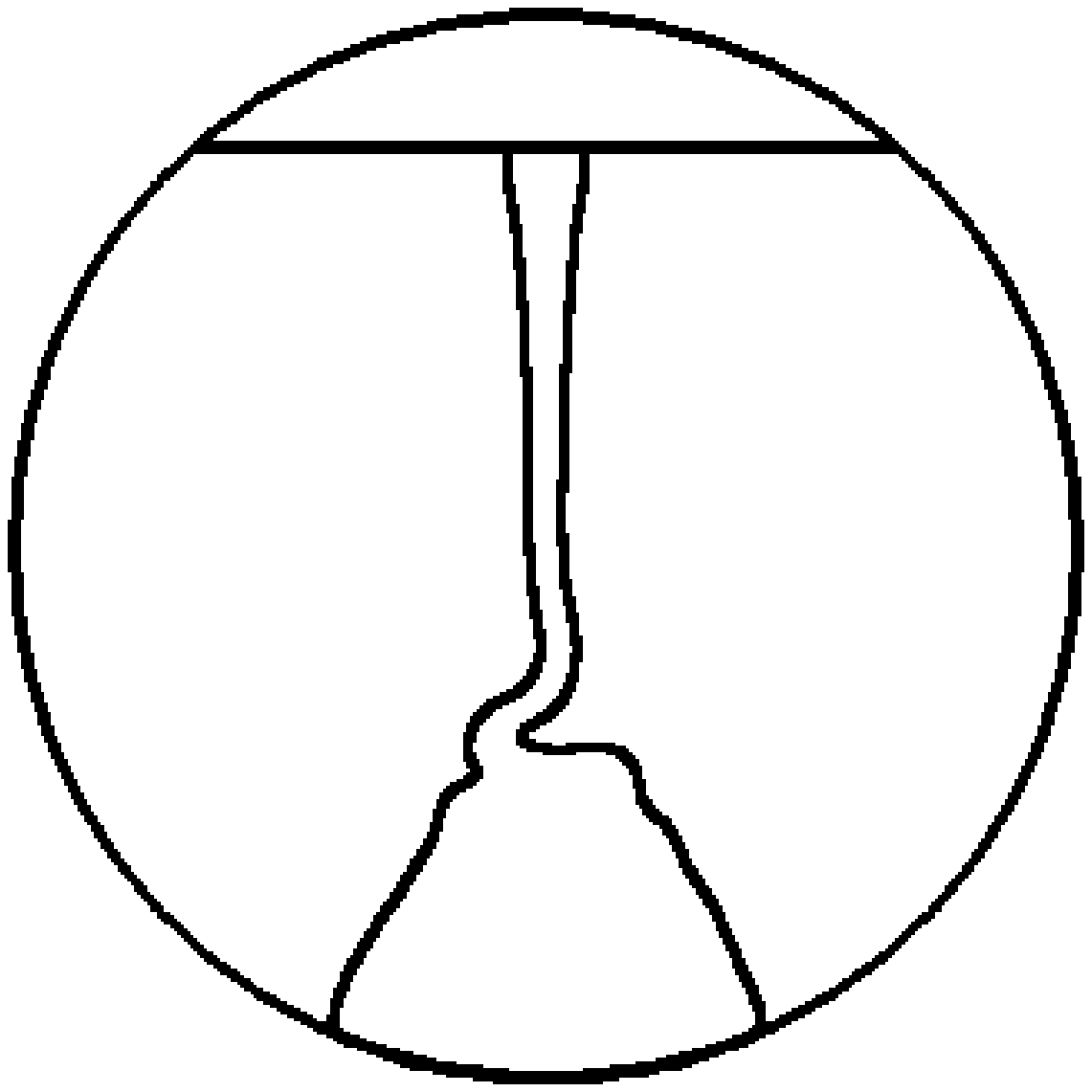}
}
\subfigure[]
{
\includegraphics[width=40mm, clip = true, trim=62mm 27mm 77mm 27mm]{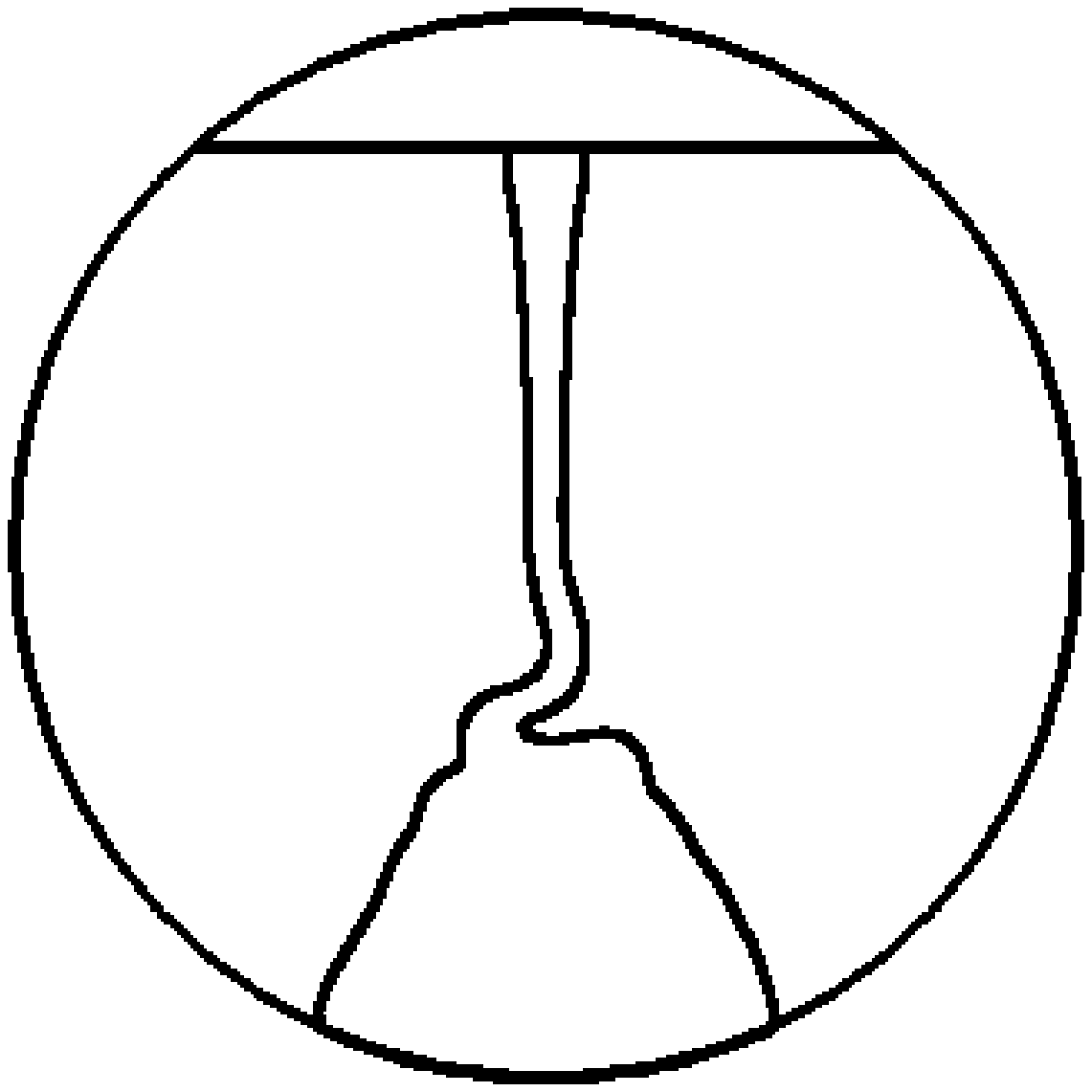}
}
\subfigure[]
{
\includegraphics[width=40mm, clip = true, trim=62mm 27mm 77mm 27mm]{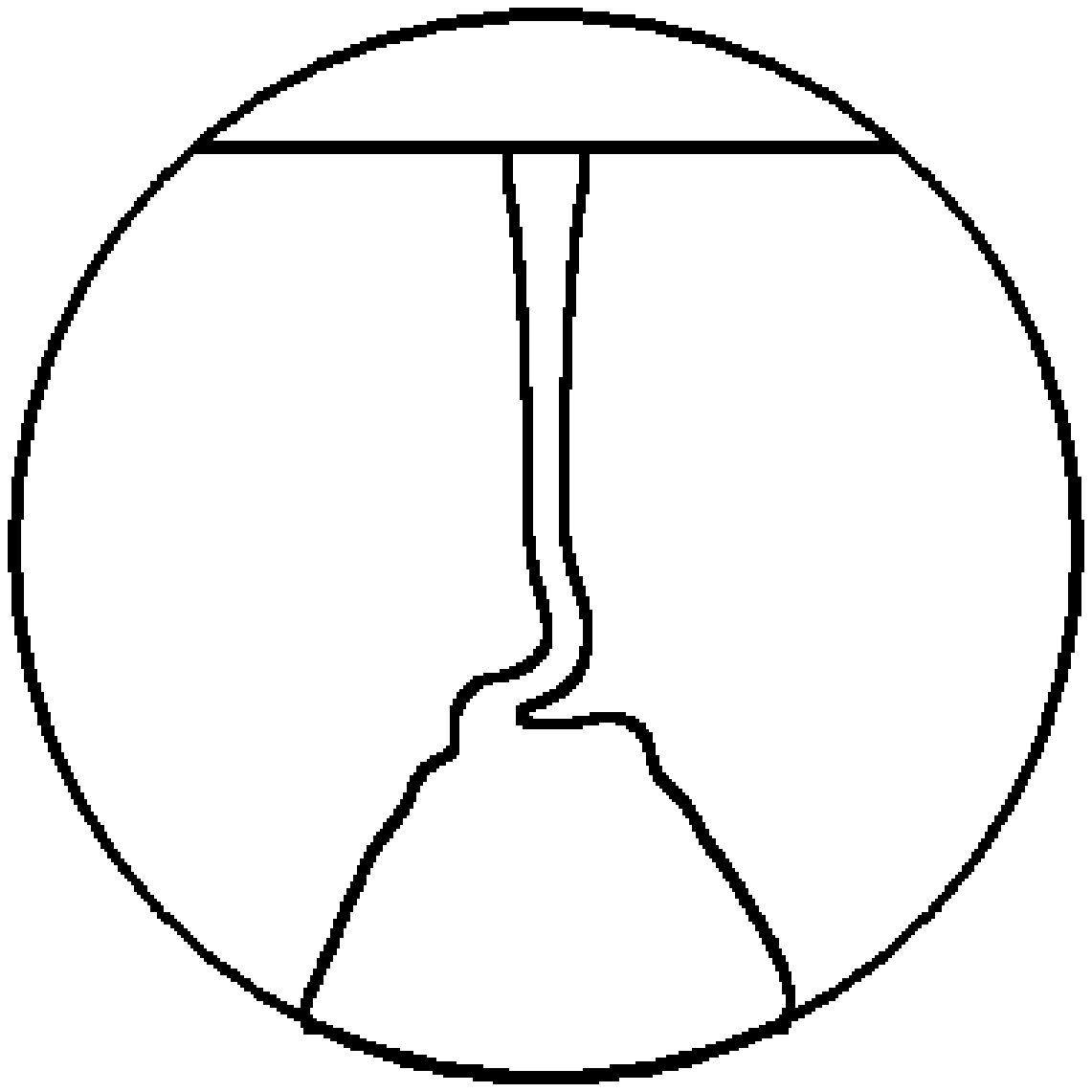}
}
\subfigure[]
{
\includegraphics[width=40mm, clip = true, trim=62mm 27mm 77mm 27mm]{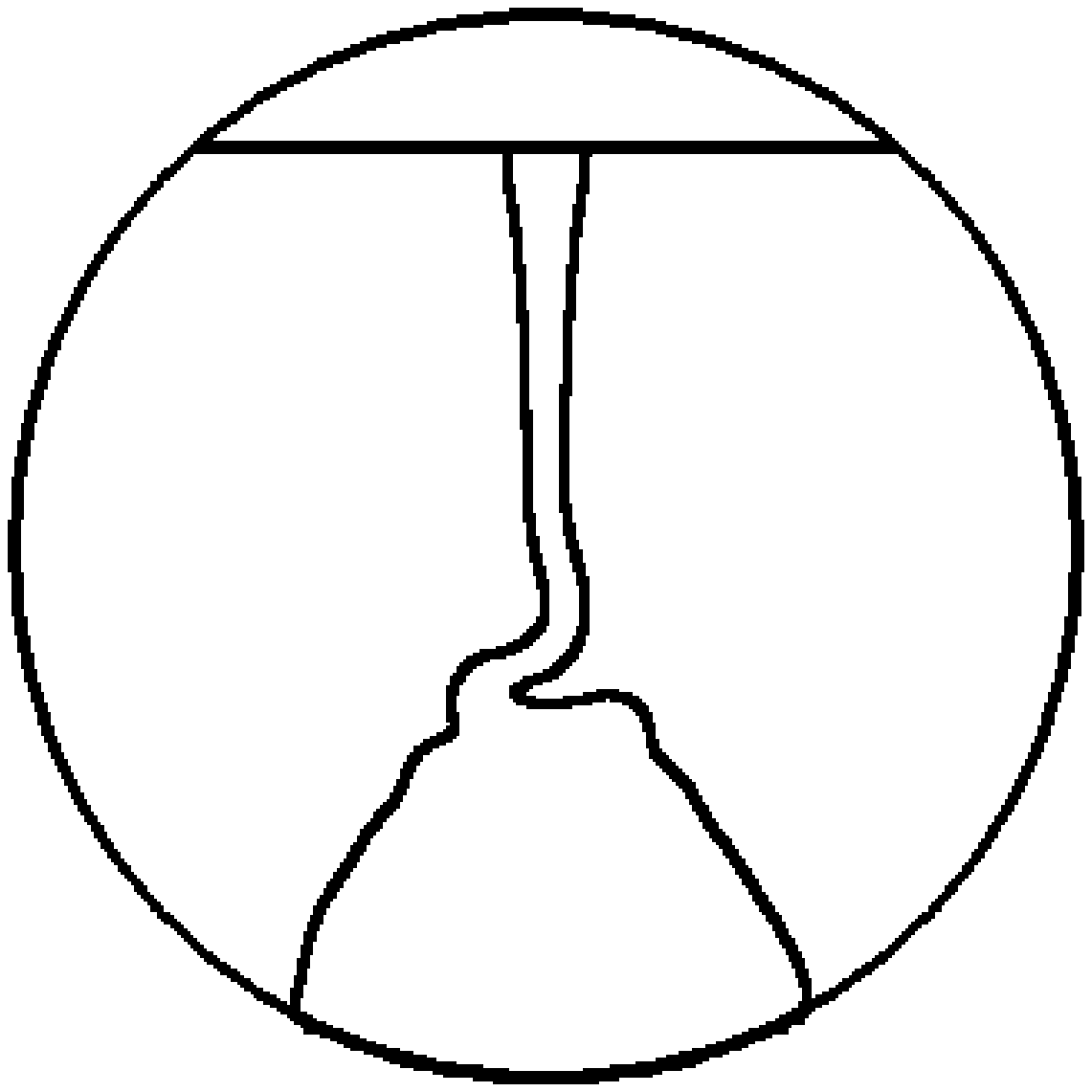}
}
\caption{A two-dimensional example of viscous jet buckling performed using our simple Navier-Stokes routine. The first image occurs 0.5 seconds into the simulation, and subsequent frames occur at 0.2 second intervals.}
\label{fig:buckling2D}
\end{figure}

Figure \ref{fig:buckling2D} presents the results of a two-dimensional simulation of planar viscous jet buckling.  The simulation domain is a circle of radius $0.4 [m]$ centred at $(0.5,0.5)$.  A horizontal ceiling is placed at $y=0.8 [m]$, featuring a liquid jet inflow centered at $x=0.5 [m]$ with a fixed vertical velocity of $U=-0.5 [m/s]$ and a width of $D=0.06 [m]$.  This configuration yields a drop height of $H=0.7 [m]$.  The density of the liquid is $\rho = 1 [kg/m^3]$ and the dynamic viscosity of the liquid is $\mu = 0.075 [Pa\cdot s]$. Gravity is set at $-9.81 [m/s^2]$.  The simulation grid used a resolution of $150\times150$ cells.

Following Tom\'e and McKee \cite{Tome1999}, this yields a Reynolds number of $Re = \frac{\rho D U}{\mu} = 0.4$ and an aspect ratio for the liquid jet of $H / D = 0.7 / 0.06 = 11.667$. This falls within the regime in which planar buckling is expected to occur ($Re < 0.5, H/D > 10$) according to Cruikshank and Munson\cite{Cruikshank1981,Cruikshank1988}; as shown our method reproduces the buckling phenomenon.

\subsection{Three Dimensional Jet Buckling}
\begin{figure}
\centering
\subfigure[]
{
\includegraphics[width=40mm, clip = true, trim=43mm 0mm 43mm 0mm]{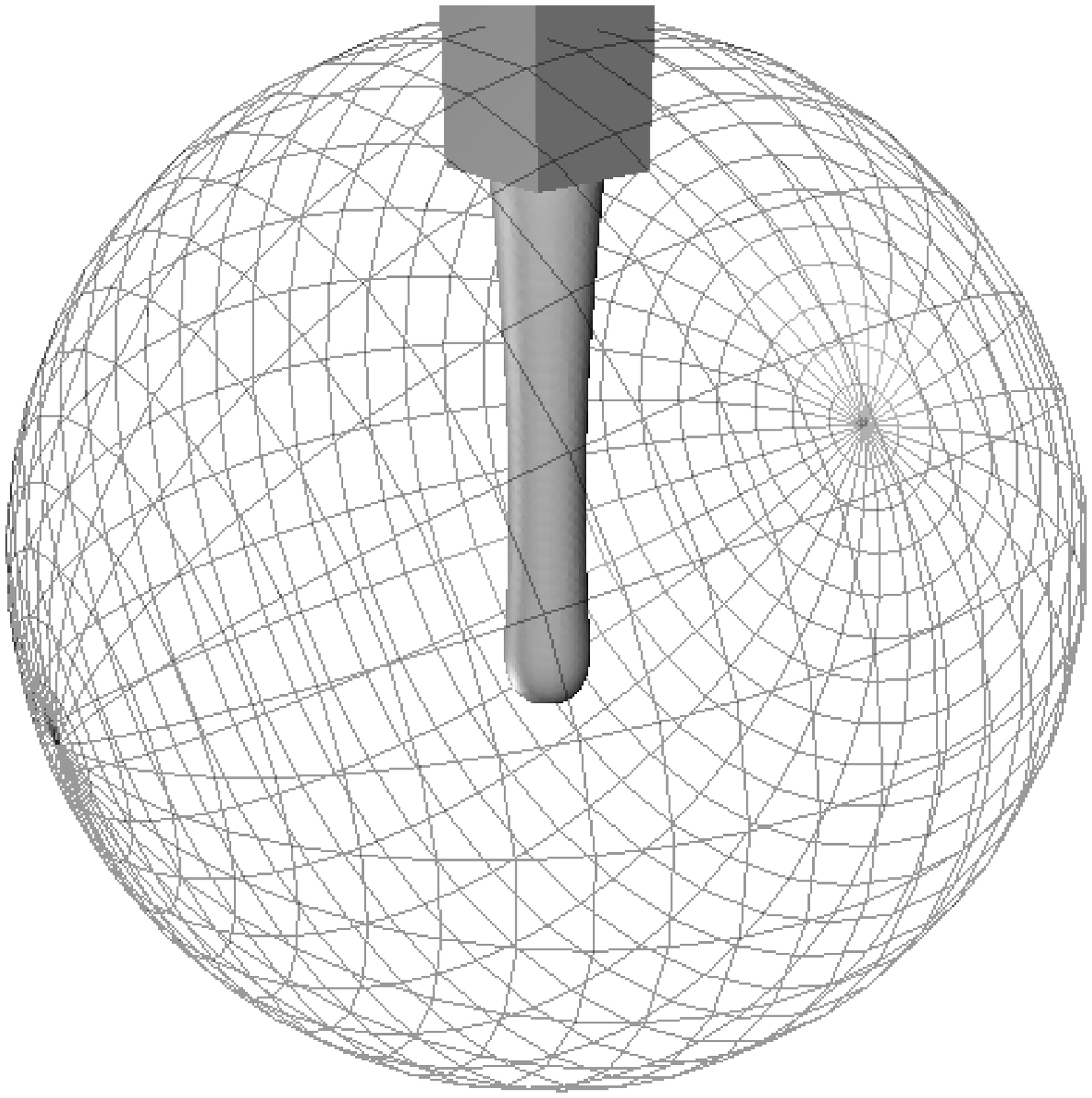}
}
\subfigure[]
{
\includegraphics[width=40mm, clip = true, trim=43mm 0mm 43mm 0mm]{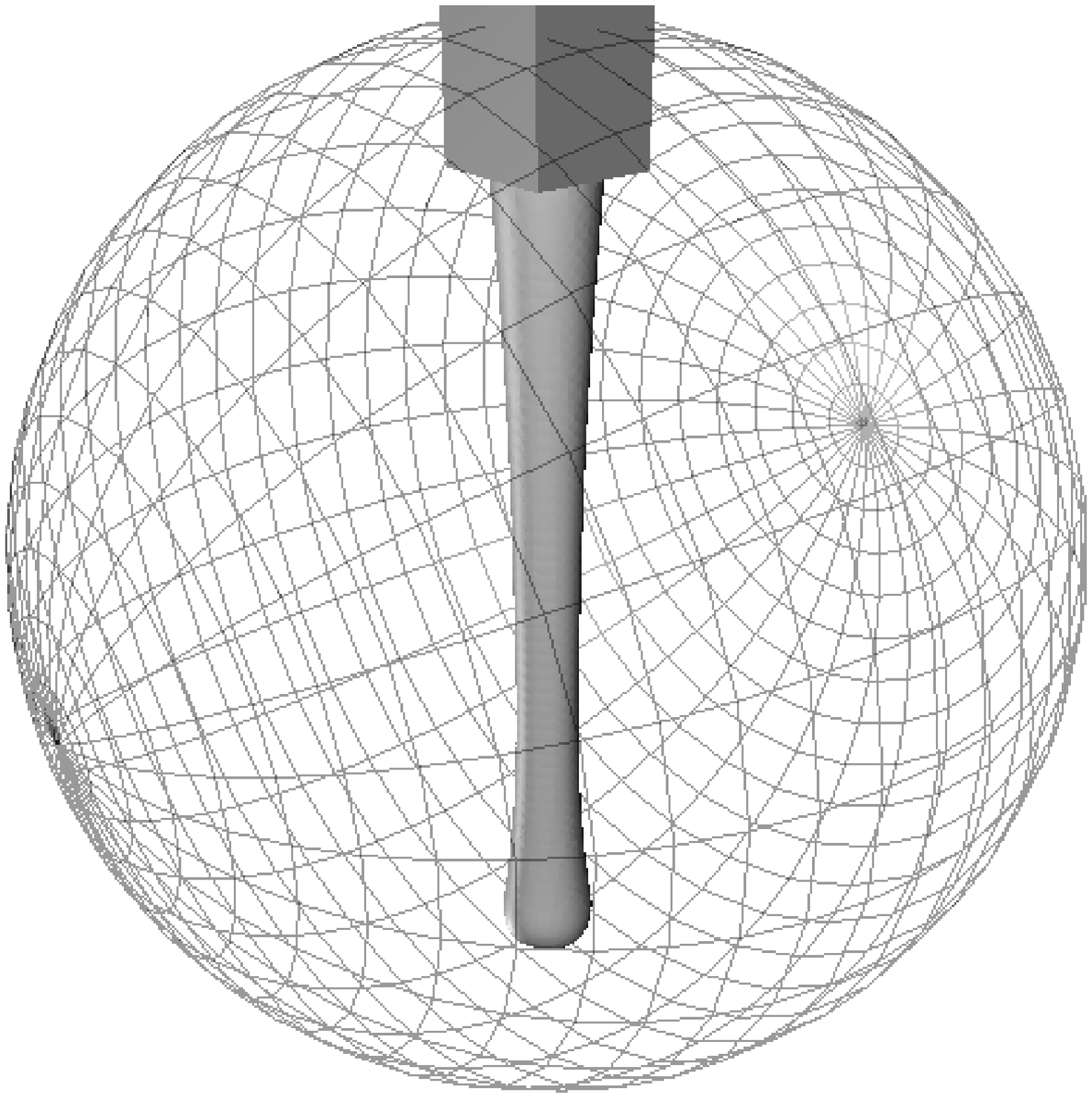}
}
\subfigure[]
{
\includegraphics[width=40mm, clip = true, trim=43mm 0mm 43mm 0mm]{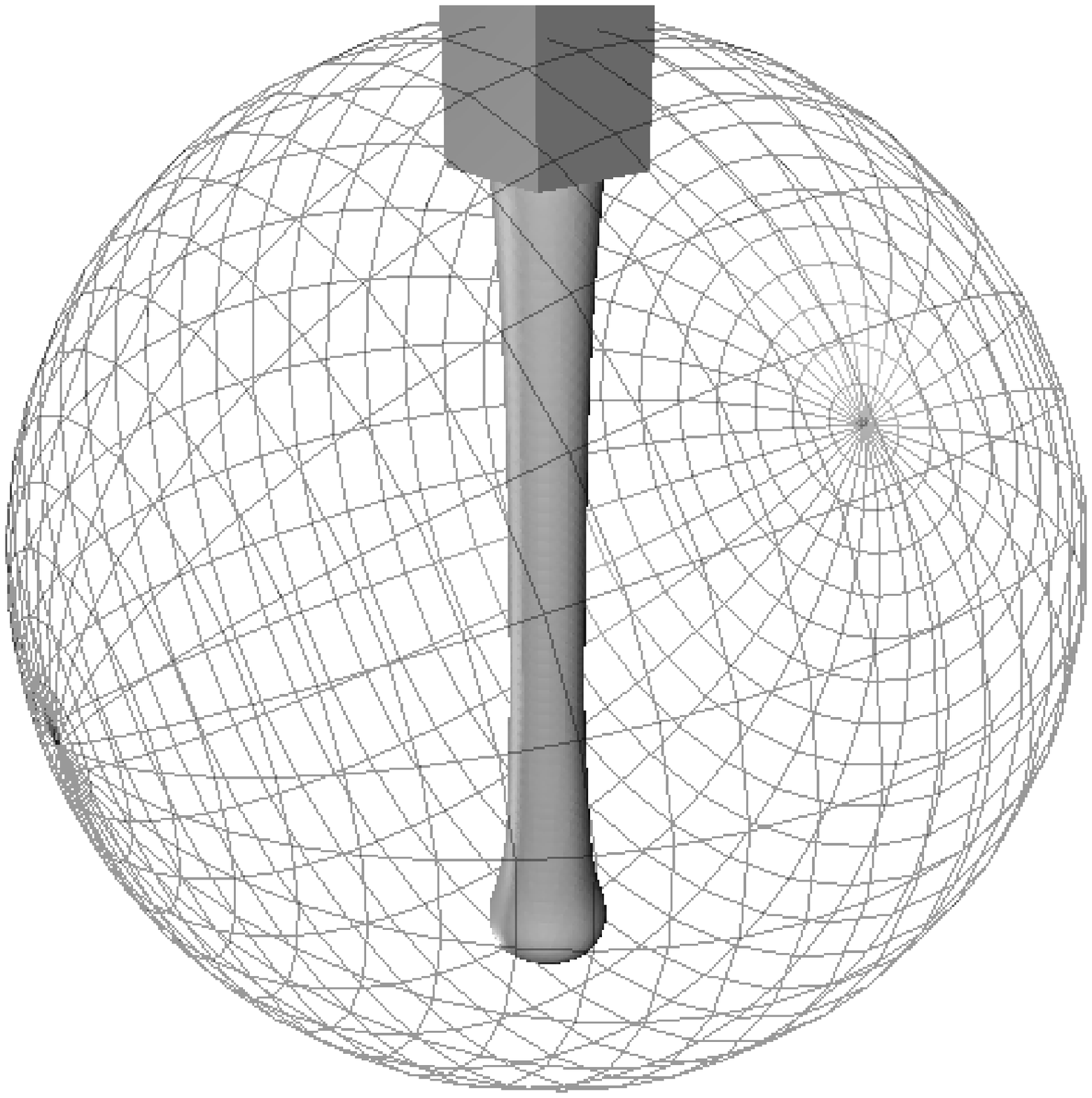}
}
\subfigure[]
{
\includegraphics[width=40mm, clip = true, trim=43mm 0mm 43mm 0mm]{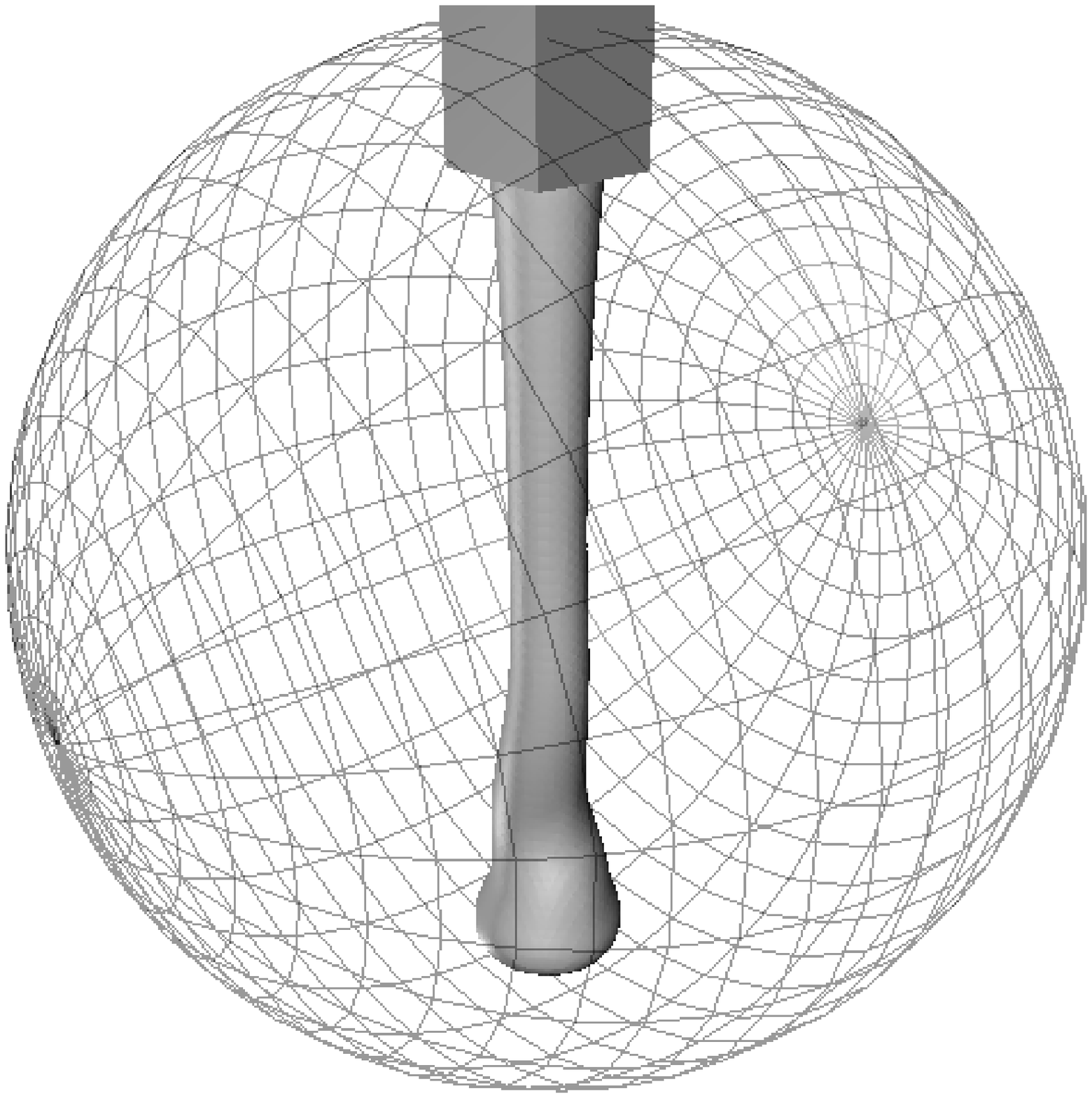}
}
\subfigure[]
{
\includegraphics[width=40mm, clip = true, trim=43mm 0mm 43mm 0mm]{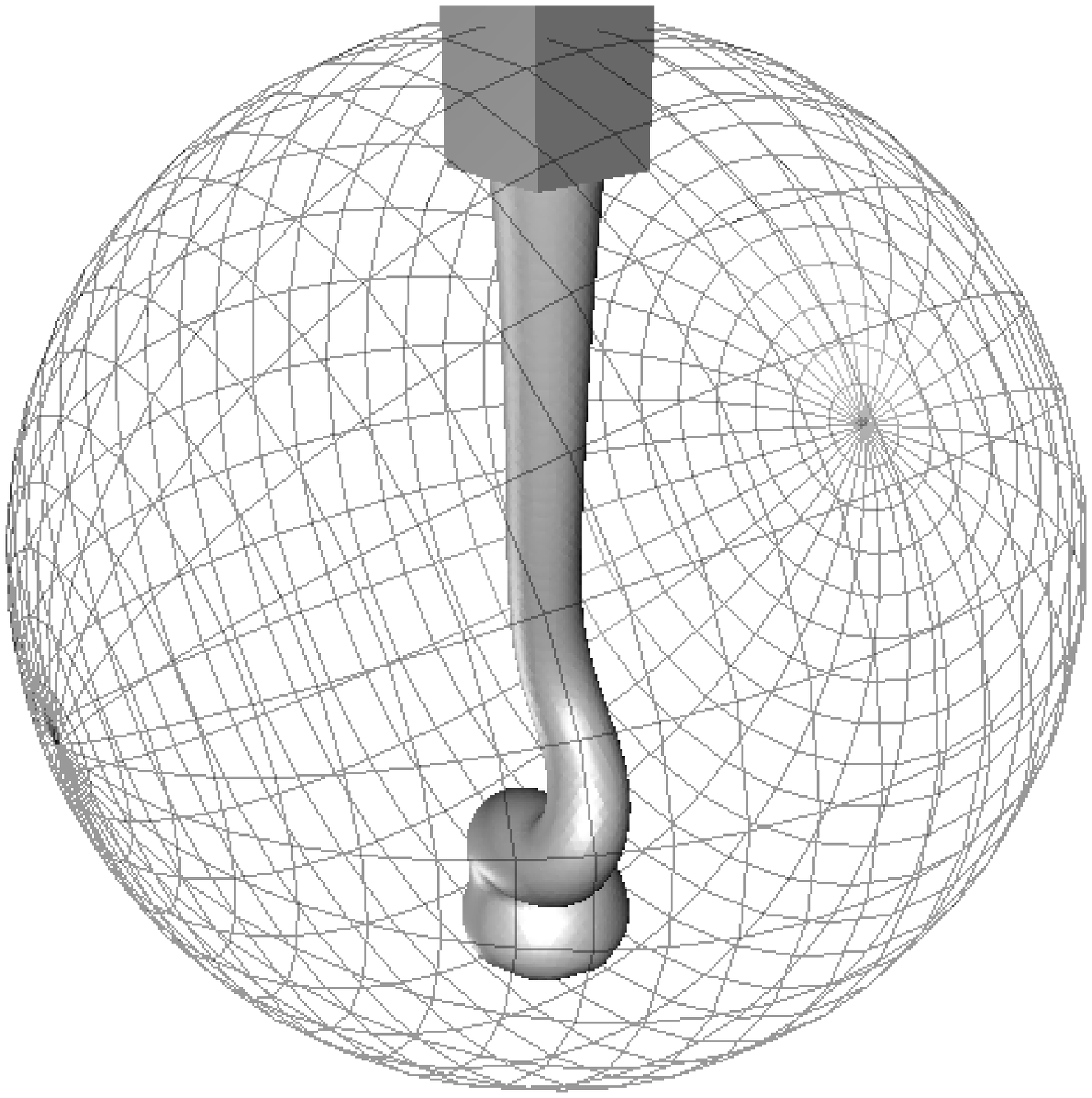}
}
\subfigure[]
{
\includegraphics[width=40mm, clip = true, trim=43mm 0mm 43mm 0mm]{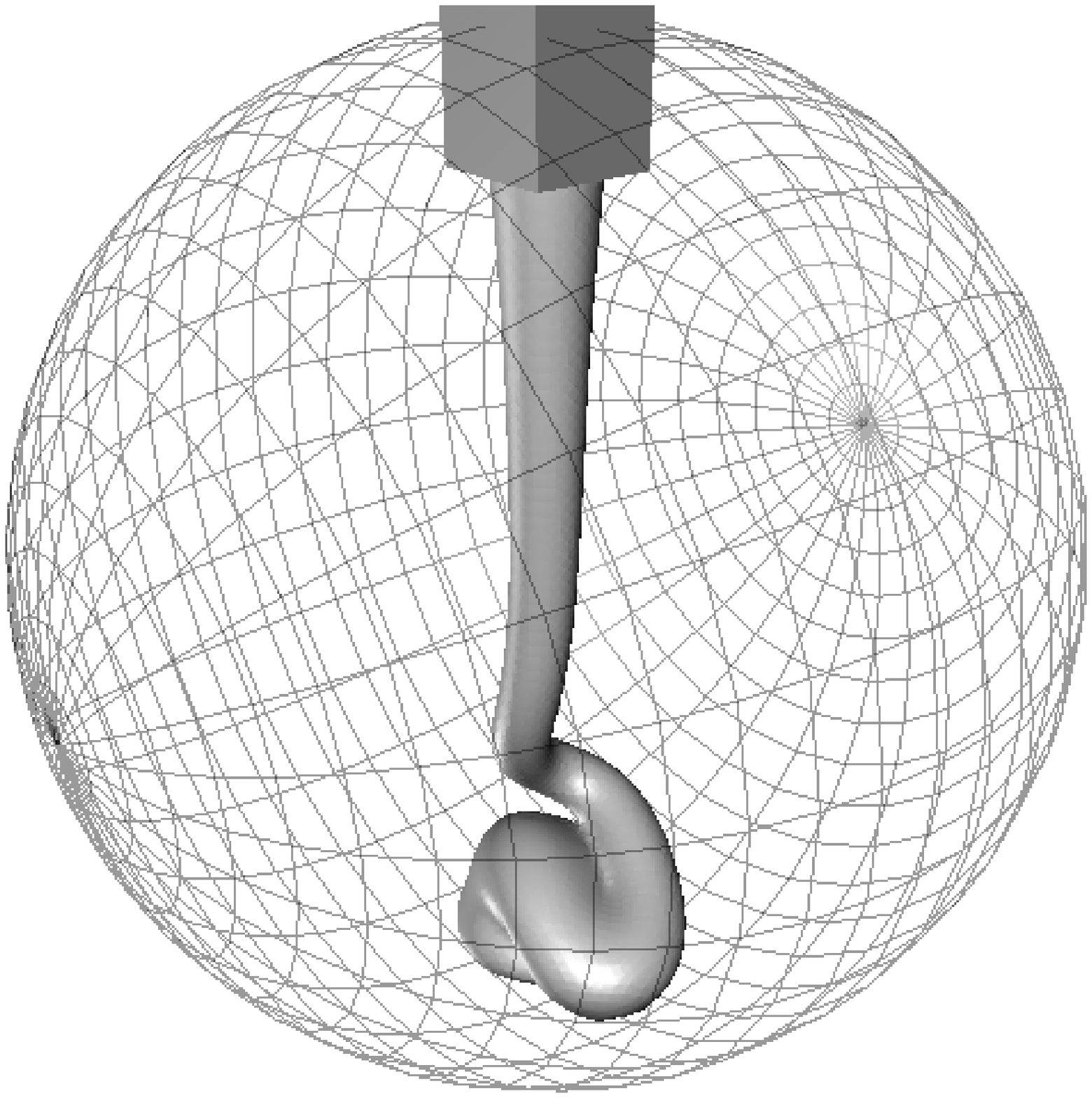}
}
\subfigure[]
{
\includegraphics[width=40mm, clip = true, trim=43mm 0mm 43mm 0mm]{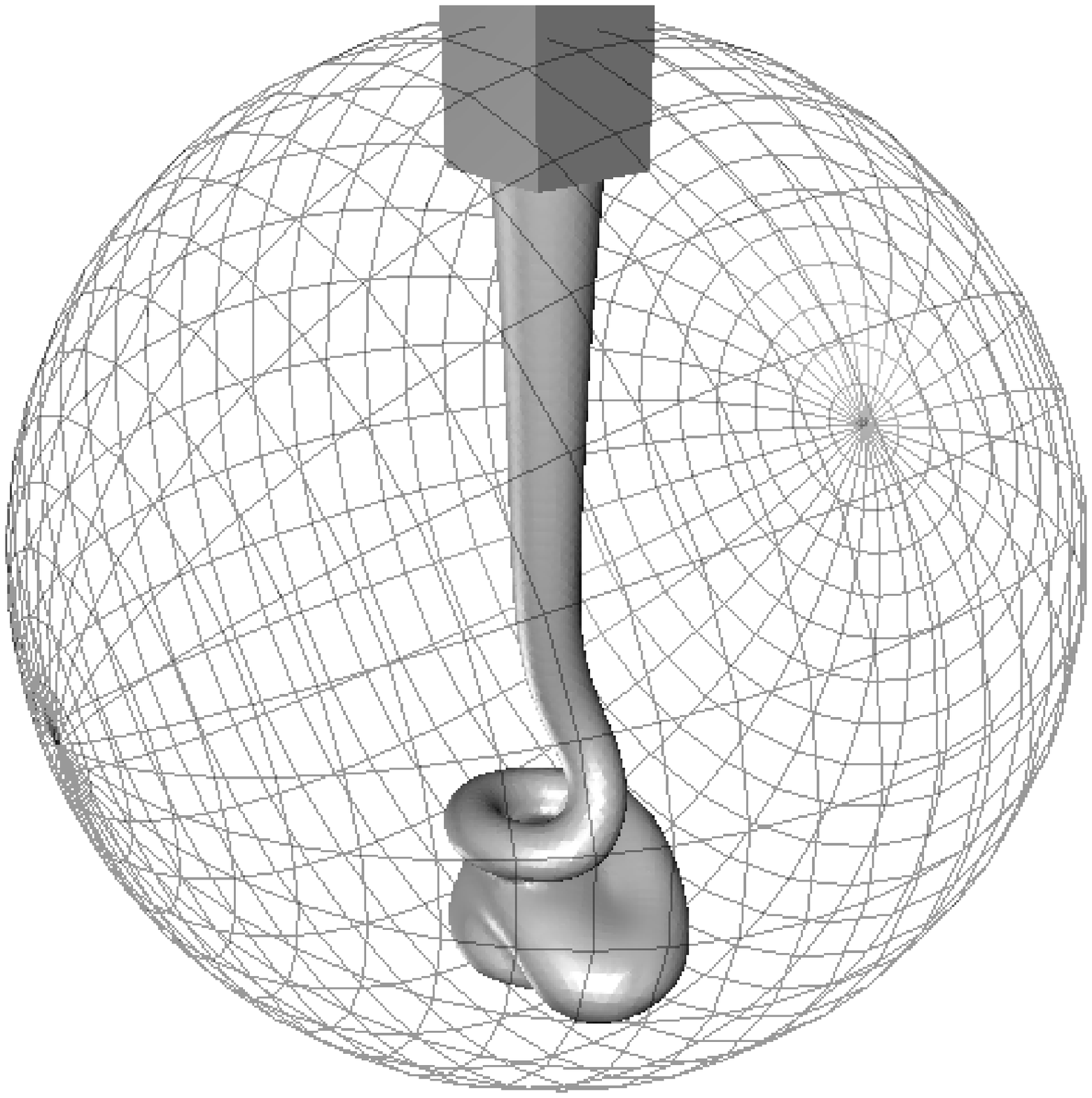}
}
\subfigure[]
{
\includegraphics[width=40mm, clip = true, trim=43mm 0mm 43mm 0mm]{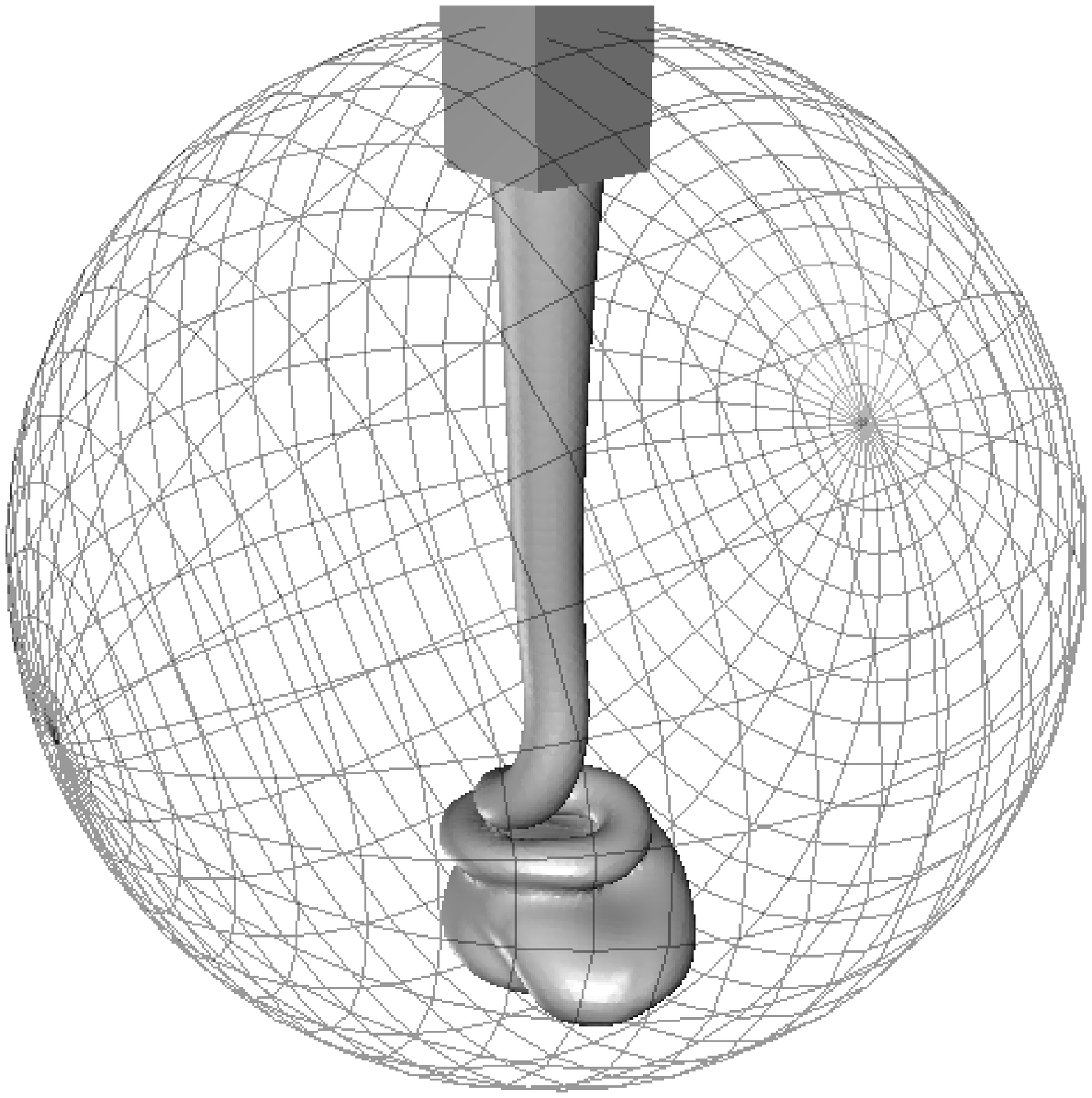}
}
\subfigure[]
{
\includegraphics[width=40mm, clip = true, trim=43mm 0mm 43mm 0mm]{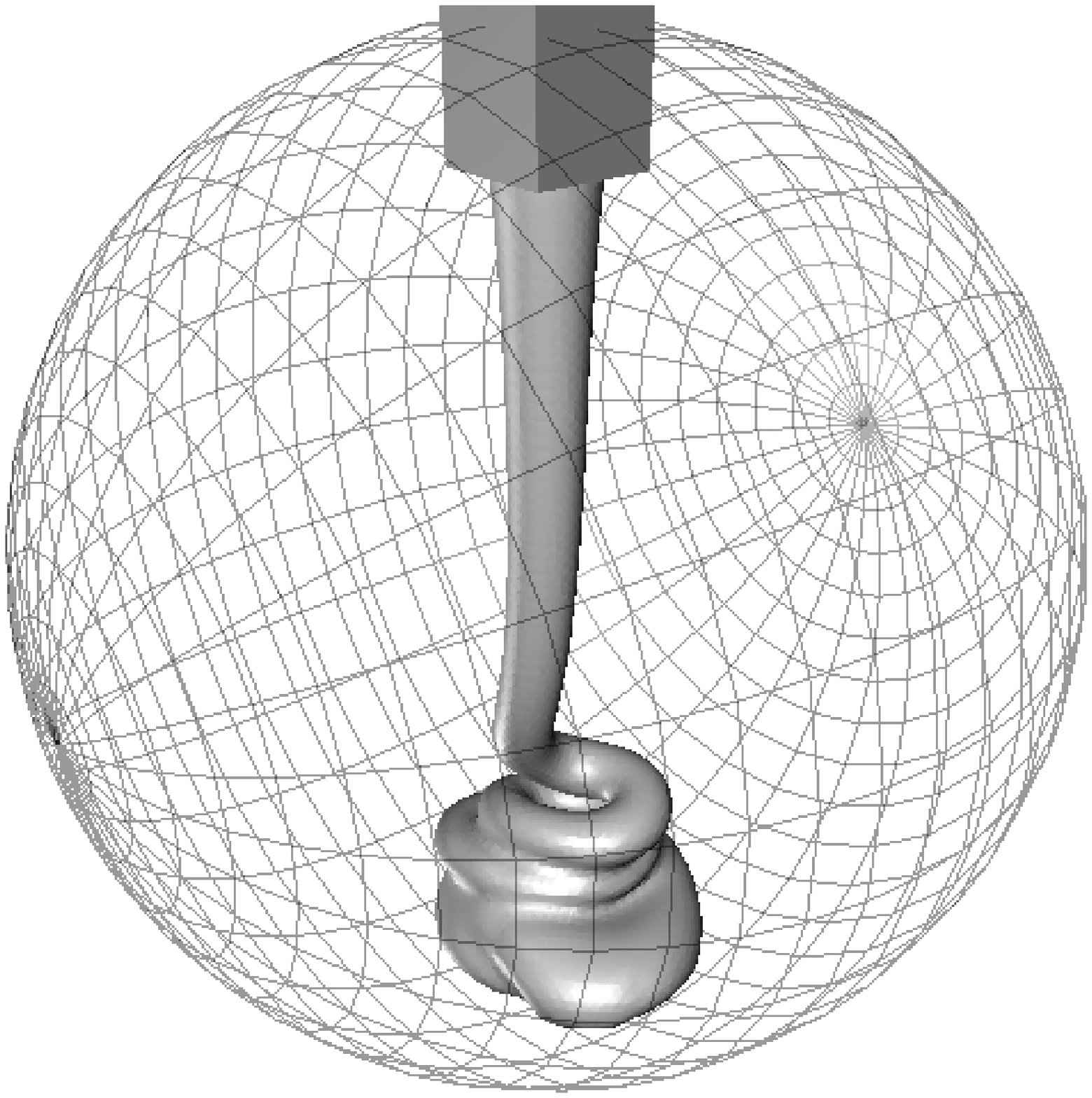}
}
\caption{A three-dimensional example of viscous jet buckling performed using our simple Navier-Stokes routine. The first image occurs 0.6 seconds into the simulation, and subsequent frames occur at 0.3 second intervals.  Additional images are shown in Figure \ref{fig:buckling3D_cont}.}
\label{fig:buckling3D}
\end{figure}

\begin{figure}
\centering
\subfigure[]
{
\includegraphics[width=40mm, clip = true, trim=43mm 0mm 43mm 0mm]{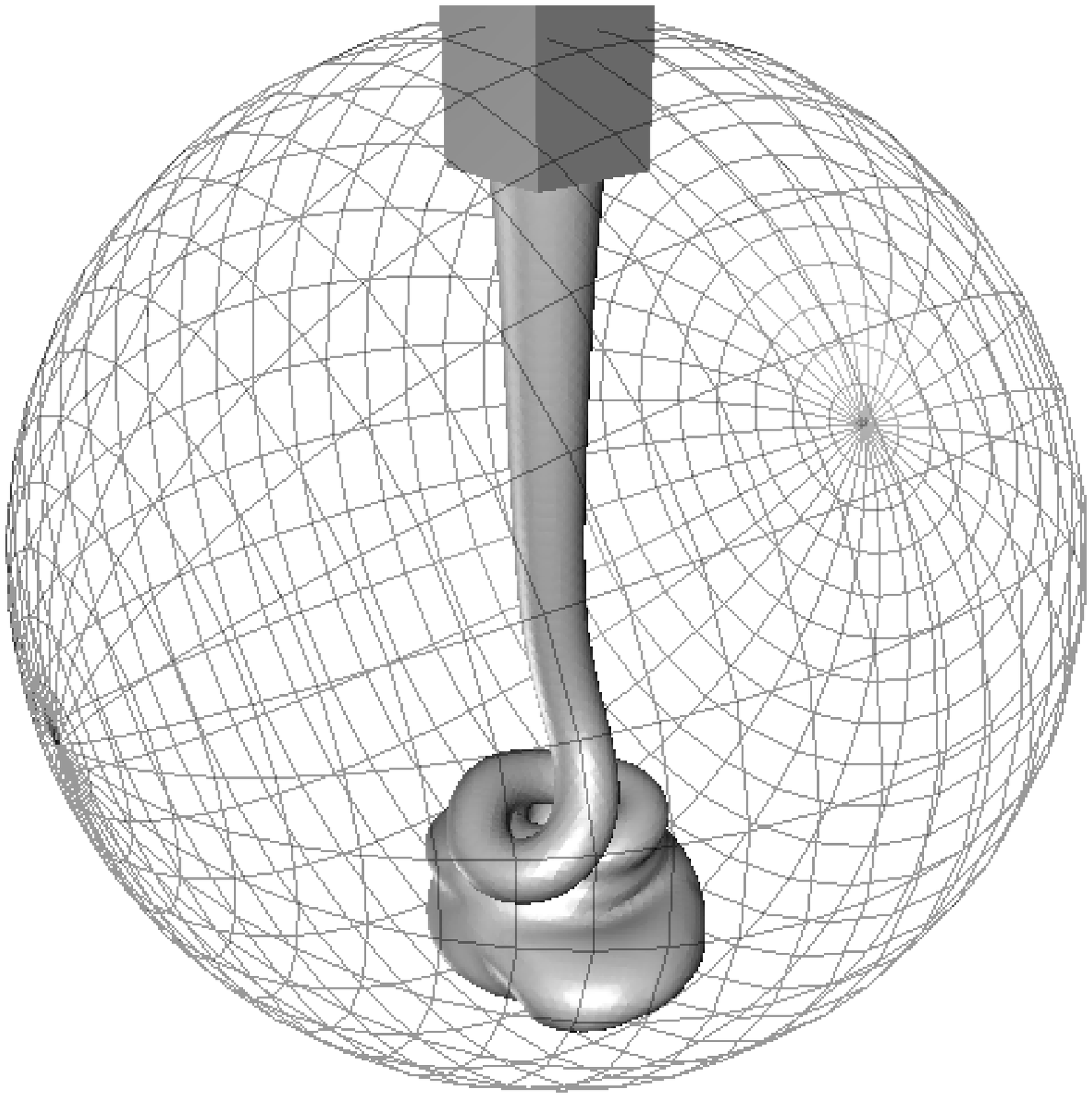}
}
\subfigure[]
{
\includegraphics[width=40mm, clip = true, trim=43mm 0mm 43mm 0mm]{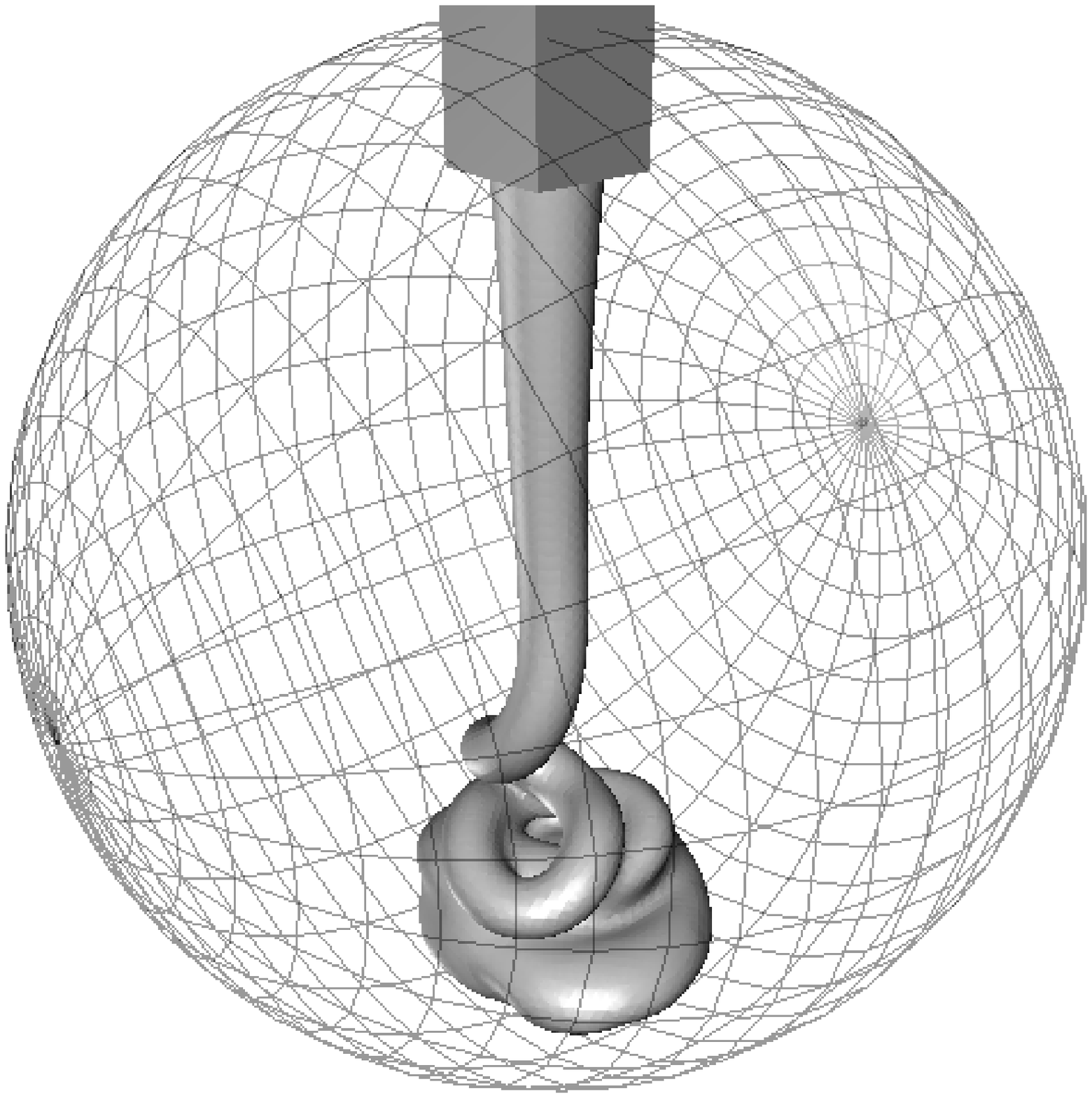}
}
\subfigure[]
{
\includegraphics[width=40mm, clip = true, trim=43mm 0mm 43mm 0mm]{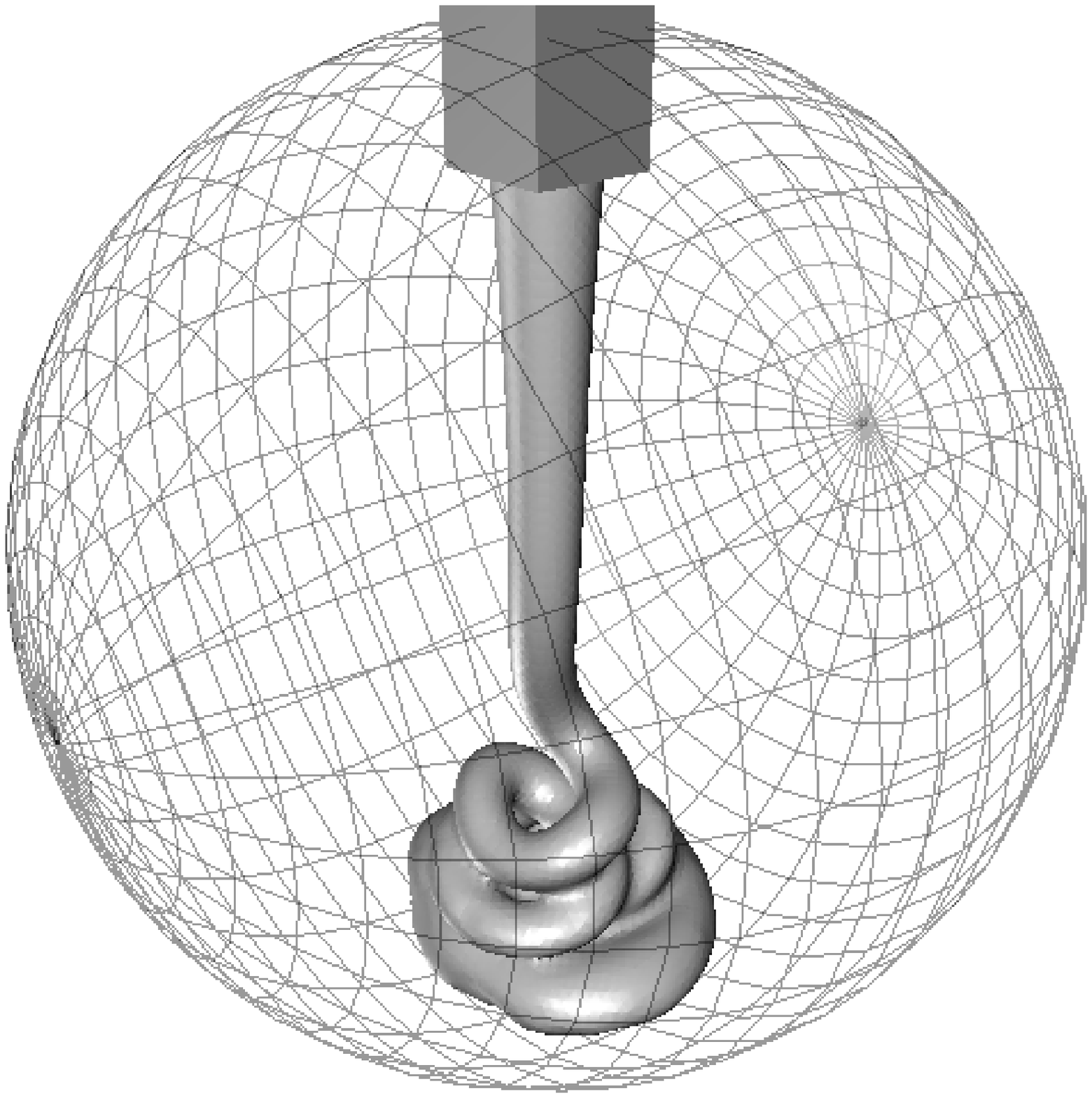}
}
\subfigure[]
{
\includegraphics[width=40mm, clip = true, trim=43mm 0mm 43mm 0mm]{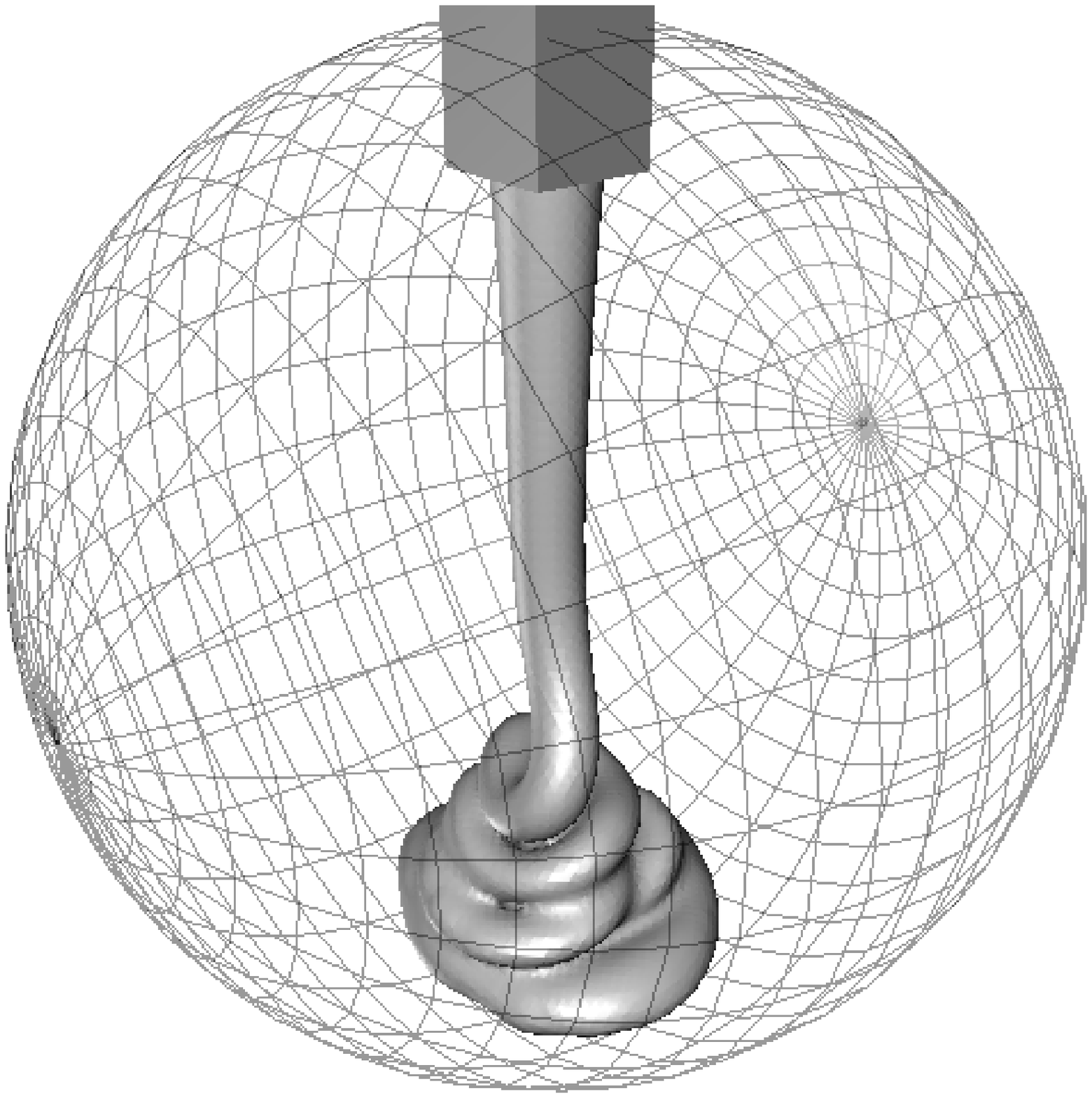}
}
\subfigure[]
{
\includegraphics[width=40mm, clip = true, trim=43mm 0mm 43mm 0mm]{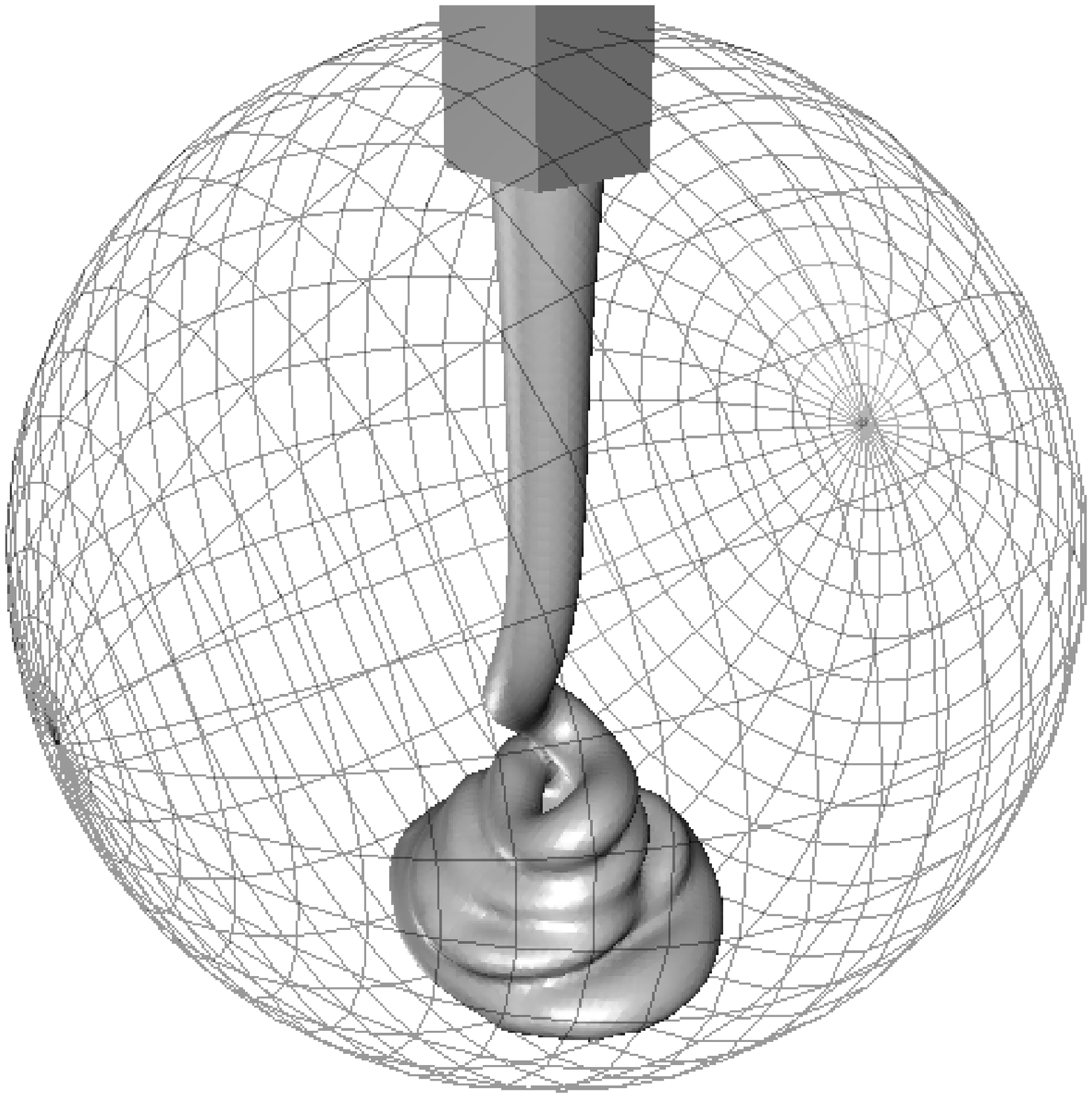}
}
\subfigure[]
{
\includegraphics[width=40mm, clip = true, trim=43mm 0mm 43mm 0mm]{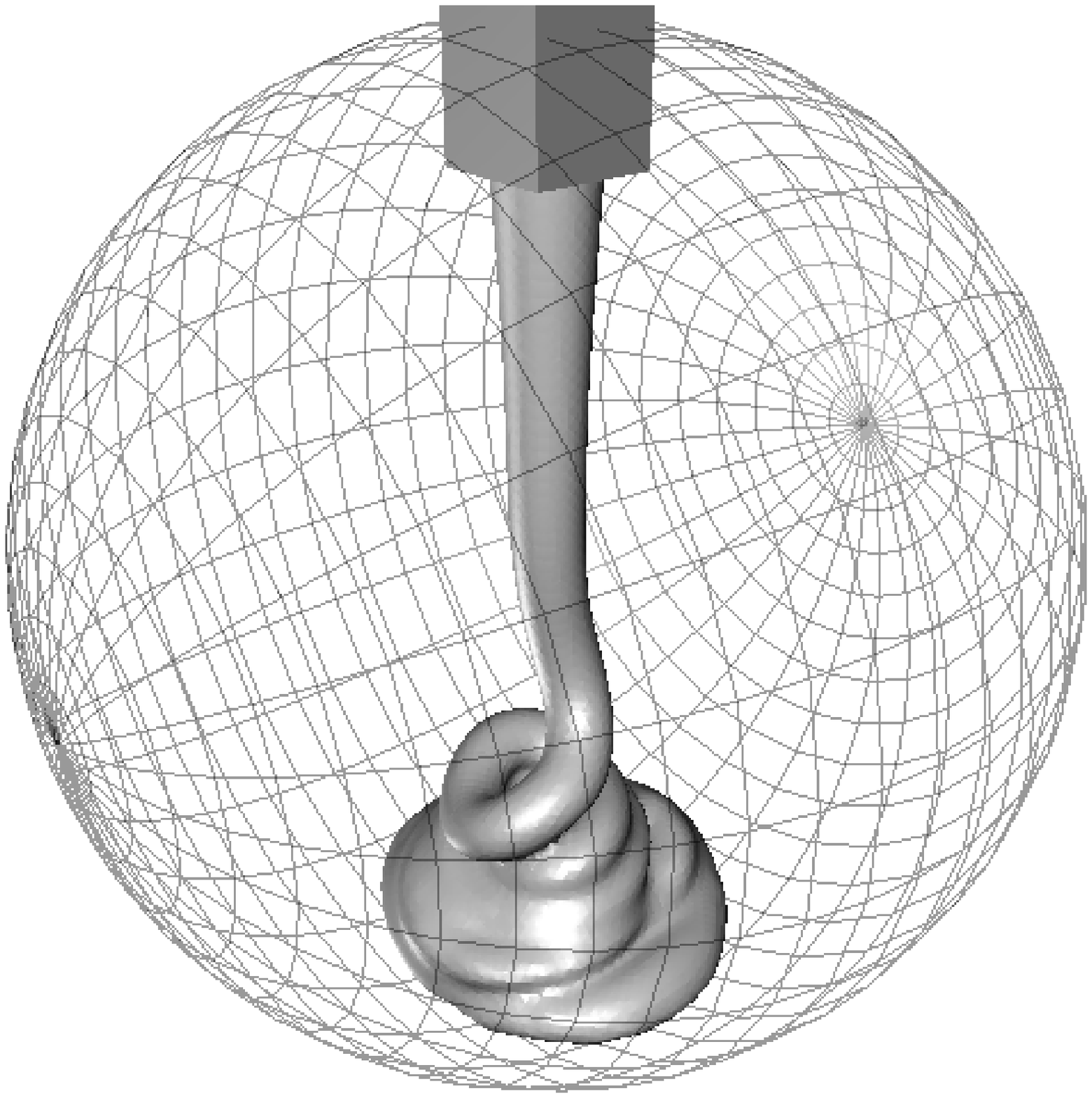}
}
\subfigure[]
{
\includegraphics[width=40mm, clip = true, trim=43mm 0mm 43mm 0mm]{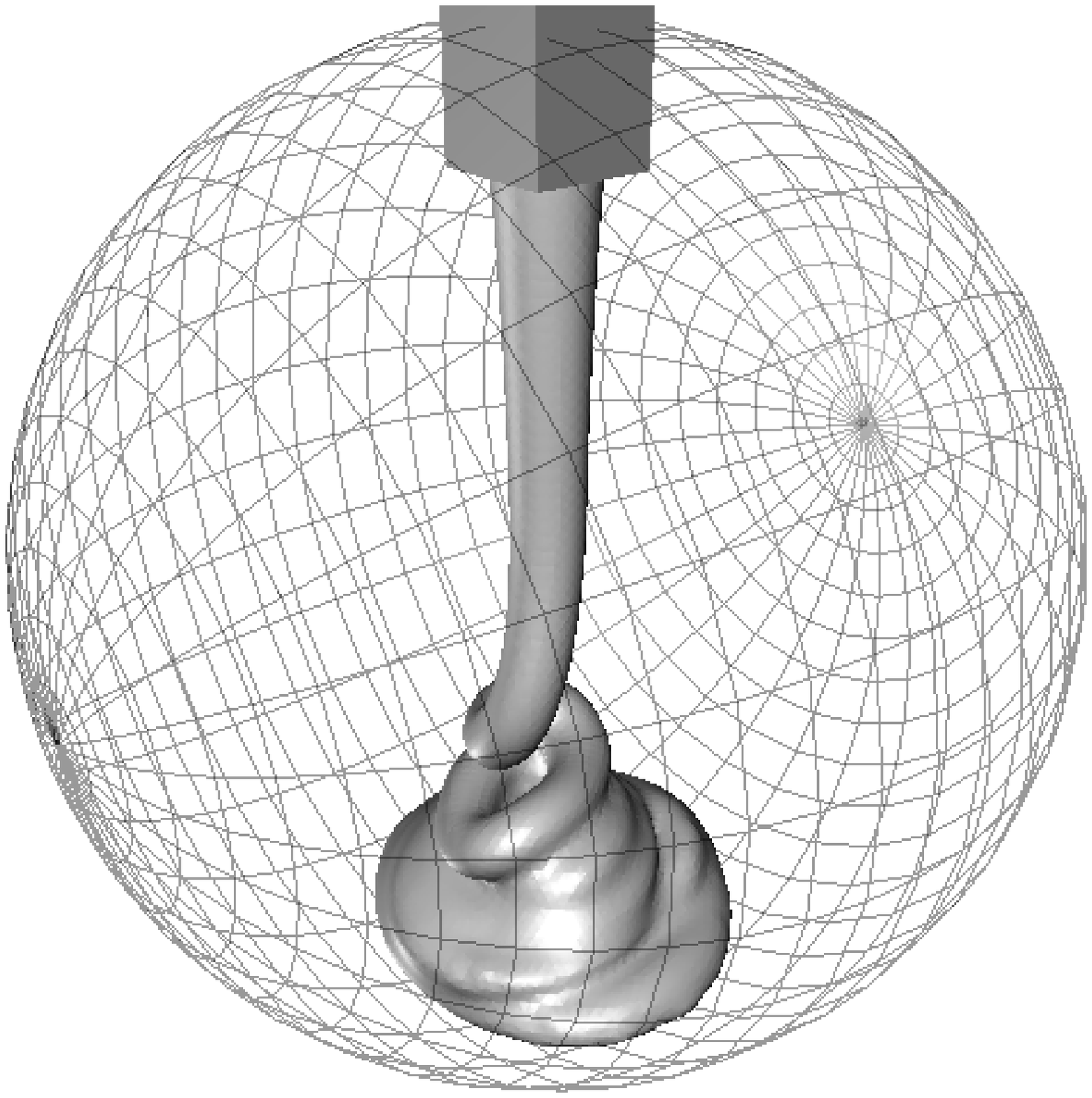}
}
\subfigure[]
{
\includegraphics[width=40mm, clip = true, trim=43mm 0mm 43mm 0mm]{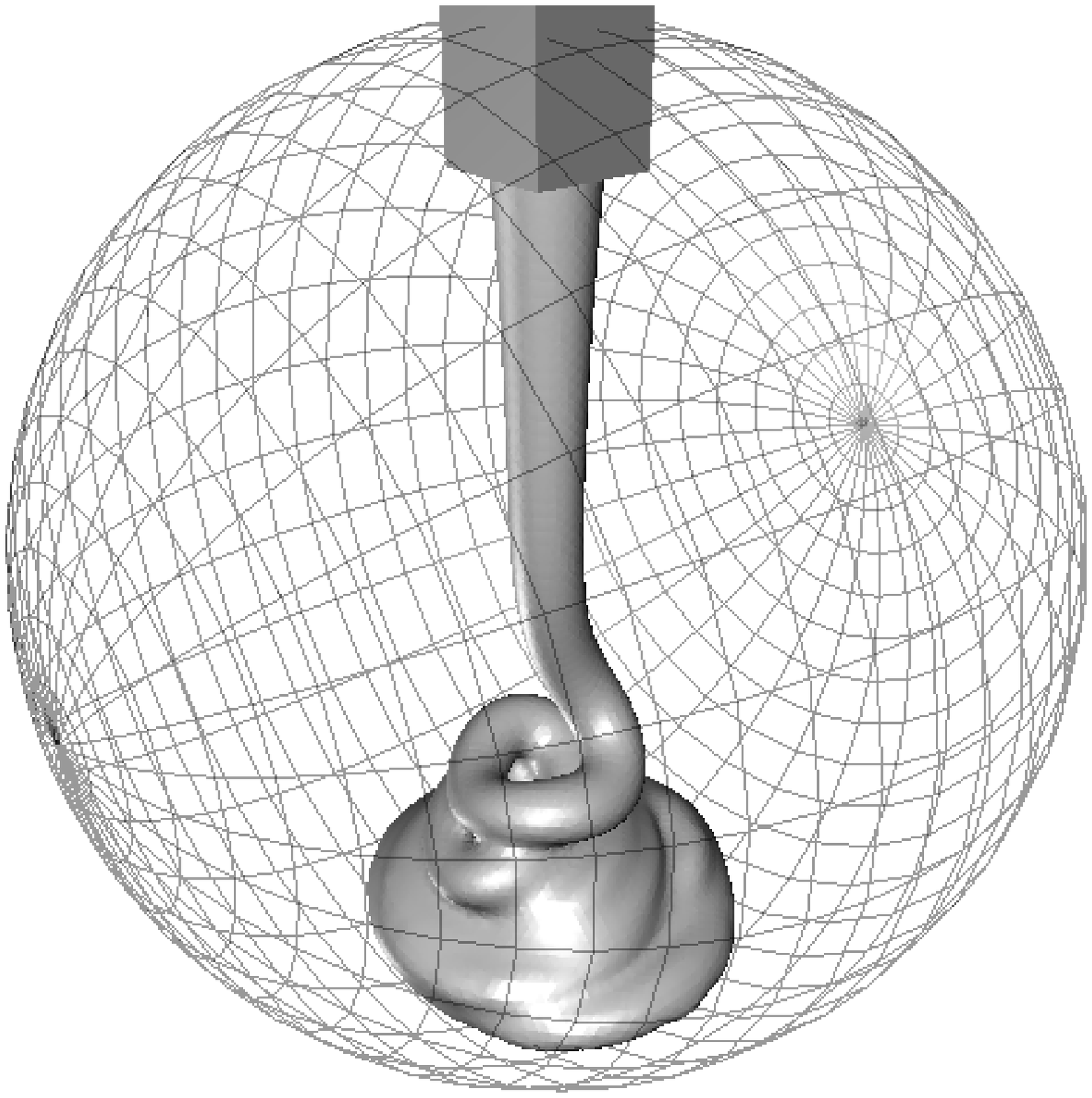}
}
\subfigure[]
{
\includegraphics[width=40mm, clip = true, trim=43mm 0mm 43mm 0mm]{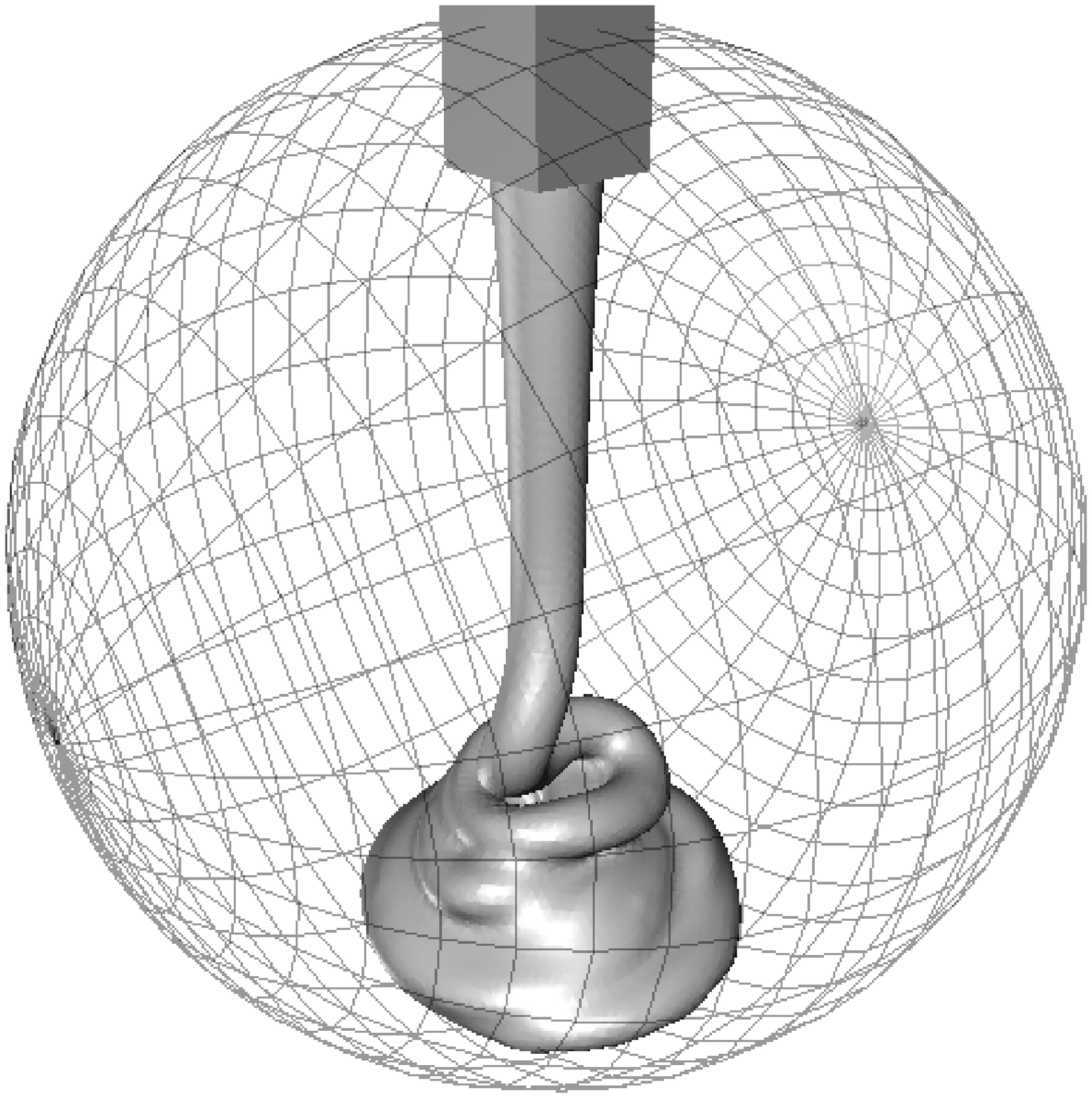}
}
\caption{Additional images of viscous jet buckling, continued from Figure~\ref{fig:buckling3D}.}
\label{fig:buckling3D_cont}
\end{figure}

Figures \ref{fig:buckling3D} and \ref{fig:buckling3D_cont} present the results of a three-dimensional simulation of cylindrical viscous jet buckling (ie. coiling).  The simulation domain is a sphere of radius $0.4 [m]$ centred at $(0.5,0.5,0.5)$.  A circular inlet is centred at $(0.5,0.8,0.5)$, with a fixed vertical velocity of $U=-0.5 [m/s]$ and a diameter $D=0.08 [m]$.  This configuration yields a drop height of $H=0.7 [m]$.  The density of the liquid is $\rho = 1 [kg/m^3]$ and the dynamic viscosity of the liquid is $\mu = 0.3 [Pa\cdot s]$. Gravity is set at $-9.81 [m/s^2]$.  The simulation grid used a resolution of $80\times80\times80$ cells.

This yields a Reynolds number of $Re = \frac{\rho D U}{\mu} \approx 0.133$ and an aspect ratio for the liquid jet of $H / D = 0.7 / 0.08 = 8.75$.  This falls within the guidelines for when axisymmetric buckling typically arises ($Re < 1.2, H/D > 7$) according to Cruikshank and Munson\cite{Cruikshank1981,Cruikshank1988}, and the result does indeed exhibit substantial buckling.

\section{Conclusions and Future Work}
\label{conclusion}

We have shown that a Cartesian grid finite difference method derived from a variational principle can correctly capture difficult irregular boundary conditions in Stokes flow problems, while providing stability for large time steps and yielding a sparse, symmetric positive-definite linear system. To do so, we have unified and extended recent work on embedded boundary methods for pressure projection and viscosity. In practice the method's implementation is remarkably simple, yet it correctly captures the challenging free surface boundary condition that enables simulation of jet buckling phenomena. This work raises a number of questions and directions for future work.

In our numerical study, we observed improved $L^\infty$ convergence in 3D compared to 2D, and plan to investigate if this truly holds---perhaps beginning by deriving a full-fledged analytic test case as we have done in 2D. The 2D convergence test cases we presented also considered only scenarios where the two different boundary conditions (solid and free surface) do not meet.  We suspect that this is an inherently more difficult problem to address, giving rise to issues analogous to those which occur in the presence of sharp boundary features, and our preliminary experiments support this conjecture. Nevertheless, such configurations occur frequently in the buckling examples we have included, illustrating that the method remains stable and provides qualitatively reasonable results.

While the viscous jet buckling example provides a practical validation of our method's boundary condition enforcement, the current underlying Navier-Stokes simulator is fairly basic. A thorough study of this phenomenon would likely need to consider improved advection and time-splitting methods in place of the first order approaches applied here.

In terms of handling related phenomena, we have noted that our work is closely related to that of Batty et al.\ who considered considered the simpler Poisson problem for incompressibility with rigid bodies \cite{Batty2007}. An obvious extension of the current work would therefore be to consider Stokes flow coupled with fully dynamic deformable structures. Two-phase flow would also be a useful direction to pursue, as would non-Newtonian fluid models. Finally, it would be interesting to consider whether exploiting variational principles in this manner might be useful for handling irregular boundaries in other problems that are commonly discretized on staggered grids. Some potential examples include vorticity-based formulations of fluid flow, diffusion problems, or porous flow.

\appendix

\section{Notation}
\label{symbols}

The following symbols and letters are used throughout the paper, with units given for dimensionful quantities:

\begin{list}{\quad}{\setlength{\topsep}{0ex}\setlength{\parsep}{0ex}\setlength{\itemsep}{0.5ex}}
\item $\mu$ : coefficient of dynamic viscosity, $Pa \cdot s$
\item $\rho$ : density coefficient, $kg/m^3$
\item $\Omega_F$ : fluid domain
\item $\Omega_S$ : solid domain (the complement of $\Omega_F$)
\item $\Omega_L$ : liquid domain
\item $\Omega_A$ : air domain (the complement of $\Omega_L$)
\item $\vec{u}$ : velocity vector, $m/s$
\item $p$ : pressure, $Pa$
\item $\tau$ : deviatoric stress tensor, $Pa$
\item $\Delta t$ : time step, $s$
\item $\vec{T}$ : traction vector, $Pa$
\item $D$ : discrete deformation rate matrix
\item $G$ : discrete gradient matrix
\item $M$ : diagonal matrix of viscosity coefficients, per velocity sample
\item $P$ : diagonal matrix of density coefficients, per pressure/stress sample
\item $W_F$ : diagonal matrix of fluid (non-solid) fraction weights
\item $W_S$ : diagonal matrix of solid fraction weights, complementing $W_L$
\item $W_L$ : diagonal matrix of liquid (non-air) fraction weights
\item $W_A$ : diagonal matrix of air fraction weights, complementing $W_L$
\item $W^u, W^p, W^\tau$ : superscripts on weight matrices indicate the associated sample position. ie. velocity $u$ (cell faces), pressure $p$ (cell centres), stresses $\tau$ (cell centres and edges).
\end{list}

\bibliographystyle{siam}
\bibliography{vfdstokes}

\end{document}